\documentclass[twocolumn]{aastex701}
\usepackage{amsmath}	
\usepackage{tikz}     

\begin{document}

\title{Eclipse Timing Variations of Circumbinary Substellar Objects in TESS Data}

\author[0000-0002-6191-459X]{Ekrem M. Esmer}
\affiliation{Department of Physics and McDonnell Center for the Space Sciences, Washington University, St. Louis, MO 63130, USA}
\affiliation{Department of Astronomy and Space Sciences, Faculty of Science, Ankara University, 06100 Ankara, T\"urkiye}
\email[show]{ekrem@wustl.edu}
\email[show]{ekremmuratesmer@gmail.com}

\author[0000-0002-6939-9211]{Tansu Daylan}
\affiliation{Department of Physics and McDonnell Center for the Space Sciences, Washington University, St. Louis, MO 63130, USA}
\email{tansu@wustl.edu}

\begin{abstract}
Circumbinary planets and brown dwarfs form in complex gravitational environments, offering insights into formation, orbital stability, and habitability prospects. However, they remain underrepresented, with only 60 confirmed or candidate systems known. In this work, we leverage TESS photometry to search for circumbinary companions through eclipse timing variations (ETVs), analyzing 152 detached eclipsing binaries. By modeling eclipse timings, we identify 37 systems with significant periodic signals, 19 of which have false alarm probabilities below 0.01.
One system, TIC 142979644, emerges as a promising candidate to host a circumbinary substellar companion, with estimated masses of $18.8~M_{\rm J}$ and $11.1~M_{\rm J}$ from different methods.
Simulations using synthetic ETVs indicate a 5\% recovery rate for circumbinary brown dwarfs and 0.1\% for Jupiter-like planets, with median masses of $56.6^{+16.5}_{-23.4}~M_{\rm J}$ and periods of $1404^{+1361}_{-953}$ days. Our simulations show that the smallest detectable mass is $1.6~M_{\rm J}$ at a period of 1860 days and confirm that ETV methods are effective in detecting misaligned systems. In the absence of a detection, we set an upper limit of 40\% on the occurrence rate of circumbinary brown dwarfs at the 2$\sigma$ confidence level. In contrast, a confirmed single detection would imply an occurrence rate of 13.08\%. These constraints are consistent with previous abundance estimates for circumbinary brown dwarfs ($\lesssim6.5\%$).
As most circumbinary substellar companions detected through ETVs are found around post-common envelope binaries, our recovery rate of 0.83\% in their progenitors implies that even a single detection would strongly favor a first-generation origin.
\end{abstract}

\keywords{Eclipsing binary stars (444), Timing variation methods (1703), Exoplanet detection methods (489), Exoplanets (498), Brown dwarfs (185)}

\section{Introduction} \label{sec:intro}


Circumbinary planets are fascinating due to their unique formation mechanisms \citep{Marzari2019Galax...7...84M, Coleman2024MNRAS.527..414C}, dynamical evolution \citep{Correia2016CeMDA.126..189C,Hamers2016MNRAS.455.3180H}, and potential for habitability \citep{Haghighipour2013ApJ...777..166H}. Despite the growing number of confirmed exoplanets, now exceeding $\sim$5800, circumbinary planets and brown dwarfs remain rare, with only 60 confirmed or candidate systems \citep[NASA Exoplanet Archive\footnote{\url{https://exoplanetarchive.ipac.caltech.edu/}}]{Basturk2023BSRSL..9211197B, Esmer2024CoSka..54b.228E}. This disparity raises questions about their abundances and detectability \citep{Armstrong2014MNRAS.444.1873A, Fleming2018ApJ...858...86F}. Despite their small numbers, the current circumbinary population spans a wide range of masses, from brown dwarfs \citep{Benedict2017AJ....153..258B} to Earth-sized planets \citep{Orosz2019AJ....157..174O}.

Circumbinary systems are also of significant theoretical interest, as they challenge traditional models of planetary formation and stability. This is particularly evident in post-common-envelope binary (PCEB) systems, where planets may form as first- or second-generation objects \citep{Zorotovic2013A&A...549A..95Z, Bear2014MNRAS.444.1698B}. Dynamical studies reveal their complex orbital behaviors, including constraints on stability, dynamical evolution, and inclination effects, which influence both the survival of planets and their detectability \citep{Foucart2013ApJ...764..106F, Chavez2015MNRAS.446.1283C, Correia2016CeMDA.126..189C, Hamers2016MNRAS.455.3180H, Quarles2018ApJ...856..150Q, Chen2023MNRAS.521.5033C, Chen2024ApJ...961L...5C, Wang2024AJ....168...31W}. Additionally, circumbinary systems may contribute to the population of free-floating planets due to planet-ejecting orbital instabilities \citep{Coleman2024MNRAS.530..630C}.

Although terrestrial planets largely underrepresent the current population of circumbinary planets, these systems remain compelling for habitability studies. Their unique orbital dynamics often result in complex insolation patterns, which influence their potential to host life \citep{Haghighipour2013ApJ...777..166H, Kane2013ApJ...762....7K, Simonetti2020ApJ...903..141S, Graham2021A&A...650A.178G}.

While other detection techniques, such as radial velocity \citep{Standing2023NatAs...7..702S} and microlensing \citep{Bennett2016AJ....152..125B}, are increasingly effective in identifying circumbinary planets, the discovery of these systems has primarily relied on two methods: transit photometry and eclipse timing variations (ETVs). Together, these techniques have contributed to the detection of 31 circumbinary planets, 17 of which were identified using ETVs. These include well-known systems such as Kepler-16 \citep{Doyle2011Sci...333.1602D}, Kepler-34 and Kepler-35 \citep{Welsh2012Natur.481..475W}, TOI-1338 \citep{Kostov2020AJ....159..253K}, Kepler-451 \citep{Baran2015, Esmer2022MNRAS.511.5207E} and Kepler-1660 \citep{Goldberg2023MNRAS.525.4628G}. A full list of circumbinary planet discoveries and their corresponding references is provided in Table \ref{tab:cbp_ref}. ETVs are particularly advantageous for detecting massive companions around eclipsing binaries, as they measure variations in stellar eclipse timings caused by the gravitational influence of unseen objects. Their effectiveness in detecting circumbinary brown dwarfs and planets has been tested by some studies \citep{Getley2021MNRAS.504.4291G, Sybilski2010MNRAS.405..657S}. Precise determination of mid-eclipse times is critical for ETV analyses, requiring high-accuracy measurements \citep{Mikulasek2014CoSka..43..382M, Deeg2017A&A...599A..93D, Deeg2020Galax...9....1D, Esmer2022MNRAS.511.5207E}.

Efforts to detect circumbinary objects employ various methods, including transit searches \citep{Windemuth2019MNRAS.490.1313W}, direct imaging \citep{Bonavita2016A&A...593A..38B}, and radial velocity \citep{Martin2019A&A...624A..68M}. Additionally, significant efforts focus on searches utilizing ETVs, which remain a prominent tool in circumbinary research \citep{Pribulla2012AN....333..754P, Borkovits2015MNRAS.448..946B, Bours2016MNRAS.460.3873B, Borkovits2016MNRAS.455.4136B, Papageorgiou2021MNRAS.503.2979P, Basturk2023BSRSL..9211197B, Esmer2024CoSka..54b.228E, Mitnyan2024A&A...685A..43M, Moharana2024MNRAS.527...53M}.

The advent of space-based observatories such as TESS \citep{Ricker2015JATIS...1a4003R} has significantly advanced the search for circumbinary planets. With its high photometric precision and extensive sky coverage, TESS has become an indispensable tool for detecting circumbinary systems through both transits \citep{Guerrero2021ApJS..254...39G} and ETVs \citep{Marcadon2024arXiv240307694M}.

\begin{figure*}
    \centering
    \includegraphics[width=1.9\columnwidth]{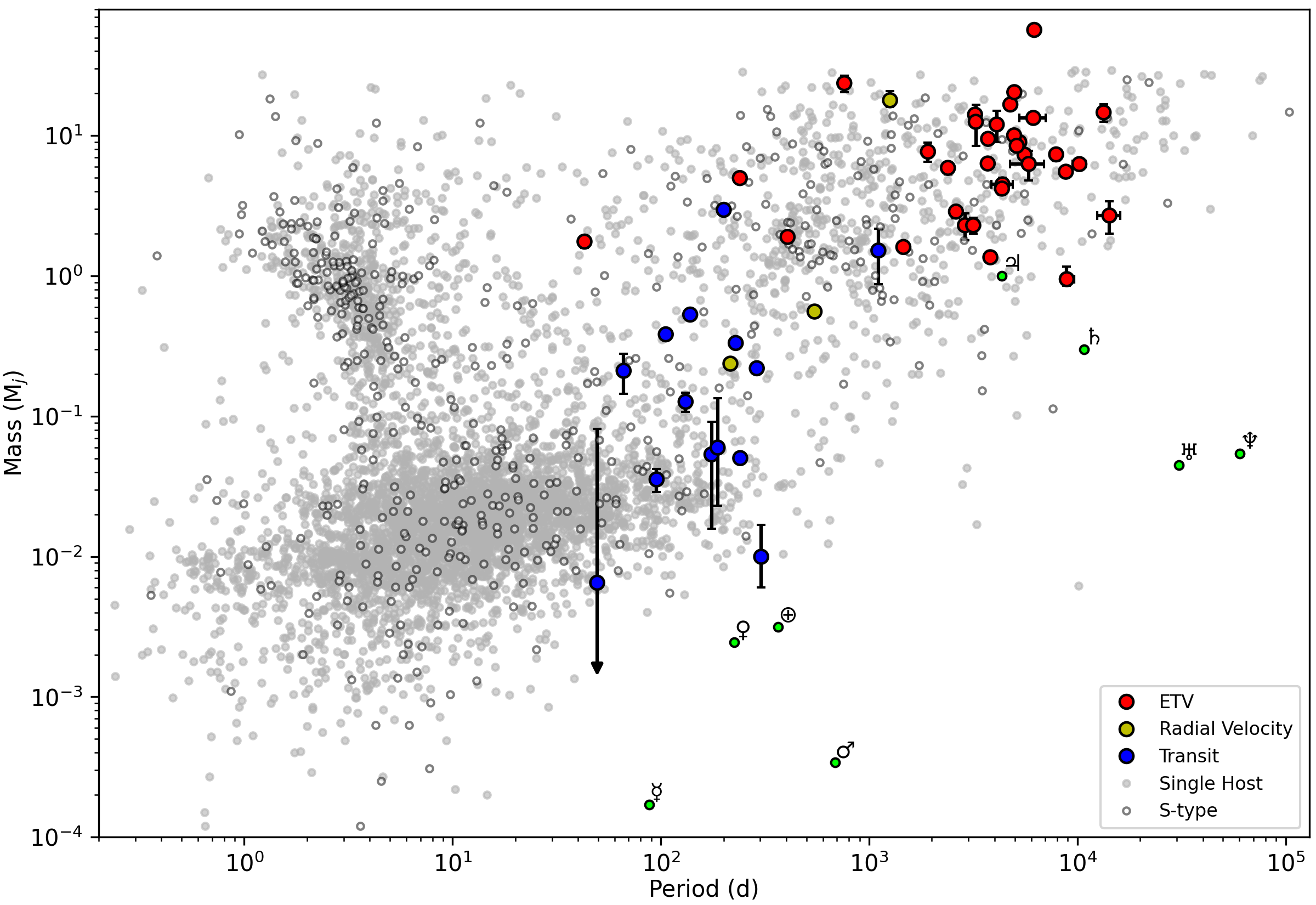}
    \caption{Mass (either projected or true) vs orbital periods of circumbinary (p-type) brown dwarf and planet population derived from NASA Exoplanet Archive, \cite{Basturk2023BSRSL..9211197B} and \cite{Esmer2024CoSka..54b.228E} including candidates along with i) all known single host planets, ii) planets with s-type orbits in multiple star systems and iii) planets of the Solar system (green markers).}
    \label{fig:cb_m_vs_p}
\end{figure*}

Previous studies on the abundance of circumbinary planets have primarily relied on transit \citep{Armstrong2014MNRAS.444.1873A, Martin2014A&A...570A..91M} and radial velocity methods \citep{Martin2019A&A...624A..68M}, which are more sensitive to shorter-period planets. In contrast, the ETV method is better suited for detecting massive companions on wider orbits, mainly through the light-travel time effect (LiTE). Figure \ref{fig:cb_m_vs_p} illustrates how circumbinary gas giants detected via different methods occupy distinct regions in mass-period space, demonstrating the unique role of ETVs in probing long-period systems. Given this sensitivity, determining occurrence rates using the ETV method provides crucial constraints on the population of massive circumbinary companions that may otherwise remain undetected.

With these perspectives, we conduct an ETV-based search for circumbinary substellar objects to characterize their properties and assess their detectability. We model ETVs for a target sample to detect potential circumbinary hosts and provide insights into ETV properties. By running simulations, we determine the efficiency of the ETV method for different system architectures and provide constraints on the population of circumbinary brown dwarfs and planets.
This study is an initial step toward a comprehensive analysis of ETV detectability and the statistical occurrence of substellar-mass circumbinary objects.

\section{Methodology}\label{sect:Methodology}

We aim to model and analyze ETVs in binary systems, first by characterizing ETVs within our target sample and then by simulating eclipsing binaries with circumbinary companions to assess detection rates.

We begin by selecting binary systems based on eclipse features for precise timing analysis, followed by processing light curves to ensure high data quality. Next, we model the primary eclipses in each target to estimate mid-eclipse times and then analyze the ETVs to detect periodicities within the data. Finally, we simulate synthetic binary systems with circumbinary objects to evaluate the detectability of these companions based on the ETV signals generated.

\subsection{Target Selection}\label{sect:TargetSelection}

We identified our target sample from the TESS Eclipsing Binary Catalog\footnote{\url{https://tessebs.villanova.edu/}} \citep[TEBC,][]{Prsa2022}, focusing on detached binaries. Detached binaries were selected because their eclipse profiles are typically deeper, narrower, and more symmetric, allowing for more precise timing measurements. In contrast, semi-detached and contact binaries exhibit more continuous light curve variations and distorted eclipse shapes due to strong binary interaction, such as mass transfer, tidal deformation, or a shared envelope. These effects not only complicate the identification of eclipse minima but can also produce intrinsic ETV trends unrelated to third-body companions, thereby reducing the reliability of timing-based detections.
We used the morphology parameter from the TEBC to filter the systems, selecting only those with \texttt{morphology} $< 0.2$. We limited the binary periods to less than 14 days to capture at least one eclipse within a given TESS sector. The minimum binary period in the sample was 1.5 days.

We filtered out targets with relative period uncertainties of $10^{-4}$ or larger to obtain precise period determinations. We also included targets with established reference eclipse times and associated uncertainties. The target sample consisted of 310 systems with TESS magnitudes ranging from 5 to 16.

\subsection{Data Preparation}\label{sect:DataPreparation}

To analyze the eclipse timing of our target binaries, we retrieved and processed their light curves from TESS observations accessed via the Mikulski Archive for Space Telescopes (MAST)\footnote{\url{https://mast.stsci.edu/}} portal. For this study, we used Simple Aperture Photometry (SAP) fluxes, as the Presearch Data Conditioning SAP (PDCSAP) fluxes can introduce artificial trends that may distort the intrinsic features of eclipsing binary light curves. We collected only 2-minute exposures for each target, with the last observation for any target occurring in TESS Sector 80.

The data preparation involved removing NaN values to exclude incomplete measurements, selecting data points with a quality flag of zero to retain only high-quality observations, and normalizing each light curve to its median value.

\subsection{Eclipse Light Curves}\label{sect:LightCurveModeling}

To determine the eclipse parameters of the target binaries, we modeled their light curves using the \texttt{allesfitter} \citep{allesfitter-paper,allesfitter-code} code with a Markov Chain Monte Carlo (MCMC) approach. Initial values for the orbital period and time of the primary eclipse were taken from the TEBC. The free parameters included the radius ratio, the sum of stellar radii over the semi-major axis, the cosine of the inclination, the time of the primary eclipse, the orbital period, the quadratic limb darkening coefficients, and the logarithm of the flux error. We fitted only the primary eclipses and iteratively selected the light curves within a phase range centered on zero. We extended $\pm$1/4 of the orbital period, ensuring that the primary eclipse was included in the selected data. To model the flux baseline for the eclipses, we applied splines to the out-of-eclipse regions surrounding each primary eclipse.

To minimize the accumulation of uncertainties on the linear ephemeris, we modeled only the eclipses in each target’s first and last available TESS sectors, limiting the analysis to 6 primary eclipses to reduce computation time.

\begin{figure}
    \centering
    \includegraphics[width=\columnwidth]{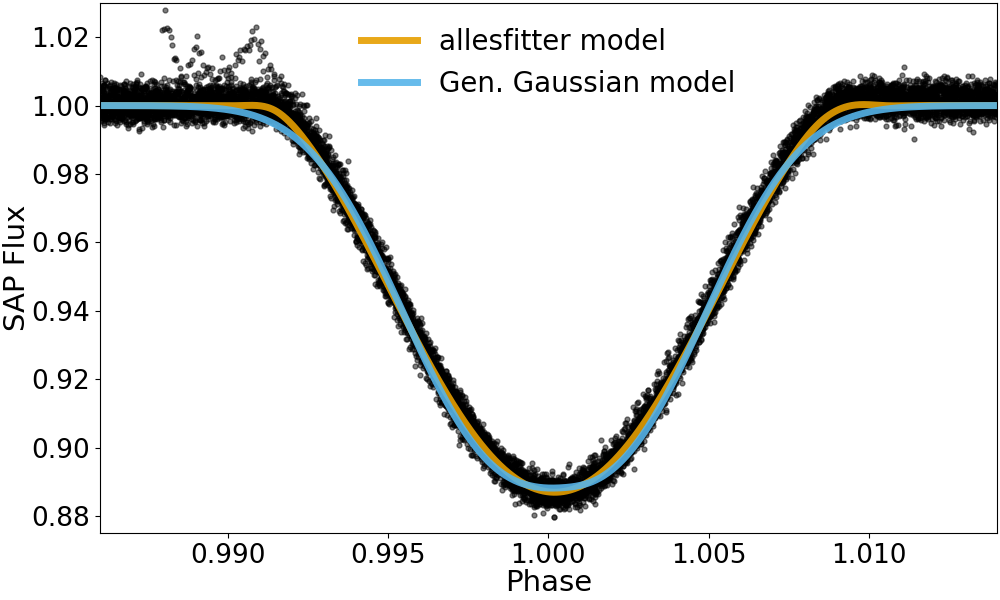}
    \caption{Phase-folded SAP primary eclipse light curve of TIC 142979644 (black points) with two fitted models overplotted. The allesfitter model (orange) provides a better overall fit, accurately following the steep ingress and egress as well as the flat mid-eclipse region. The generalized Gaussian model (blue) is computed using the median values of the fitted parameters; it reproduces the overall eclipse shape but performs less accurately at the sharp ingress/egress transitions and in the mid-eclipse region where flux changes occur rapidly.}
    \label{fig:lc_model_example}
\end{figure}

When sampling from the posterior of the model parameters using MCMC, we used 300 walkers, discarded 27,000 samples as burn-in, and retained 5,000 steps. This process generated distributions for each free parameter, allowing us to quantify median values and uncertainties. We used these eclipse models in the next step to determine eclipse mid-times. As an example of our light curve models, Fig. \ref{fig:lc_model_example} shows the primary eclipse of one of our targets, TIC 142979644, fitted with the \texttt{allesfitter} model and overplotted with the generalized Gaussian model used for mid-eclipse time determinations, which we describe in the following section.

\subsection{Eclipse Timing Calculations}\label{sect:calculatingEclipseTimes}

To calculate the mid-times of the eclipses, we used the primary eclipse model of each target, relying on parameters obtained from our previous light curve modeling and using Markov Chain Monte Carlo (MCMC) implemented via \texttt{allesfitter}. Initial values for parameters were taken from the previous modeling phase: the radius ratio, the sum of stellar radii over the semi-major axis, the cosine of the inclination, the time of the primary eclipse, the orbital period, the quadratic limb darkening coefficients, and the logarithm of the flux error. For the mid-time calculation, we fixed the time of the primary eclipse and orbital period, keeping the remaining parameters free to capture light curve variations across cycles. A spline was used to model the out-of-eclipse regions to maintain a stable baseline, consistent with the previous flux baseline modeling step.

When estimating the posterior mid-time via MCMC, we used a number of walkers set to four times the number of eclipses, with a minimum of 40 walkers. We discarded 15,000 steps as burn-in and retained the subsequent 5,000 steps to sample from the posterior. This fitting process yielded mid-times for each primary eclipse, allowing us to quantify deviations with associated uncertainties.

We calculated mid-eclipse times by fitting a generalized Gaussian function, using a method similar to that employed by \citet{Kostov2022ApJS..259...66K}. In Eq. \ref{eq:genGaussian}, \( A \) is the baseline flux level, \( B \) is the eclipse depth, and \( t_0 \) is the mid-eclipse time. The width and shape of the eclipse are controlled by \( \omega \) and \( \beta \), respectively; \( \beta = 2 \) gives a Gaussian shape. The terms \( C_1 \) and \( C_2 \) account for linear and quadratic trends in the baseline flux.

\begin{equation}\label{eq:genGaussian}
F(t) = A - B \, e^{ -\left( \frac{ |t - t_0| }{ \omega } \right)^{\beta} } + C_1 (t - t_0) + C_2 (t - t_0)^2
\end{equation}

\begin{figure}
    \centering
    \includegraphics[width=\columnwidth]{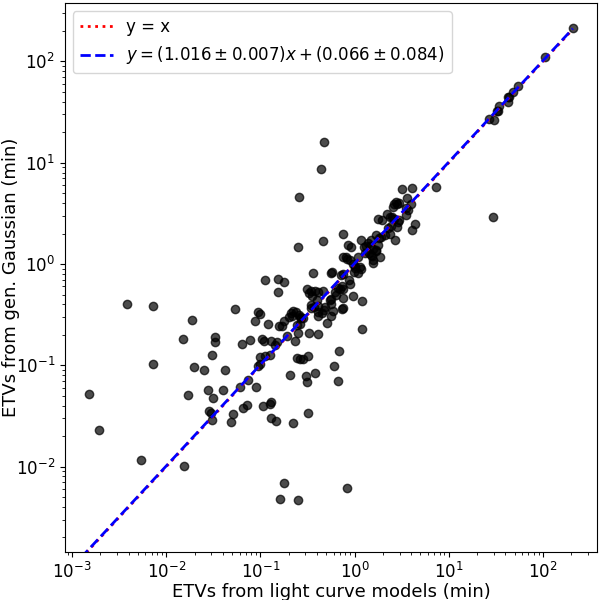}
    \caption{Comparison of ETV data points for all targets derived from light curve models using \texttt{allesfitter} and from generalized Gaussian fits. Each point represents an individual eclipse event. The red dotted line represents the identity line (y = x), indicating perfect agreement, while the blue dashed line shows the best-fit linear relation, with the equation $y = (1.016 \pm 0.007)x + (0.066 \pm 0.084)$. The two lines are indistinguishable from each other.}
    \label{fig:alles_gauss_comp}
\end{figure}

We fitted this function using the Levenberg–Marquardt least-squares minimization method \citep{Levenberg1944, Marquardtdoi:10.1137/0111030} as implemented in the \texttt{LMFIT} package \citep{newville_2025_16175987}, separately to each primary and secondary eclipse in both SAP and PDCSAP light curves. For targets lacking SPOC light curves, we searched for full-frame images (FFIs), extracted light curves using \texttt{lightkurve} \citep{Lightkurve2018ascl.soft12013L}, and applied the same fitting procedure when possible. This method is significantly faster than the MCMC routine used in our main light curve modeling.
To assess the reliability of the generalized Gaussian approach, we compared the resulting ETVs to those derived from MCMC-based light curve modeling with \texttt{allesfitter}. As shown in Figure~\ref{fig:alles_gauss_comp}, the two methods produce consistent mid-eclipse times across a wide range of eclipse depths and data quality, supporting the use of generalized Gaussian fits for trend comparisons and ETV analysis. The full machine-readable tables containing the mid-eclipse times, uncertainties, and cycle numbers are provided in the online material.

\subsection{Lomb-Scargle Periodograms of ETVs}\label{sect:lombscargleETVs}

To investigate periodic signals in the eclipse timing variations (ETVs) of our target binaries, we conducted a Lomb-Scargle frequency analysis \citep{Lomb1976Ap&SS..39..447L, Scargle1982ApJ...263..835S}, allowing us to identify and characterize potential periodic trends. For this purpose, we utilized relevant packages in \texttt{astropy} \citep{astropy:2013,astropy:2018,astropy:2022}.

We began by filtering the ETV data to exclude mid-time measurements with uncertainties on the order of hours, as these were indicative of unreliable mid-time determinations. We performed a least squares fit to determine a linear ephemeris for each eclipse type and light curve source. Residuals exceeding $\pm3\sigma$ from this fit were removed using a single-pass clipping. We also excluded data points whose calculated uncertainties exceeded the mean by more than $3\sigma$ of the overall uncertainty distribution, to remove low-quality measurements. This standard procedure helps eliminate poorly constrained mid-times caused by shallow or distorted eclipses. The vertical gray dotted lines in Figure~\ref{fig:etv_142979644} (and subsequent figures) denote the $\pm1\sigma$ range, showing that retained points lie well within the clipping threshold. Lomb-Scargle analysis was performed for each target only if more than four mid-time measurements remained after these filtering steps.

Following this process, we successfully analyzed mid-eclipse times for 152 targets, a reduction from the initial sample of 310. This decrease was primarily due to an insufficient number of eclipses in the TESS data, while in some cases, complex light curve features also contributed to exclusion.

To prepare the data for analysis, we fitted a linear model to ETVs to update the linear ephemerides. We then used the Lomb-Scargle periodogram on the ETV residuals, scanning frequencies up to a maximum limit set at four times the binary period. Since our data are separated by multiples of the binary period ($P_{\rm bin}$), the Nyquist frequency of our dataset is $\frac{1}{2P_{\rm bin}}$ (for further explanation, see \citealt{VanderPlas2018ApJS..236...16V}). From the periodograms, we identified the frequencies with the highest power. We calculated the corresponding periods, amplitudes, and false alarm probabilities (FAP) of the detected signals, quantifying the statistical significance of the observed periodicity.

We separately computed Lomb-Scargle periodograms for the secondary eclipse timings, when a secondary eclipse was present in the TESS light curves. For each target, we selected the strongest peak in the periodogram of the secondary mid-eclipse times to investigate whether the observed ETV variations could be attributed to mechanisms such as apsidal motion or stellar spot migration. These effects are known to produce periodic but anti-correlated timing variations between the primary and secondary eclipses, whereas the presence of a third body induces correlated variations in both. By comparing the dominant periodicities and phase behavior of the primary and secondary ETVs, we aimed to distinguish between these different physical origins of the observed signals.

We also investigated the potential influence of instrumental systematics by comparing the ETV signals derived from SAP and PDCSAP light curves. Lomb-Scargle periodograms were computed for both data products, and the resulting signals were examined for consistency. This comparison is particularly important for low-amplitude ETV signals, where instrumental trends or residual systematics may dominate. When SAP and PDCSAP signals were in strong agreement, we interpreted the variation as more likely astrophysical in origin. Conversely, significant discrepancies between the two pointed toward an instrumental cause.

\subsection{Simulating ETV Systems}\label{sect:SimulatingETVs}

\begin{figure*}
    \centering
    \includegraphics[width=1.9\columnwidth]{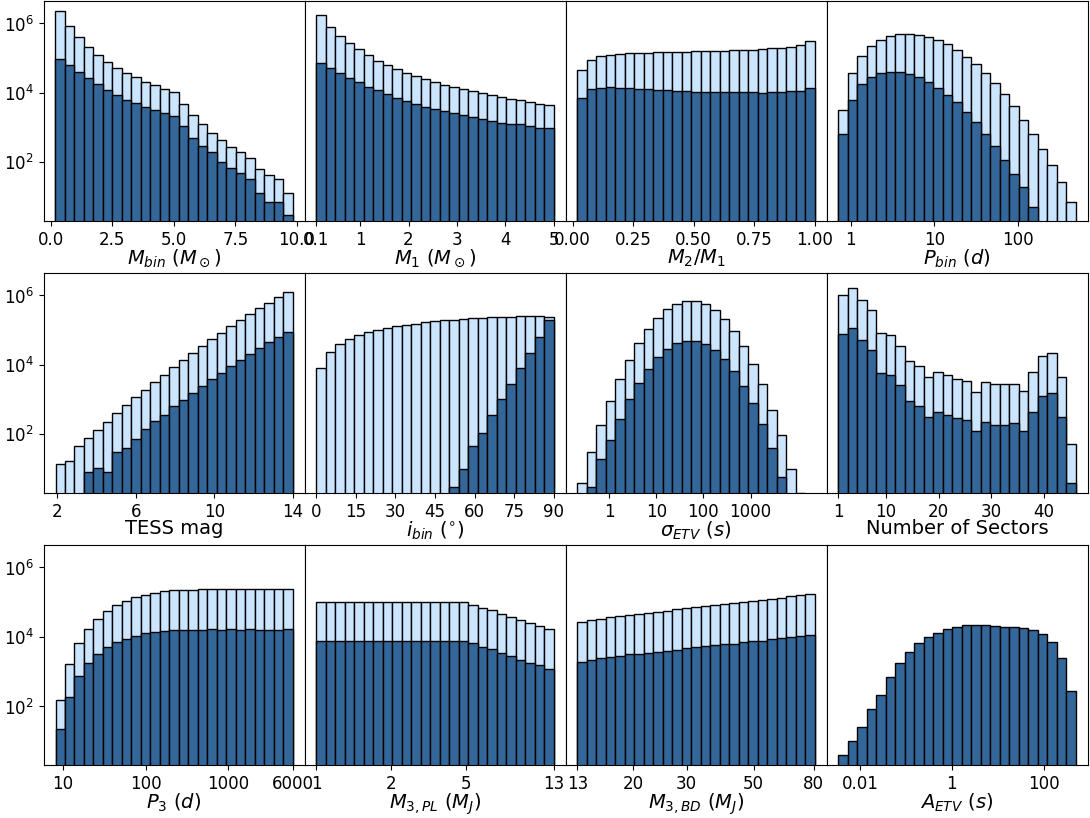}
    \caption{Distributions of the simulated binary and circumbinary parameters. The subscripts \textit{bin}, \textit{1}, \textit{2}, \textit{3}, \textit{PL}, and \textit{BD} correspond to parameters related to the binary system, primary star, secondary star, circumbinary object, planet mass, and brown dwarf mass, respectively. Note that all vertical axes are logarithmic, while some horizontal axes are also logarithmic. The full set of $4 \times 10^6$ samples is represented in light color, while the eclipsing binary sample (287,472 systems, or 7.2\%) is shown in dark color. The ETV amplitude ($A_{\rm ETV}$) values were calculated only after a binary eclipse condition was satisfied; therefore, this parameter is displayed exclusively in dark colors. Other binary and circumbinary parameters were sampled from uniform distributions and are not shown here. See Sections \ref{generateBinaryParameters} and \ref{generateCircumbinaryParameters} for further details.}
    \label{fig:simulatedParameters}
\end{figure*}

In this section, we describe our approach to simulating binary systems with circumbinary companions, aiming to assess the detectability of planets and brown dwarfs using ETVs. We generated realistic parameters for binary systems observable by TESS, assigned properties to circumbinary objects with various orbital and mass distributions, and calculated the resulting ETVs. The simulated binary and circumbinary parameter distributions can be seen in Figure \ref{fig:simulatedParameters}, and further explanations can be found in the following two sections. We focused exclusively on timing variations caused by the presence of an additional body, without including secular trends, apsidal motion, or other timing variations related to stellar activity. Using Lomb-Scargle periodogram analyses, we evaluate the effectiveness of ETV methods in detecting these simulated circumbinary objects.

\subsubsection{Generating Binary Parameters}\label{generateBinaryParameters}

We generate the following parameters for binary systems: equatorial coordinates (RA \& Dec), brightness, masses of the primary and the secondary, orbital period, reference eclipse time, orbital inclination, and ETV uncertainty.

\begin{figure*}
    \centering
    \includegraphics[width=2\columnwidth]{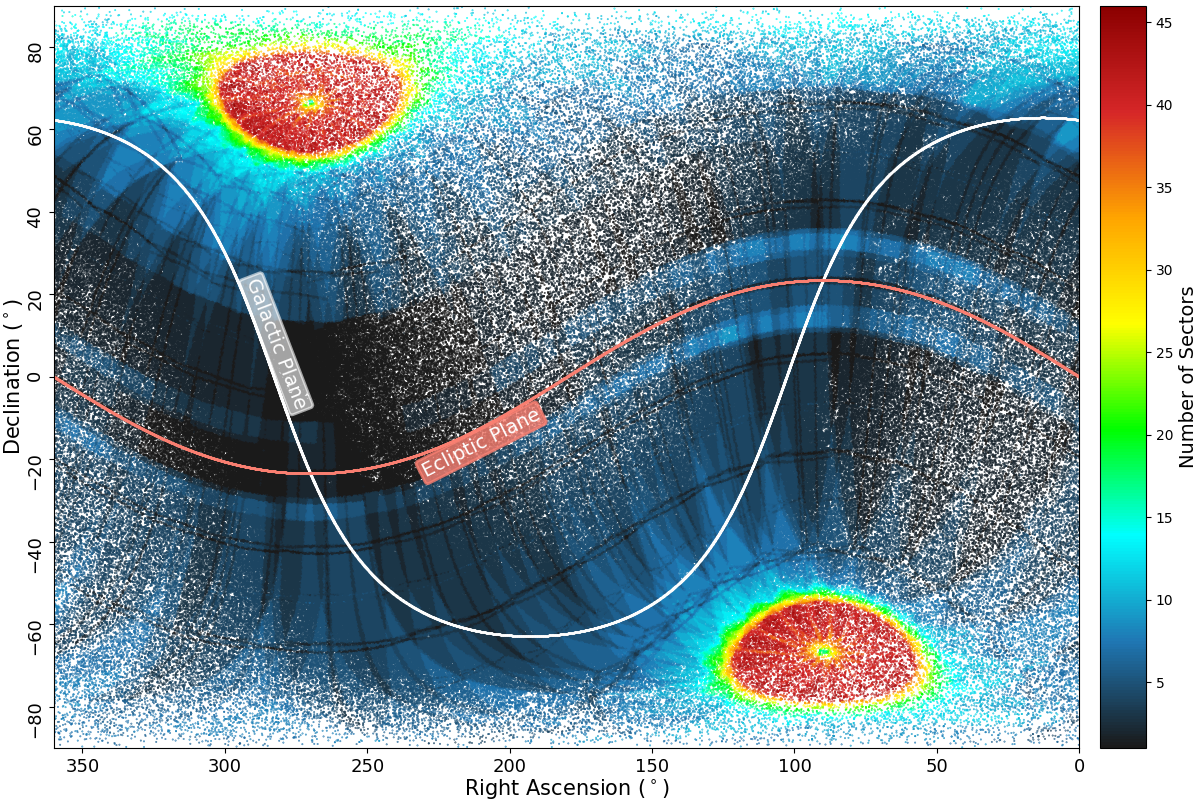}
    \caption{The distribution of the simulated binary sample in the equatorial coordinates with color coding based on the number of TESS sectors. Warmer colors represent areas with higher observational coverage. The TESS Continuous Viewing Zones (CVZs) appear as the red regions near the ecliptic poles.}
    \label{fig:galacticDist}
\end{figure*}

To simulate realistic ETV data that reflects the TESS sky coverage as a function of celestial coordinates, we gathered coordinate and brightness values for $10^{7}$ stellar objects with $RP < 14$ from the Gaia Archive \citep{GaiaCollaboration2016A&A...595A...1G, GaiaCollaboration2023A&A...674A...1G}. We filtered out galaxy and quasar candidates and selected only sources classified as stars, setting the discrete source classifier probability for stellar sources at \texttt{classprob\_dsc\_combmod\_star} $>$ 0.9. To ensure the sample consists of stellar sources, we limited our selection to objects with a parallax signal-to-noise ratio (SNR) greater than 5. To generate a binary, we randomly selected RA and Dec, along with $RP$ brightness as a proxy to TESS magnitude. We checked if any of these random coordinates were observable with TESS and determined their observation frequency based on the number of sectors they were covered in, using the TESS-Point tool \citep{TESSPointBurke2020ascl.soft03001B} (Figure \ref{fig:galacticDist}). We did not account for contamination from nearby sources due to the TESS pixel size. This effect is especially pronounced near the galactic plane, where most of our sample is located.

We sampled the masses of the companions in a binary system from the initial mass function (IMF) derived for stellar systems by \cite{Chabrier2003}. After sampling a primary star mass, we used this value as an upper limit for the secondary mass and sampled it from the same IMF to ensure the mass ratio $M_{\rm 2}/M_{\rm 1}$ remains between (0,1]. The individual masses were selected between 0.1 and 5 solar masses. Therefore, in some cases, the total binary mass exceeds 5$M_\odot$. We calculated the radii of the binary components from the mass-radius relationship (Eq. \ref{eq:massradiusrelation}) from \cite{Gorda1998}.

\begin{equation}\label{eq:massradiusrelation}
\log_{10}(R) =
\begin{cases}
0.049 + 0.993 \log_{10}(M) & \text{if } M \leq 1.4 \\
0.096 + 0.652 \log_{10}(M) & \text{if } M > 1.4
\end{cases}
\end{equation}

We generate binary periods from the log-normal period distribution in TEBC, and for the reference eclipse times, we used uniform sampling between $\pm$ binary period. We calculated the eclipse times for each binary and determined the eclipses that were observable in any Sectors based on the beginnings and ends of TESS observations\footnote{\url{https://tess.mit.edu/observations/}}.

We uniformly sampled $\cos i$ between 0 and 1 to obtain randomly oriented orbits. For each simulated binary system, we checked the eclipse condition (Eq. \ref{eq:eclipsecondition}) and continued to ETV calculation if satisfied.

\begin{equation}\label{eq:eclipsecondition}
\cos(i) \leq \frac{R_{\rm pri} + R_{\rm sec}}{\sqrt[3]{(M_{\rm pri} + M_{\rm sec})\times P_{\rm bin}^{2}}}
\end{equation}

To assign timing uncertainties to each generated binary system, we first modeled the logarithm of ETV uncertainties as a function of brightness using a two-dimensional Gaussian distribution. This model was constructed using the brightness values and ETV uncertainties derived from the MCMC-sampled mid-times obtained with \texttt{allesfitter} (see Section~\ref{sect:calculatingEclipseTimes}). Based on this distribution, we sampled uncertainties corresponding to the brightness values of each simulated system.

\subsubsection{Generating Circumbinary Objects}\label{generateCircumbinaryParameters}

We generated the following parameters for the circumbinary objects: mass, orbital period, mean anomaly, and inclination. We assumed circular orbits, hence keeping eccentricity, the argument of periastron, and the longitude of ascending nodes fixed to zero. We separated this stage into two main groups: i) coplanar orbits between the binary (C) and ii) uniform circumbinary orbital inclinations (U). Each of these parts was further divided into two groups based on their masses: planets (PL; 1-13 $M_{\rm J}$) and brown dwarfs (BD; 13-80 $M_{\rm J}$).

We used the exoplanet mass function from \cite{Mordasini2018} to generate masses less than $13\ M_{\rm J}$. This function is a power law and is $\propto M^{-1}$ for masses between $1-5\ M_{\rm J}$, and $\propto M^{-2}$ for masses larger than $5\ M_{\rm J}$. To generate the brown dwarf masses ($13 - 80\ M_{\rm J}$), we used the IMF from \cite{Chabrier2003}.

Due to their low ETV amplitudes, we did not extensively simulate masses smaller than $1\ M_{\rm J}$. However, we calculated the SNR for the masses of Saturn, Neptune, and Earth on coplanar orbits around random binaries as $10^{-2.4}$, $10^{-3.1}$, and $10^{-4.3}$ $\pm 10^{0.5}$, in respective order. The maximum SNR for Saturn analogs was 0.16 within 2000 random samples.

We sampled the orbital periods of the circumbinary objects from a log-uniform distribution between 10 times the binary period and 6000 days. The inner range should reflect the minimum stable period ratio of a circumbinary, and for our sample, we set the inner limit based on the largest value calculated by \cite{Quarles2018ApJ...856..150Q}. The larger limit was selected to be approximately three times the data time span. The smaller limit was chosen to ensure the dynamical constraints (though not included in any calculations) would be considered. In the case of large mass and short periods for the circumbinary object, as well as small binary mass, the gravitational interactions between the objects would introduce additional timing variations. We did not simulate this interaction in this study, and the recovery and occurrence regarding these configurations should differ from our results.

For the coplanar orbits, we adopted the binary inclination as the inclination of the circumbinary object. In contrast, we sampled $\cos i$ uniformly between $0-1$ for the groups of uniform circumbinary orientation.
We sampled the mean anomaly from a uniform distribution between $\pm 360^{\circ}$.

\subsubsection{ETVs of Simulated Systems}\label{etvsOfSimulatedSystems}

By using the generated binary and circumbinary parameters, we performed orbital simulations with \texttt{rebound} \citep{Rein2012_REBOUND} and the IAS15 integrator \citep{Rein2015_IAS15}. Each binary system was treated as a single object with its total mass, reducing the system to a two-body problem. Eclipse timing variations (ETVs) were calculated from numerical simulations following the method described in \cite{Esmer2023MNRAS.525.6050E}.

For each of the four groups categorized by mass and inclination, we generated $10^6$ samples. Among these, 287,472 (7.2\%) satisfied the eclipse condition, while the remaining systems were excluded before ETV calculations. Additionally, systems with fewer than four observable eclipses in TESS data were excluded from further analysis.

To simulate ETVs, we added the LiTE model corresponding to each circumbinary system to the eclipse times. We introduced Gaussian scatter to the data based on the timing uncertainties generated for the systems where eclipses are observable with TESS. Similar to analyzing observed data, we fitted a line to ETVs to update the linear ephemeris and used the ETV residuals in the further steps. While we did not add a linear ephemeris difference of any kind - not to binary period nor to reference eclipse times, we implemented the prescribed linear fit to investigate the deviation of the detected periods and amplitudes from the original values, especially for the cases of undersampled ETVs. We performed Lomb-Scargle analyses of the ETVs and recorded the periods, amplitudes, and FAPs corresponding to the peak frequency in the periodograms. We limited the periodogram to frequencies between four times the binary period and twice the available data range for each system. All of the steps explained in this paragraph were repeated three times, and only the results with the highest FAP values were saved to decrease the number of false positives.

\begin{figure*}
\centering
\includegraphics[width=\linewidth]{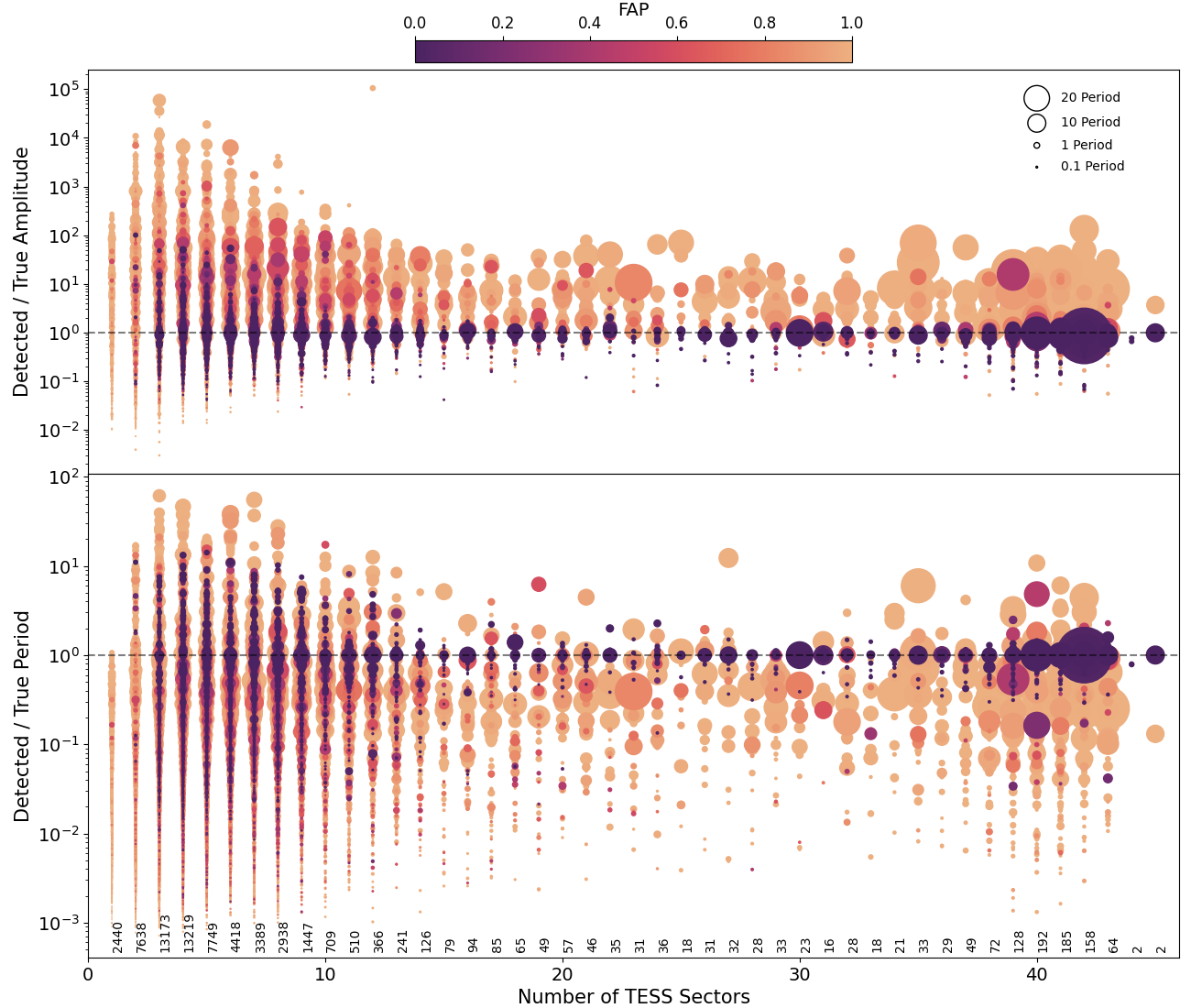}
\caption{Ratios of detected to true ETV amplitudes (top) and periods (bottom) in the coplanar brown dwarf (CBD) simulations, plotted against the number of observed TESS sectors. Point sizes are proportional to the phase coverage, with the scale shown at the top right in units of circumbinary periods, and colors indicate the false alarm probability (FAP) of the detected signal. Numbers below each bin indicate the number of systems in that bin. Large deviations from unity are common for short baselines, but both amplitudes and periods converge toward the true values as the baseline increases, with minimal scatter beyond $\sim$20 sectors.}
\label{fig:amp_per_ratio_vs_sectors}
\end{figure*}

Undersampled ETVs, arising from short observational baselines, show large deviations between the recovered and injected parameters in our simulations. As illustrated in Figure~\ref{fig:amp_per_ratio_vs_sectors}, when fewer than $\sim$10--15 TESS sectors are available, both the detected amplitudes and periods exhibit substantial scatter, often differing from the true values by orders of magnitude. This behavior reflects the limited phase coverage of the injected signal, particularly for long-period companions, and the increased susceptibility to spurious low-FAP peaks. With longer baselines, the phase coverage improves (indicated by the larger points in the figure), and the recovered values converge toward the injected ones. By about 20 sectors, both amplitude and period ratios cluster around unity with minimal dispersion, indicating that accurate recovery is only achievable once a sufficient fraction of the orbit is sampled. We also calculated the k-sample Anderson-Darling test \cite{Scholz1987} for all the sector bins, with the null hypothesis that they all come from the same parent distribution. We used related functions in \texttt{scipy}, and got $p$-values around $10^{-5}$, which is interpreted as strong evidence against the null hypothesis.

\section{Results}
\label{section:Results}

\subsection{Results of ETV Analyses of the Target Sample}\label{sect:etv_results}

We analyzed ETVs for 152 eclipsing binaries using both MCMC-based eclipse models with allesfitter and generalized Gaussian fits. To assess signal significance, we used the FAP of the strongest peak in Lomb-Scargle periodograms. A total of 37 targets exhibited FAP $<$ 0.1 in at least one method, and 19 of these had FAP $<$ 0.01. Below, we discuss representative systems grouped by the likely origin of their ETVs, including possible circumbinary companions, stellar activity, apsidal motion, or ambiguous trends. Systems without compelling evidence for astrophysical periodicities are summarized at the end of the section.

To estimate the masses of hypothetical circumbinary companions, we first derived total binary masses from stellar parameters. We adopted the effective temperature ($T_{\mathrm{eff}}$) from Gaia DR3 or the TEBC, assuming it corresponds to the primary star. Using empirical mass-temperature-radius (MTR) relations for main-sequence stars \citep{Eker2018MNRAS.479.5491E}, we interpolated the primary’s mass and radius. The secondary radius was computed from the \texttt{allesfitter}-derived radius ratio, and its mass was estimated using the same MTR relations. For secondaries falling below the valid MTR range (0.217~$M_\odot$), we applied a linear extrapolation from the lower end. We then calculated the projected mass of a third body assuming a coplanar orbit, based on the ETV period, amplitude, total binary mass, and binary inclination, following the method of \citet{Esmer2021A&A...648A..85E}.

Based on the ETVs derived from \texttt{allesfitter} modeling, three targets (TIC~142979644, TIC~166090445, and TIC~270648838) yielded projected masses of 18.8, 25.9, and 41.7~$M_{\rm J}$, respectively, all below the stellar–substellar boundary \citep{Baraffe2002A&A...382..563B}. When using mid-times from generalized Gaussian fits, the same three targets also produced substellar-mass solutions (11.1, 19.2, and 42.1~$M_{\rm J}$, respectively), in addition to three others: TIC~180412528 (28.3~$M_{\rm J}$), TIC~66355834 (31.9~$M_{\rm J}$), and TIC~350297040 (41.7~$M_{\rm J}$). Among these, the most promising candidate is TIC~142979644, which exhibits consistent ETV signals across methods. In contrast, the remaining targets show signs of flaring, anti-correlated ETVs, or inconsistent signals across data types, suggesting that their observed variations are more likely due to stellar variability or dynamical effects unrelated to a third body.

TIC 142979644 shows a flare-dominated light curve with clear out-of-eclipse variability. The \texttt{allesfitter} ETV analysis of the primaries yields a 535-day signal with a 0.33-minute amplitude and a FAP of $2\times10^{-7}$. Generalized Gaussian fits to SAP and PDCSAP light curves recover consistent periods, 516 and 518 days, with slightly lower amplitudes near 0.2 minutes and low FAPs $10^{-6}$ and $1.6\times10^{-4}$, which argues against an instrumental origin. The secondary ETVs are not consistent, 412 days with a 0.30-minute amplitude for SAP and 42 days with a 0.24-minute amplitude for PDCSAP, both with low FAPs $1.9\times10^{-6}$ and $9\times10^{-4}$. The PDCSAP periodogram also shows a weaker peak near 400 days, which is less prominent in SAP. From the visually measured eclipse phase widths, $\Delta\phi_{1}=0.0177$ for the primary and $\Delta\phi_{2}=0.0183$ for the secondary, and the secondary mid-phase $\phi_{\rm sec}=0.5183$, we obtain $e\sin\omega \simeq 0.0167$ and $e\cos\omega \simeq 0.0575$, hence $e\simeq0.060$ and $\omega\simeq16^\circ$. The first-order per-eclipse apsidal amplitude is $A_{\rm aps}\simeq (P/\pi)|e\cos\omega|=P|\phi_{\rm sec}-0.5|=0.0183\,P\approx 110$ minutes, which is far larger than the observed $0.2$–$0.33$ minutes. Thus, the measured ETVs cannot be explained by apsidal motion. While third-body fits imply masses of $0.018\, M_\odot$ (\texttt{allesfitter}) and $0.011\, M_\odot$ (SAP), the strong out-of-eclipse variability can also induce timing shifts through spot evolution and baseline changes, and further investigation is required to distinguish between them.

The remaining targets either have traces of stellar mass companions, apsidal motion, secular trends that exceed the baseline for a precise classification, or stellar activity-related variations in their ETVs. Some of them have indications that the significant variations are likely from instrumental effects on the data, rather than having an astrophysical origin. For further remarks on the remaining systems with significant ETVs, see Appendix \ref{sect:etv_results_appendix}.

\subsection{Results for ETV Simulations}\label{sect:etvSimulation_results}

The successful detection of a simulated circumbinary companion is characterized primarily by a detected period and amplitude that closely match the true parameter values or their harmonics or subharmonics. To evaluate this, the results of our simulations for each sample were analyzed using ratio distributions of detected and true values for periods and amplitudes. Additionally, FAP must be low (FAP $\ll$ 1), as it quantifies the likelihood that the detected signal arises from random noise (i.e., a false positive).

\begin{figure*}
    \centering
    \includegraphics[width=2\columnwidth]{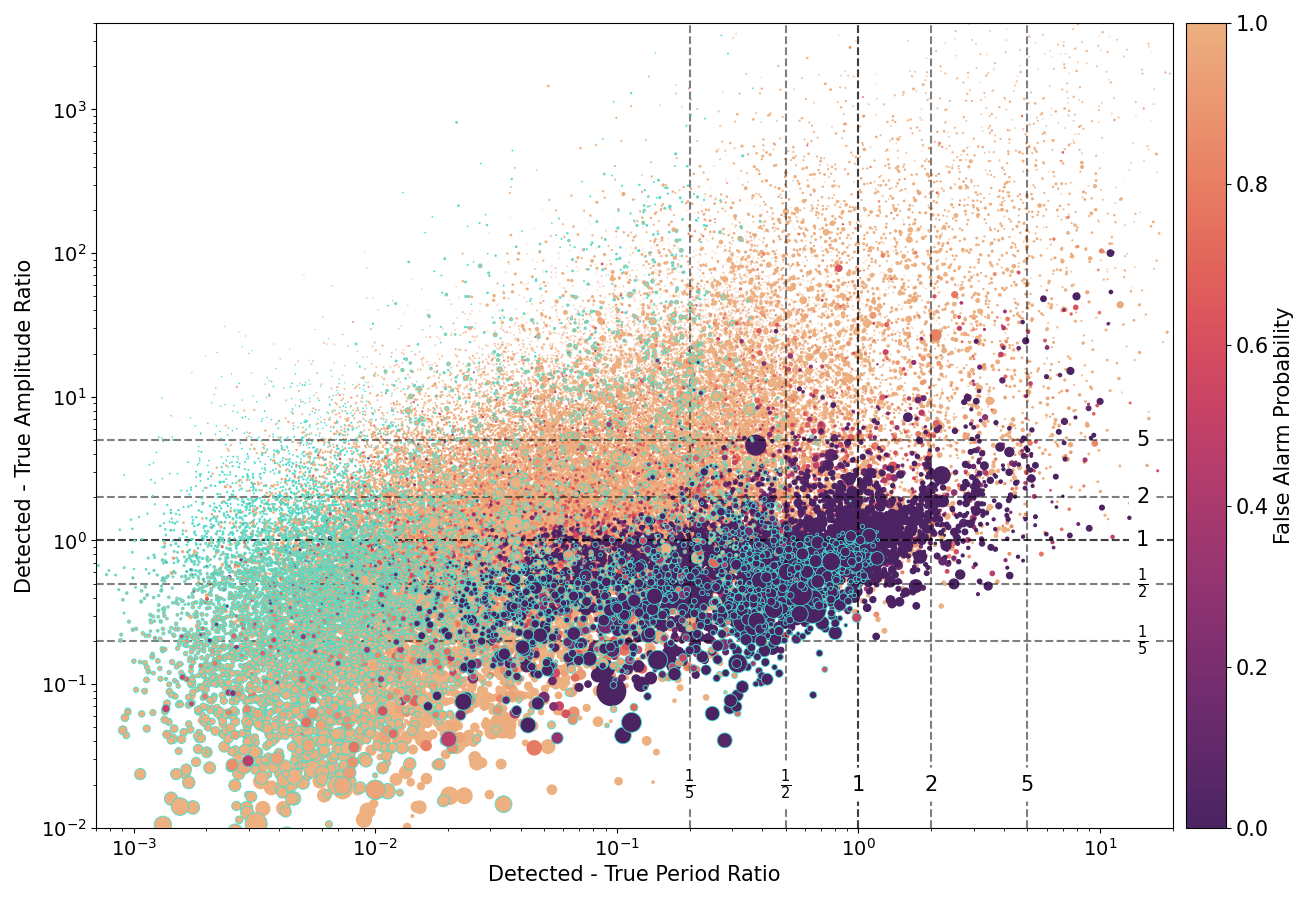}
    \caption{Correlation between detected-to-true period ratio and amplitude ratio across data points for the coplanar \& brown dwarf (CBD) group. The color scale represents false alarm probability (FAP), with darker points indicating lower FAP values. Markers with turquoise edges are for true periods longer than 3000 days. Point sizes are proportional to SNR.}
    \label{fig:ratioPlotsCBD}
\end{figure*}

The distributions of detected vs true period and amplitude ratios for the CBD group are shown in Figure \ref{fig:ratioPlotsCBD}. Most of the recovered periods and their amplitudes with FAP $\ll$ 1  pile up around 1:1 ratio regions. However, a considerable number of them form a tail towards smaller ratios. The tail-forming data are primarily for true periods larger than the total data time span for any individual target (maximum of $\sim$1800 days). The tail becomes pronounced especially when the true period reaches $\sim$3000 days, and an upper limit for detected vs true values of period and amplitudes arises around 1:1 (turquoise circles in Figure \ref{fig:ratioPlotsCBD}). The overall features for the UBD group are similar to the CBD group, while the number of detections diminishes considerably for planetary masses (UPL \& CPL), along with distinguishable features. Therefore, we construct the final detection conditions and thresholds based on the CBD and UBD groups.

\begin{figure*}
    \centering
    \includegraphics[width=1.9\columnwidth]{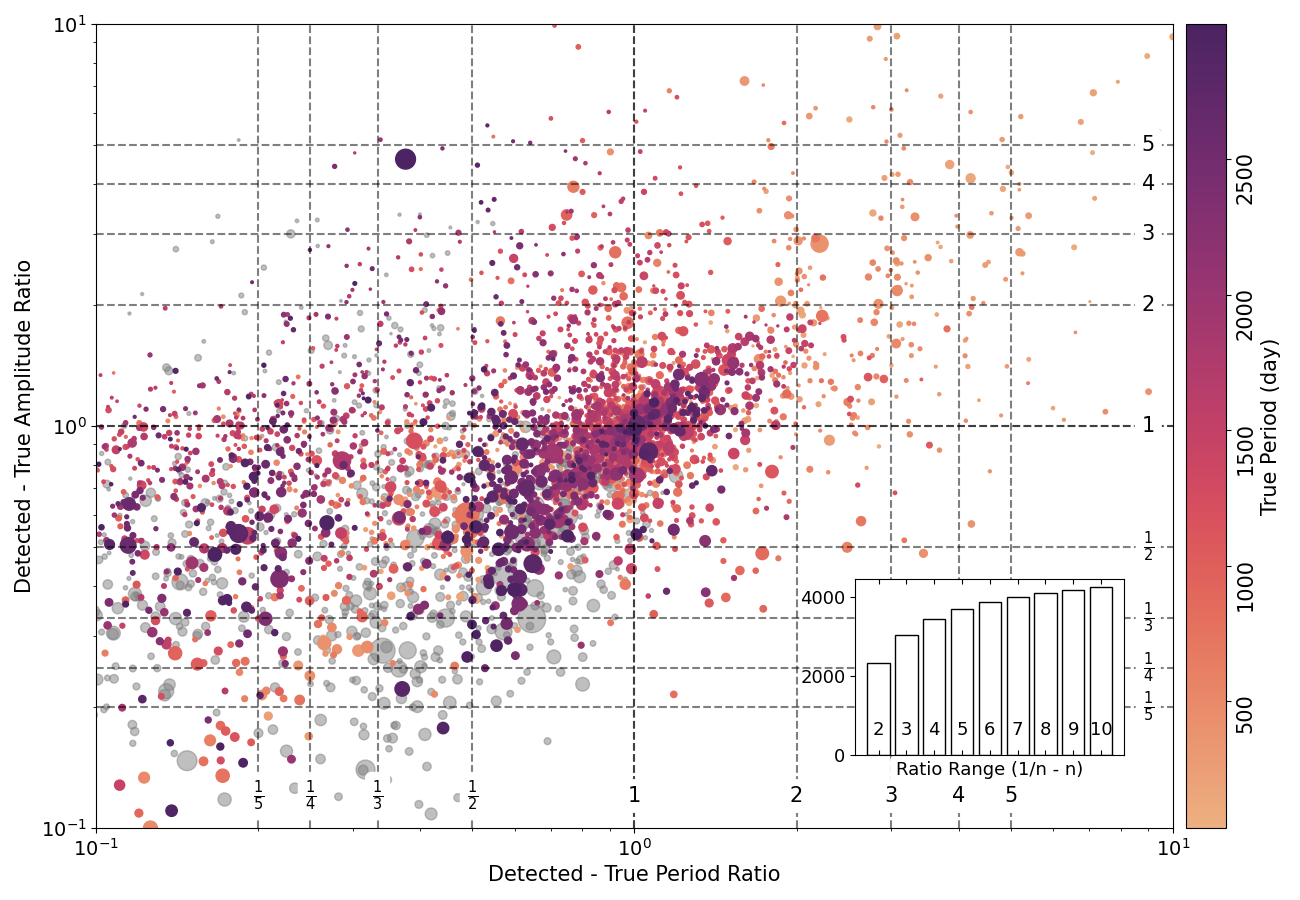}
    \caption{Similar to Figure \ref{fig:ratioPlotsCBD} but for the SNR, FAP, and detected-to-binary period ratio thresholds-applied coplanar \& brown dwarf (CBD) group, zoomed to 0.1 - 10 ratio range on both axes. The color scale represents true periods, with darker points indicating longer periods. Grey markers are for true periods longer than 3000 days. Point sizes are proportional to SNR. Lower-right: The subplot represents the data within the $1/n$ - $n$ ratio range on both axes. There are 2317 data points within the 1/2 - 2 ratio and 3022 points within the 1/3 - 3 ratio ranges, including true periods longer than 3000 days (grey).}
    \label{fig:ratioPlots_filtered_CBD}
\end{figure*}

A threshold for the detected-to-binary period ratio is essential to address stability concerns in circumbinary systems. As shown in \cite{Quarles2018ApJ...856..150Q}, the inner limit for a stable circumbinary orbit depends on parameters such as stellar masses, mass ratio, eccentricity, and other binary properties. We adopted a minimum value of four times the binary period limit in the Lomb-Scargle period search of simulated systems for the shortest period limit. However, we observed a significant clustering of detected signals near this limit, which we interpret as failed circumbinary detections. To eliminate these false positives, we applied a threshold that requires the detected-to-binary period ratio to exceed five, thereby retaining only signals with higher values.

We set an SNR threshold of 1 to filter out spurious results, selecting only data above this limit. Specifically, the signal is defined as the amplitude calculated from the Lomb-Scargle analysis, while the noise is characterized by the ETV scatter used to generate the simulated data noise. Data with SNR $<$ 1 consistently corresponded to FAP $\gg$ 0, and to further filter out false positives, we set a maximum FAP limit of 0.1, focusing on statistically significant variations.

In addition to applying the detected-to-binary period ratio, SNR, and FAP thresholds, we investigated the detected-to-true amplitude and period ratios, focusing on values around 1. As previously mentioned, most data concentrate near a ratio of 1:1, while there are (i) less populated clusters around harmonics and subharmonics and (ii) a linear correlation between the two ratios. Both behaviors become more pronounced when filtering out the data corresponding to true periods longer than approximately 3000 days. To define a region that avoids being overly conservative and excludes potential successful detections while not being too broad and rendering the filtering process ineffective, we examined the number of data points within subharmonic and harmonic ranges from 2 to 10.

At least half of the data for all groups lay within detected-to-true amplitude and period ratios of 2. For the CBD group, the detected-to-true amplitude and period ratio plot is presented in Figure \ref{fig:ratioPlots_filtered_CBD}, while for the groups of UBD, CPL, and UPL, the same plots are presented in Figures \ref{fig:ratioPlots_filtered_UBD}, \ref{fig:ratioPlots_filtered_CPL} and \ref{fig:ratioPlots_filtered_UPL} in respective orders. We focused on two key ratio values: 2, where the data clustered around the 1:1 region, and 3, which includes harmonics and subharmonics.

Table \ref{tab:detectionCounts} summarizes the simulated eclipsing binary sample sizes and detection counts across the four groups. Each group contains over 60,000 simulated systems, ensuring comparability across groups. Detection rates were highest for the CBD and UBD, ranging from 2.92\% to 5.03\%, while CPL and UPL groups exhibited significantly lower detection rates, with a maximum of 0.10\%. The median circumbinary mass across the entire sample is $56.6^{+16.5}_{-23.4} M_{\rm J}$. Additionally, the detection rates of CBD systems are approximately 30\% higher than in UBD systems. Furthermore, the results indicate that BDs are about 50 times more likely to be detected using the ETV method than Jupiter-like planets in the mass range of 1–13~$M_{\rm J}$, highlighting the sensitivity of this technique when applied to TESS data.

The total binary mass in the detected sample spans from 0.2 $M_\odot$ to 5.6 $M_\odot$, with a median value of $0.54^{+0.68}_{-0.25} M_\odot$. The corresponding binary periods range from 0.75 to 33.1 days, with a median period of $3.1^{+3.0}_{-1.4}$ days. The detected sample is skewed toward lower binary masses and shorter binary periods, as these conditions lead to higher ETV amplitudes and a greater number of timing measurements within the observational baseline.

\begin{figure}
    \centering
    \includegraphics[width=\columnwidth]{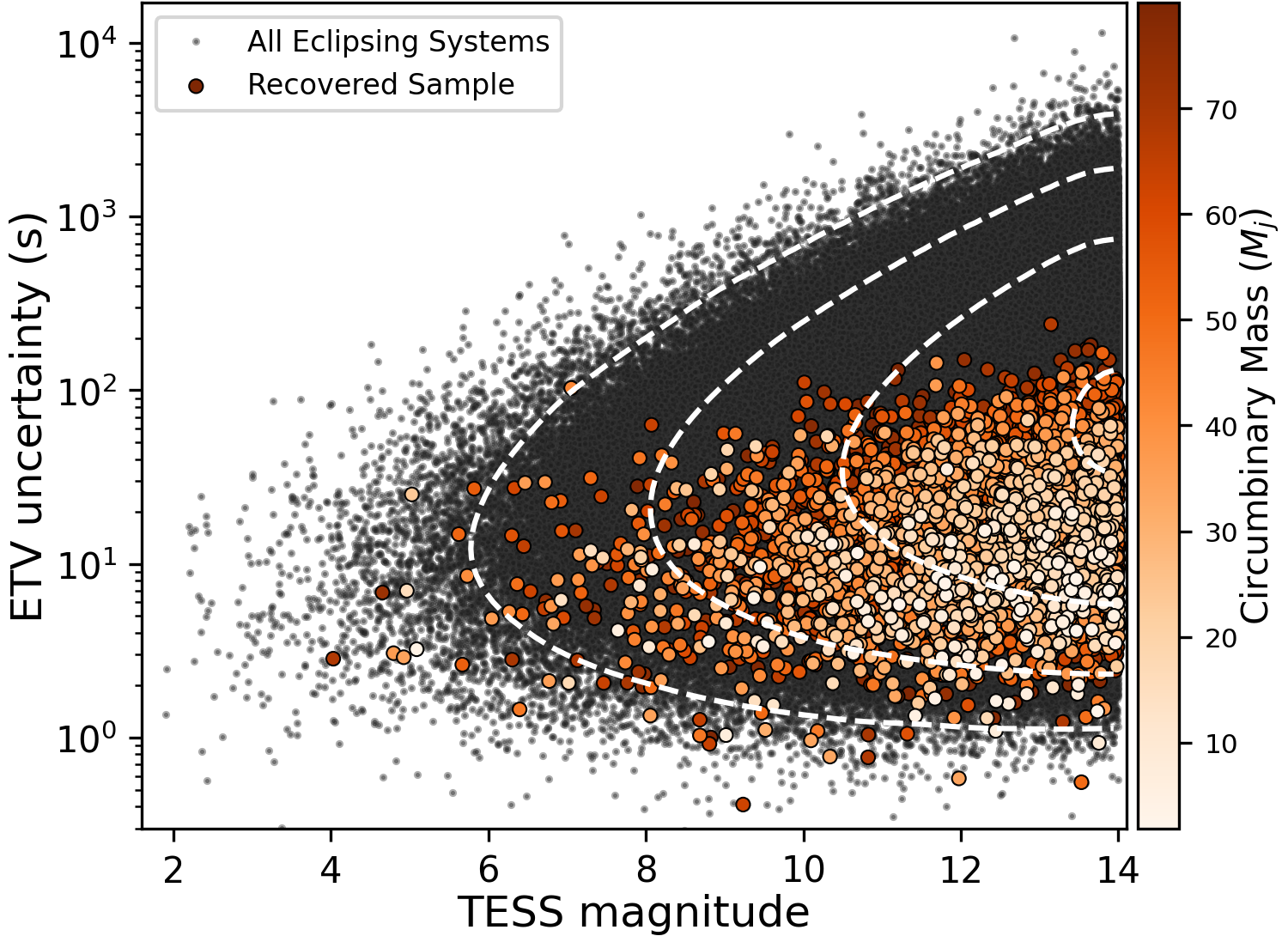}
    \caption{TESS magnitude versus ETV uncertainty (\(\sigma_{\rm ETV}\)) for the entire sample of eclipsing binaries (black points) and the recovered sample (orange circles). Point colors indicate the inferred circumbinary companion masses, and the dashed contours highlight density variations in the entire sample.}. The recovered systems are concentrated in regions with lower timing uncertainties, while the overall distribution shows increasing uncertainty with decreasing brightness.
    \label{fig:mag_sigmaetv}
\end{figure}

The TESS magnitudes of the recovered systems range from 4 to 14 mag. The number of both simulated and recovered eclipsing binaries increases with decreasing system brightness. In contrast, the number of recovered systems peaks where timing uncertainties remain low enough for detections, with a median TESS magnitude of $12.82^{+0.87}_{-1.79}$ mag. The ETV uncertainties of the recovered sample span 0.4 to 240 seconds, with a median value of $16.4^{+20.7}_{-9.7}$ seconds. Figure \ref{fig:mag_sigmaetv} shows these distributions for the entire sample of eclipsing binaries and the recovered systems.

The distribution of recovered systems across the sky reveals a concentration near the Galactic plane, particularly in regions neighboring the TESS CVZ on both hemispheres (cyan or warmer-colored regions in Figure \ref{fig:galacticDist}). While the CVZ benefits from long-term observational coverage, it is not the most densely populated region for recoveries due to the lower number of eclipsing binaries compared to the Galactic plane. Additionally, most recovered systems are found in regions with fewer than 20 observed TESS sectors, outnumbering those with more than 20 sectors by a factor of ten. However, TESS data for sources near the Galactic plane are expected to suffer from significant contamination due to the large pixel sizes. As a result, real detections in these regions are more likely to be either bright sources, where the impact of contamination is reduced, or systems located at higher Galactic latitudes compared to our recovered sample.

\begin{table}[!ht]
\caption{Simulated eclipsing binary sample and detection counts across CBD, UBD, CPL, and UPL groups. The 1/n - n values represent the thresholds for detected-to-true period and amplitude ratios, with detections defined as data within these limits. We used thresholds of 2 and 3. Additional detection thresholds for SNR, FAP, and detected-to-binary period ratio were also applied. Labels: C = coplanar orbit, U = uniform inclination, PL = planet mass, BD = brown dwarf mass.}\label{tab:detectionCounts}
\begin{center}
\begin{tabular}{ccccc}
& CBD & UBD & CPL & UPL \\ \hline
Total Number & 60102 & 59933 & 65636 & 65541 \\ \hline
Detection & 2317 & 1751 & 52 & 38 \\ 
1/2 - 2 & (3.86\%) & (2.92\%) & (0.08\%) & (0.06\%) \\ \hline
Detection & 3022 & 2364 & 66 & 45 \\ 
1/3 - 3 & (5.03\%) & (3.94\%) & (0.10\%) & (0.07\%) \\
\end{tabular}
\end{center}
\end{table}

\begin{figure*}
    \centering
    \includegraphics[width=1.4\columnwidth]{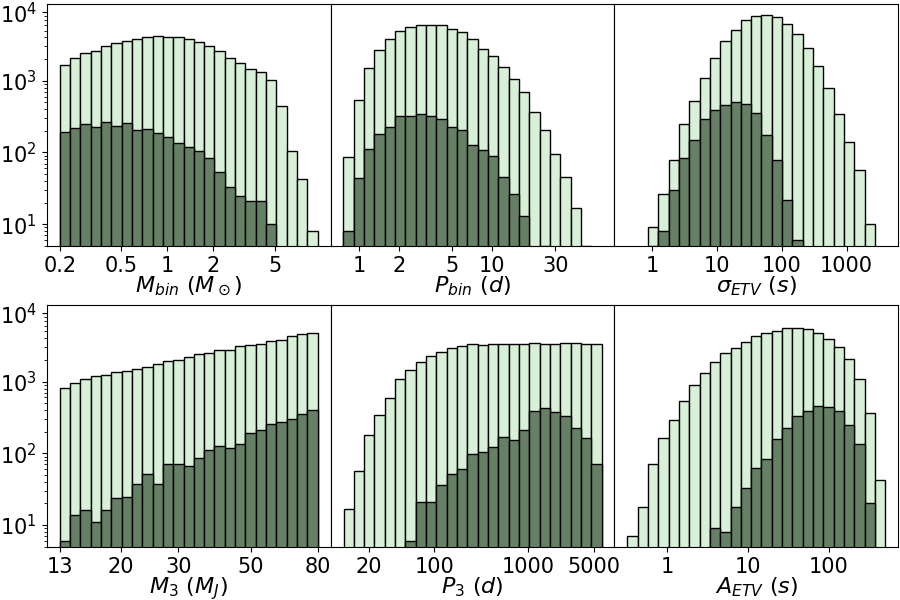}
    \caption{The distributions of parameters for coplanar \& brown dwarf (CBD) group. The light colors are for all the samples where the binary is eclipsing, while the dark colors are for the sample where circumbinary objects are detected between 1/3 and 3 range for amplitude and period ratios. Note that both axes for all distributions are on a logarithmic scale.}
    \label{fig:detectedHistCDB}
\end{figure*}

\begin{figure*}
    \centering
    \includegraphics[width=2\columnwidth]{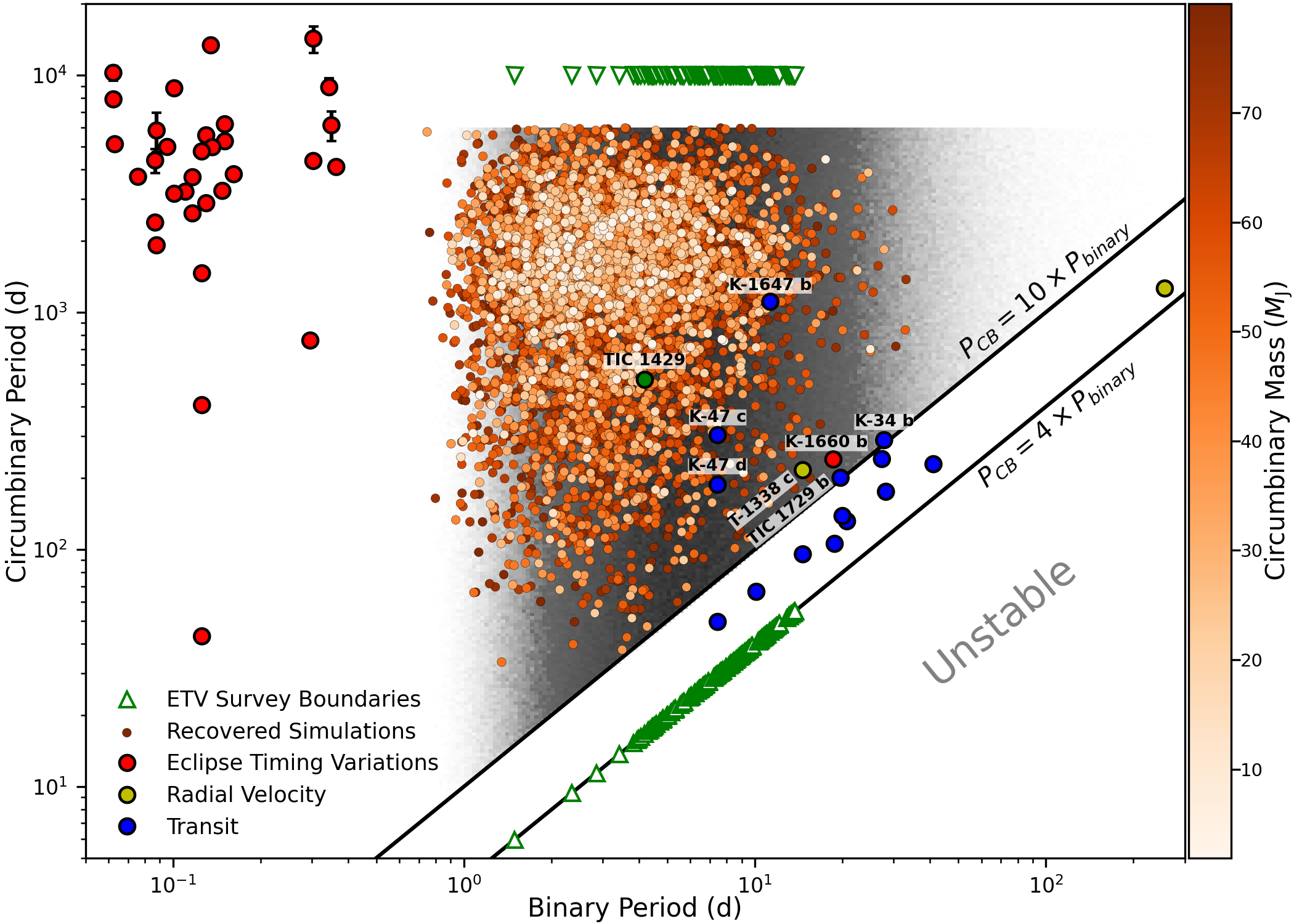}
    \caption{Recovered sample from our ETV simulations (small orange markers, shaded by circumbinary mass) overlaid on the known circumbinary planet population. The gray shading shows the full simulated population. Green triangles mark the Lomb--Scargle search ranges for our ETV survey of 152 targets. The black lines indicate circumbinary period constraints, with orbits below $4 \times P_{\rm binary}$ considered dynamically unstable. In comparison, periods up to $10 \times P_{\rm binary}$ were excluded from our recovery methodology due to susceptibility to Nyquist aliases and dynamical ETVs rather than LiTE. Several circumbinary planets (e.g., Kepler-1660 b, Kepler-1647 b, Kepler-34 b, Kepler-47 c/d) that broadly overlap with our simulations are labeled for reference, along with our candidate TIC 142979644 (TIC 1429).}
    \label{fig:mp_plot}
\end{figure*}

\subsection{Occurrence Rates}

In our ETV modeling of the target sample, assuming a given binary mass, we calculated the hypothetical circumbinary masses corresponding to the peak frequencies in the periodograms. For six systems, these masses were found to be below 0.08~$M_\odot$. Among them, variations in five systems, excluding TIC 142979644, are likely due to out-of-eclipse variations, apsidal motion, or sampling effects. TIC 142979644, despite showing flare activity and out-of-eclipse oscillations, is retained for occurrence rate calculations because its ETV signal is consistent across methods and light curve products, has very low FAP values ($\leq 1.6 \times 10^{-4}$), and cannot be explained by apsidal motion. The corresponding hypothetical third-body mass lies within the range for brown dwarfs or gas giant planets. Consequently, we proceed with occurrence rate calculations under two scenarios. One assumes no confirmed planetary detections in the sample ($k = 0$) to provide an upper limit, and the other includes TIC~142979644 as a candidate system ($k = 1$). For all scenarios, we consider only the coplanar cases.

For both scenarios, we use the relation:
\begin{equation}
    f = \frac{k}{R \cdot N}
\end{equation}
where $f$ is the occurrence rate, $k$ is the number of detections, $R$ is the recovery rate, and $N$ is the total number of systems analyzed. 

In the first scenario, with $k = 0$, the upper limit on $f$ is calculated using binomial statistics. The probability of observing zero detections in $N$ systems is:
\begin{equation}
    P(k = 0 | f) = (1 - f)^N
\end{equation}
which at a confidence level of $\alpha$ gives an upper limit to $f$ as,
\begin{equation}
    f_\text{upper} = 1 - (1 - \alpha)^{1/N}
\end{equation}
For $N = 152$ targets, we find $f_\text{upper} \approx 0.020124$ for $\alpha = 0.9545$.
Correcting for the recovery rate of the CBD group, $R = 5.03\%$, the true upper limit on the occurrence rate can be calculated as,
\begin{equation}
    f_\text{upper, corr} = \frac{f_\text{upper}}{R}
\end{equation}
which implies that the occurrence rate of detectable brown dwarfs in the sample is constrained to be less than $40.01\%$ at a $2\sigma$ confidence level. In the second scenario, assuming $k = 1$, the occurrence rate of circumbinary brown dwarfs is estimated at $13.08\%$.

For planetary-mass objects (\(< 13 \, M_{\rm J}\)) and a recovery rate of the CPL group (\(R = 0.0010\)), the calculated occurrence rate upper limit (\(f\)) exceeds 1, indicating that detections based solely on TESS data are not sufficiently sensitive for such masses. This reflects the low recovery rate and observational limitations of TESS. Complementary datasets or improved techniques are needed to enhance sensitivity to planetary-mass objects around binary systems.

\section{Summary and Discussion}
\label{section:SummaryAndDiscussion}

We investigated the eclipse timing variations (ETVs) of detached eclipsing binaries observed by TESS to identify potential circumbinary companions. Our target selection was based on the TEBC, focusing on detached binaries that allow for precise timing measurements. We modeled their light curves and ETVs using \texttt{allesfitter} and performed Lomb-Scargle analyses on 152 systems. Among these, 26 exhibited significant periodicities in their ETVs, though for some, the observed variations could be attributed to out-of-eclipse modulations. Assuming that the observed ETV signals originate from the LiTE due to circumbinary companions, we estimated their hypothetical masses. We found that six systems could host objects in the brown dwarf mass range.
One system, TIC 142979644, shows a robust primary-eclipse ETV signal detected consistently across independent methods with very low false-alarm probabilities. The secondary timings are mutually inconsistent, and simple apsidal motion cannot account for the observed variations. Third-body fits place any companion in the brown dwarf or planet regime, but out-of-eclipse variability could bias the timings. We therefore retain it as a promising yet tentative candidate for a circumbinary brown dwarf or planet.
For the remaining five systems, the observed ETV signals are likely caused by stellar variability, apsidal motion, or sampling effects.

To further assess the detectability of circumbinary brown dwarfs and planets, we conducted simulations incorporating various parameters related to binary star properties and potential circumbinary objects. The synthetic ETVs used in these simulations were generated using noise properties derived from our TESS ETV modeling (see Figure \ref{fig:mag_sigmaetv}). These simulations allowed us to calculate the recovery rates of the ETV method, yielding approximately 5\% for brown dwarfs and 0.1\% for Jupiter-like planets.

The detection of simulated circumbinary objects in our study reveals trends that align with theoretical expectations while providing quantitative insights into the parameter space where detections are most likely. Systems with smaller binary masses ($M_{\rm bin}$) are more likely to yield detectable signals, allowing the circumbinary object to induce a more pronounced wobble in the binary. On the other hand, shorter binary periods are favored because they ensure a sufficient number of eclipses can be observed within the TESS data. Similarly, systems with smaller ETV uncertainties ($\sigma_{\rm ETV}$) are more sensitive to detecting circumbinary companions. For circumbinary objects, higher masses ($M_{\rm 3}$) and longer periods ($P_{\rm 3}$) are more frequently associated with detections, even when the periods exceed the observational time span of the data. The peak periods detected for circumbinary objects are between 1000 and 2000 days. The corresponding distributions for the CBD group are shown in Figure~\ref{fig:detectedHistCDB}, which compares the entire eclipsing binary sample (light colors) to the detected circumbinary objects (dark colors). The overall features are common for UBD, CPL, and UPL groups, while the detected sample size is considerably smaller for the latter two. The only notable difference for PL groups is that the $M_{\rm 3}$ parameter for the detected sample becomes uniform for masses larger than $5 M_{\rm J}$, an effect of the planet mass function of \cite{Mordasini2018}.

Figure \ref{fig:mp_plot} illustrates our recovered ETV sample in the circumbinary period vs. binary period space, highlighting how different detection methods populate this region. A clear separation emerges between transiting and radial velocity detections, which cluster around binaries with periods of 10 days or more. In contrast, current ETV detections are primarily found around shorter-period binaries. Our simulated sample bridges this gap, overlapping more with transiting and RV-detected planets, suggesting potential confirmation of ETV signals using these complementary methods. However, the sensitivity of the ETV method biases our recovered sample toward shorter binary periods (3.09$^{+3.04}_{-1.37}$ days) and longer circumbinary periods (1404$^{+1361}_{-953}$ days), a region currently devoid of known circumbinary substellar objects. The figure also provides insight into survey completeness, comparing our recovered sample with known circumbinary planets and simulated populations. While our survey spans binary periods of 1.5 to 13.7 days, simulations suggest greater sensitivity at shorter binary periods, indicating that increasing survey density in this range would improve detectability. TESS observations may be crucial for identifying long-period circumbinary companions and refining occurrence rate estimates, though expanding the survey is essential to capture the diverse nature of these systems. The ETV planet associated with the longest binary period is Kepler-1660ABb \citep{Getley2017MNRAS.468.2932G, Goldberg2023MNRAS.525.4628G}, detected via dynamical interactions rather than the LiTE mechanism explored here. Our survey and simulations focus exclusively on the LiTE mechanism and do not incorporate dynamical models. In addition, binary and circumbinary orbits are modeled as circular, and each system is assumed to host only a single circumbinary object. Consequently, scenarios in which ETVs arise from gravitational perturbations by close-in circumbinary companions (e.g., Kepler-1660ABb), as well as multiplanetary architectures or eccentricity effects on the recovery of cyclic ETV signals, are not addressed in this study.

Circumbinary disks tend to be misaligned when the binary period exceeds approximately 30 days \citep{Czekala2019ApJ...883...22C}, suggesting that circumbinary objects in such systems are also more likely to have misaligned orbits. This period threshold is close to the upper limit of binary periods in our simulations (33.1 days for recovered samples), suggesting that our simulations are more representative of the coplanar scenario than of a uniform or randomly misaligned inclination distribution. While the coplanar case is more favorable for ETV detections, our results show that the ETV method remains efficient enough to recover misaligned circumbinary objects, albeit with a lower detection rate (coplanar brown dwarfs are 30\% more likely to be detected via ETVs than uniformly inclined counterparts). However, there is an inherent degeneracy in determining inclination when relying solely on ETVs. This degeneracy can be mitigated through simultaneous modeling with complementary data, such as transits, astrometry, or direct imaging, which together can constrain the inclination distribution of circumbinary objects in large surveys.

Using the recovery rates, we estimated an upper limit on the occurrence rate of circumbinary brown dwarfs. We found that it should not exceed approximately 40\% at a two-sigma confidence level. This limit is notably higher than the $<$6.5\% constraint placed by the BEBOP radial velocity survey \citep{Martin2019A&A...624A..68M} over a similar mass and period range. If we instead assume a single detection, our estimated occurrence rate is 13.08\%, which is approximately twice the upper limit set by BEBOP. Due to the low recovery rate for planetary-mass objects, we were unable to place meaningful constraints on their occurrence rates.
Although the photometric precision of TESS introduces some limitations, it has already yielded $\sim$10,000 eclipsing binaries \citep{Kostov2025ApJS..279...50K}. It has the potential to uncover up to $\sim$300,000 such systems \citep{Kruse2021tsc2.confE.163K}, providing a valuable resource for large-scale ETV studies like this one.
This work lays the foundation for future searches and further investigations into the population of circumbinary objects.

The majority of current circumbinary substellar discoveries via eclipse timing variations (ETVs) are found around short-period ($<$0.5 days) evolved eclipsing binaries, where the primary stars are either subdwarf B (sdB) stars or white dwarfs (WDs) with low-mass main-sequence companions. The progenitors of these primaries are estimated to be $\lesssim 2\,M_{\odot}$ \citep{Arancibia-Rojas2024MNRAS.52711184A} for sdBs and up to $8\,M_{\odot}$ for WDs \citep{Cunningham2024MNRAS.527.3602C}. In our ETV simulations, we identified 78,589 binaries with primary and secondary masses within the expected progenitor range of currently known ETV hosts ($1\,M_{\odot} < M_{\rm 1} < 8\,M_{\odot}$, $M_{\rm 2} < 1\,M_{\odot}$), yet we successfully recovered only 651 systems, yielding a recovery rate of just 0.83\%.

The formation pathways of circumbinary substellar companions around PCEBs remain an open question, with two competing scenarios: first-generation objects that formed within the protoplanetary disk of the young binary and second-generation objects that formed within a circumbinary disk during or after the common-envelope phase \citep{Zorotovic2013A&A...549A..95Z}. Despite their relative rarity, evolved post-common envelope binaries frequently host circumbinary substellar companions, implying that if these objects are of first-generation origin, their progenitor binaries must have hosted them in significant numbers. Given the intrinsically low recovery rate of 0.83\% in our simulations, even a single discovery of a circumbinary substellar companion around a PCEB progenitor in TESS data would strongly favor the first-generation scenario. Therefore, a systematic search for these objects across a larger sample of progenitor binaries is crucial to constrain their formation mechanisms.

While we were not able to put constraints on the occurrence rates of Jupiter-like planets, our findings highlight the potential of ETVs in detecting massive circumbinary companions, with the smallest mass detected from synthetic ETV data being $1.6~M_{\rm J}$ at an orbital period of 1860 days, demonstrating the method’s reach under optimistic conditions. We plan to expand our target sample and incorporate additional data from complementary surveys to improve detection limits and extend our sensitivity to lower-mass objects. While the substellar companions detectable via ETVs are typically massive, their gravitational influence can play a crucial role in shaping the orbital evolution of terrestrial siblings and even provide stable environments for habitable exomoons. These results emphasize the need for larger datasets and multi-method approaches to fully uncover the diversity of planets in circumbinary systems, potentially revealing new worlds that challenge our understanding of planetary formation and habitability.

\begin{acknowledgments}
This paper includes data collected with the TESS mission, obtained from the MAST data archive at the Space Telescope Science Institute (STScI). Funding for the TESS mission is provided by the NASA Explorer Program. STScI is operated by the Association of Universities for Research in Astronomy, Inc., under NASA contract NAS 5–26555. This work has made use of data from the European Space Agency (ESA) mission {\it Gaia} (\url{https://www.cosmos.esa.int/gaia}), processed by the {\it Gaia} Data Processing and Analysis Consortium (DPAC, \url{https://www.cosmos.esa.int/web/gaia/dpac/consortium}). Funding for the DPAC has been provided by national institutions, in particular, the institutions participating in the {\it Gaia} Multilateral Agreement.
\end{acknowledgments}

\software{Allesfitter \citep{allesfitter-paper,allesfitter-code}, Astropy \citep{astropy:2013, astropy:2018, astropy:2022}, Corner \citep{Foreman-Mackey2016_Corner}, Lightkurve \citep{Lightkurve2018ascl.soft12013L}, Matplotlib \citep{Matplotlib_Hunter:2007}, NumPy \citep{Numpy_harris2020array}, Pandas \citep{Pandas_mckinney-proc-scipy-2010,the_pandas_development_team_2024_10697587}, Rebound \citep{Rein2012_REBOUND}, SciPy \citep{2020SciPy-NMeth}, TESS-Point \citep{TESSPointBurke2020ascl.soft03001B}}

\appendix

\begin{figure*}
    \centering
    \includegraphics[width=0.9\columnwidth]{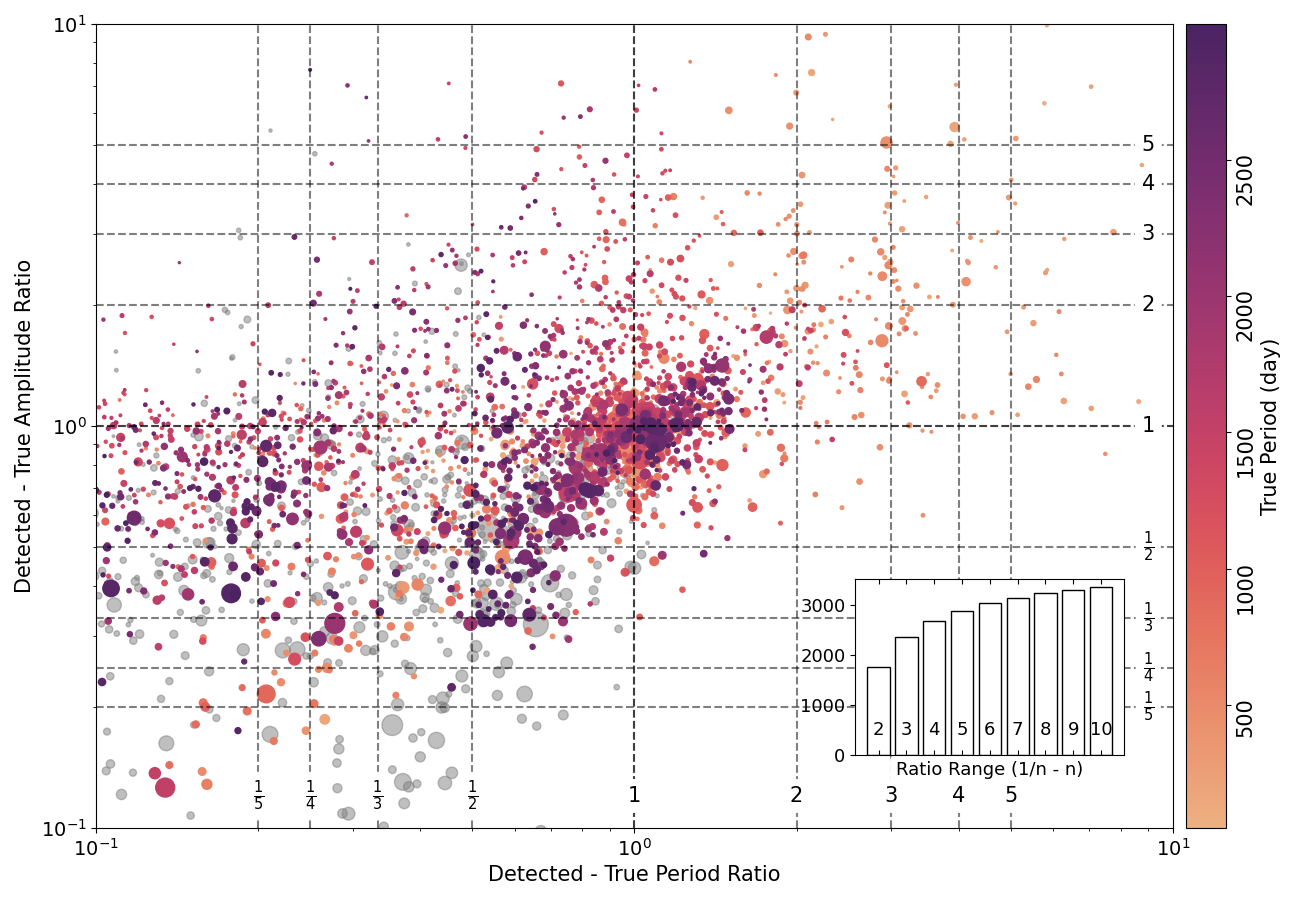}
    \caption{Similar to Figure \ref{fig:ratioPlots_filtered_CBD} but for uniform inclination \& brown dwarf (UBD) group.}
    \label{fig:ratioPlots_filtered_UBD}
\end{figure*}

\begin{figure*}
    \centering
    \includegraphics[width=0.9\columnwidth]{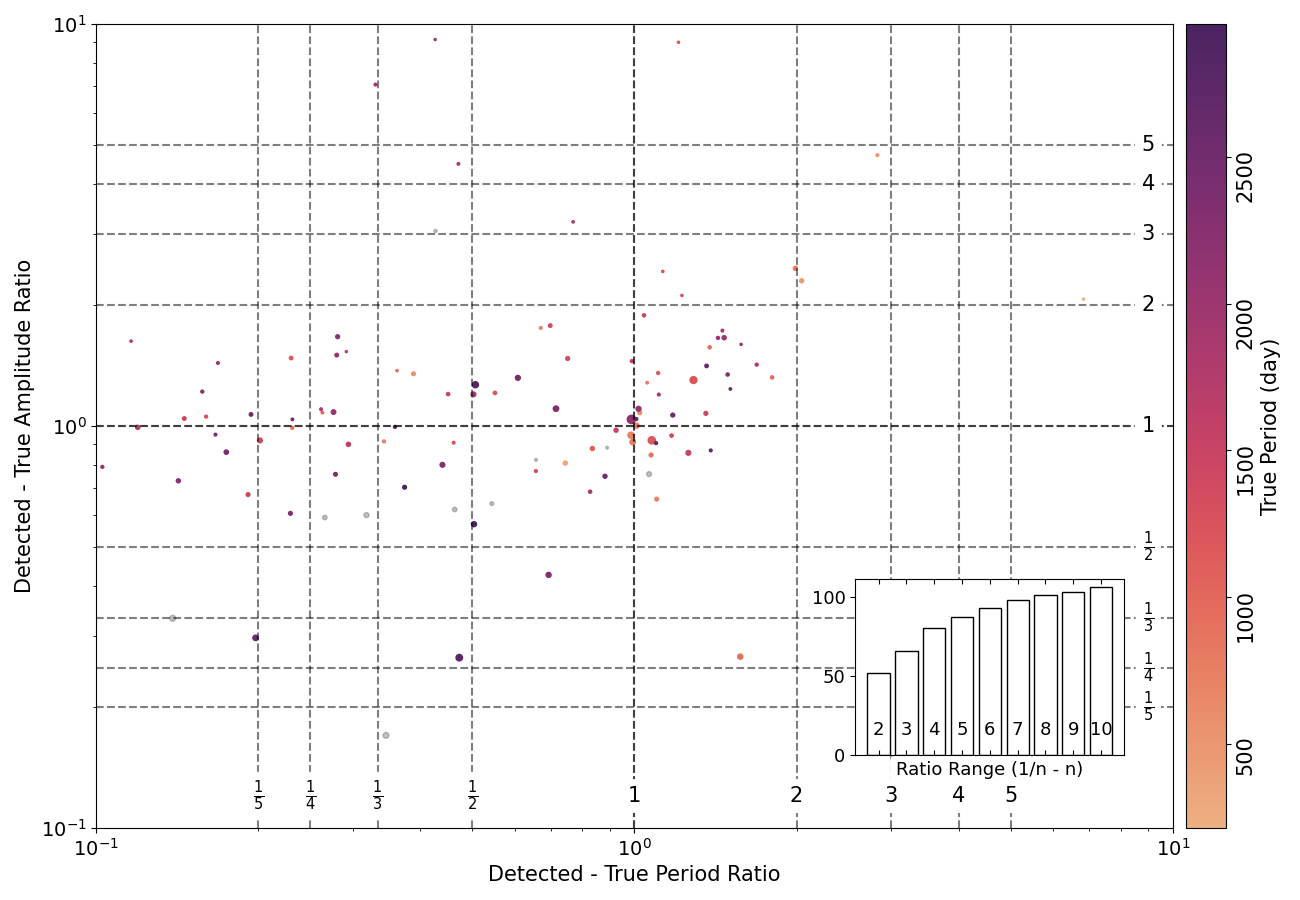}
    \caption{Similar to Figure \ref{fig:ratioPlots_filtered_CBD} but for coplanar \& planet (CPL) group.}
    \label{fig:ratioPlots_filtered_CPL}
\end{figure*}

\begin{figure*}
    \centering
    \includegraphics[width=0.9\columnwidth]{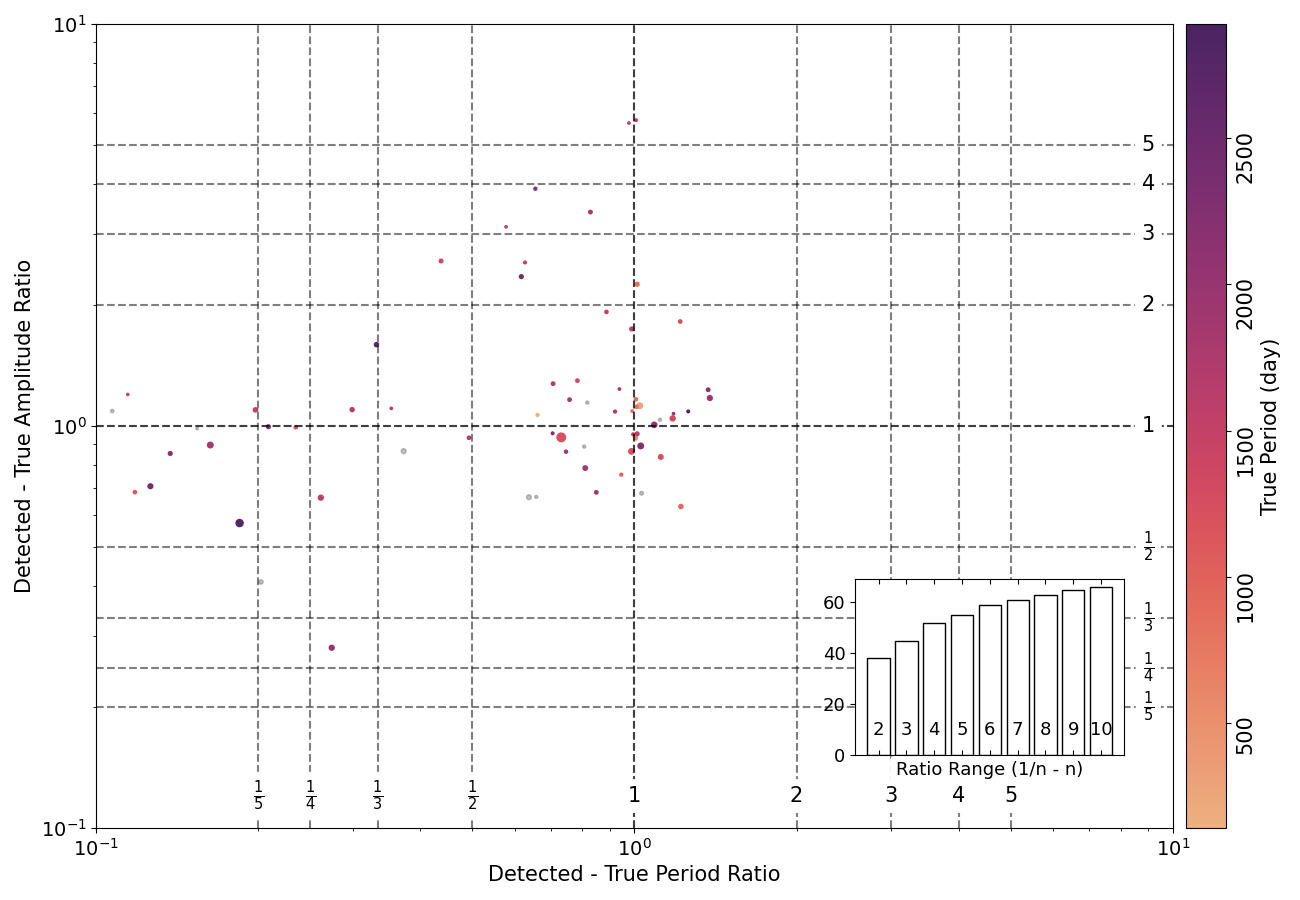}
    \caption{Similar to Figure \ref{fig:ratioPlots_filtered_CBD} but for uniform inclination \& planet (UPL) group.}
    \label{fig:ratioPlots_filtered_UPL}
\end{figure*}

\twocolumngrid

\section{Overview of the Results of ETV Analyses of the Target Sample}\label{sect:etv_results_appendix}

TIC 350297040 shows a 67.7-day ETV signal (9 seconds, FAP = 0.071) from \texttt{allesfitter}, but Gaussian fits return inconsistent periods (SAP: 121 days, PDCSAP: 77 days) with higher FAPs ($>$0.2), even after including FFI data. Secondary timings exhibit substantial scatter, with a 207-day signal of $\sim$40 seconds amplitude. No strong out-of-eclipse variability is present. While the derived companion masses ($0.092, M_\odot$ from \texttt{allesfitter}, $0.062, M_\odot$ from SAP) are in the substellar range, the low significance and inconsistencies across methods and datasets make a third-body interpretation doubtful.

TIC 66355834 shows an insignificant 115.8-day ETV signal from \texttt{allesfitter} (13 seconds, FAP = 0.61), while Gaussian fits yield inconsistent periods: 944.8 days (SAP, FAP $\approx$ 0) and 205.3 days (PDCSAP, FAP = 0.003). Secondary timings are unreliable due to poor coverage, and no anti-correlation is observed. With only three data segments, the ETV signal remains inconclusive despite some low-FAP peaks.

TIC 166090445, TIC 407584737, and TIC 260502102 all show anti-correlated primary and secondary ETVs consistent with apsidal motion. TIC 166090445 and TIC 260502102 additionally exhibit flaring, out-of-eclipse variability, or eccentric orbits, suggesting that stellar activity or secular trends may also contribute to the observed timing variations.

Several targets show low-amplitude, short-period ETV signals likely caused by stellar activity or geometric effects, rather than orbiting third bodies. TIC 270648838, TIC 180412528, and TIC 230063769 all exhibit substantial out-of-eclipse variability or flaring behavior, inconsistent or insignificant secondary eclipse timing signals, and inferred companion masses in the stellar or borderline substellar regime. In these systems, Gaussian fits often yield differing periods across data types, further weakening the case for third-body origins. TIC 140659980 similarly shows substantial variability and orbital eccentricity, with inconsistent secondary timings and a modest ETV signal (37 seconds), possibly arising from apsidal motion or activity-induced phase shifts.

TIC 170344769 (Kepler-1647) is an eclipsing binary with a known 1100-day circumbinary planet. Our ETV analyses failed to recover this signal: \texttt{allesfitter} detected a 48.7-day peak (1.5 min, FAP = 0.63), and Gaussian fits gave a 213.5-day signal (1.0 min, FAP = 0.52). This non-detection is consistent with \citet{Kostov2016}.

Several systems in our sample exhibit ETV signals that are best explained by either the presence of stellar-mass third companions or by secular variations. TIC 184298625 (KOI-6400), TIC 26542657, TIC 198537349, TIC 373915220, TIC 167692429, TIC 300161053, TIC 278988794, and TIC 167795859 show well-sampled, high-significance periodic variations, with inferred stellar companion masses ranging from 0.16 M$_\odot$ to several solar masses. In contrast, TIC 102929927 and TIC 237944385 display long-term trends that may reflect either secular timing drifts or periodic variations with timescales exceeding the TESS baseline.

Several targets in our sample show no compelling evidence for astrophysical ETV variations. These include TIC~364120439, TIC~258918831, TIC~103452621, TIC~323301918, TIC~3816260, TIC~118313102, and TIC~366072761, all of which exhibited high false alarm probabilities (FAP $\gtrsim 0.1$), inconsistent results across different light curve products, or signs of contamination from stellar activity, pulsations, or instrumental noise. While some targets (e.g., TIC~323301918 and TIC~3816260) presented long-term trends or large amplitudes, discrepancies in timing solutions and companion mass estimates across methods suggest these variations are not due to circumbinary objects. Others (e.g., TIC~118313102) displayed variability more consistent with known stellar pulsations. Additionally, a subset of targets such as TIC~336405638, TIC~293480903, TIC~360661624, TIC~182469311, TIC~371706494, TIC~148611095, and TIC~272086869 lacked coherent periodicities or suffered from low data quality, making them unlikely candidates for circumbinary companions. The results from our ETV analyses can be seen in Table \ref{tab:etv_results}, while the ETV diagrams, along with the Lomb-Scargle models for the \texttt{allesfitter} models, can be seen in the Appendix.

\section{Results for TOIs and KOIs in Our Targets}\label{sect:toikoi_results}

Our study examined several TESS Objects of Interest (TOIs) and Kepler Objects of Interest (KOIs) within the target sample. Many of these objects displayed characteristics indicative of stellar binaries rather than exoplanetary systems based on their light curve properties, radius ratios, and other inferred parameters. At the same time, only TIC 184298625 (KOI-6400) shows significant ETV trends. Below, we summarize the notable findings for each of these objects.

TIC 300871545 (TOI-184), currently listed as an exoplanet candidate. Its Gaia parameters ($T_{\mathrm{eff}} \sim 6400$ K, $\log g \sim 4.2$) and our derived radius ratio ($R_{\rm 2} / R_{\rm 1} \sim 0.15$) suggest it is more likely a stellar binary.

The companion of TIC 236387002 (TOI-2119) was confirmed as a brown dwarf by \cite{Canas2022}, and our light curve results align well with their findings.

For TIC 238197709 (TOI-646), we found $R_{\rm 2} / R_{\rm 1} \sim 0.26$, suggesting a stellar-size companion around a primary star with $T_{\text{eff}} \sim 5700$ K and $\log g \sim 3.8$.

TIC 1400770435 (TOI-1344) has a high radius ratio ($R_{\rm 2} / R_{\rm 1} \sim 0.49$) around a solar-like star ($T_{\text{eff}} \sim 5900$ K, $\log g \sim 4.2$), indicating a stellar companion.

TIC 350743714 (TOI-165, EBLM J0555-57) was classified by \cite{vonBoetticher2017} as a Saturn-sized stellar-mass object orbiting a Sun-like star. Our findings are consistent, suggesting a small secondary with a radius ratio of $R_{\rm 2} / R_{\rm 1} \sim 0.07$.

Based on our light curve models for TIC 382188882 (TOI-276) and TIC 149990841 (TOI-167), we derived radius ratios of $\sim 0.48$ and $\sim 0.47$, respectively, indicating stellar companions around Sun-like primaries.

TIC 77951245 (TOI-450) was identified as a low-mass, pre-main-sequence binary by \cite{Tofflemire2023}.

TIC 233390838 (TOI-1341) has an actual orbital period of 6.45 days, half of the period listed in the TEBC \citep{Schanche2019}.

For TIC 317507345 (TOI-1615), the primary star has Gaia parameters of $T_{eff} \sim 7200$ K and $\log g \sim 4.1$, and our derived radius ratio of $R_{\rm 2} / R_{\rm 1} \sim 0.35$ suggests a stellar-sized companion.

TIC 101395259 (TOI-623) includes a $1.17 M_\odot$ primary and a $0.098 M_\odot$ secondary as noted by \cite{vonBoetticher2019}.

Some of our remaining targets are also TOIs, for which we either only modeled their light curves; TIC 280206394 (TOI-677), TIC 321857016 (TOI-1420), and TIC 232967440 (TOI-1173) - or primarily analyzed their timing data, as in the case of TIC 277683130 (TOI-138). However, no significant variations or notable features were identified for these targets.

Some of our targets were identified as KOIs and classified as false positive candidates for transiting exoplanets: TIC 137549183, TIC 274129522, TIC 159720778, TIC 27006880, TIC 27915909, TIC 159047480, and TIC 184298625. Among these, TIC 159720778 and TIC 184298625 exhibited the strongest frequency signals in our Lomb-Scargle analysis, with FAP values of 0.099 and 0.0005 and corresponding periods of 48 days and 413 days, respectively. The ETV signals of TIC 184298625 (KOI-6400) are significant across all data and eclipse types, with amplitudes of $\sim$2.2 minutes and an inferred circumbinary companion mass of $0.4 M_\odot$. In contrast, the remaining targets showed their strongest frequency signals with FAP values close to 1, indicating no significant periodic signals.

\begin{deluxetable}{lcc}
\tabletypesize{\scriptsize}
\tablecaption{List of currently known circumbinary planets, their detection methods, and discovery references.}\label{tab:cbp_ref}
\tablehead{
\colhead{Planet} & \colhead{Method} & \colhead{Reference}
}
\startdata
Kepler-16 b	&	Tr	&\citet{Doyle2011Sci...333.1602D}	\\
Kepler-1647 b	&	Tr	&\citet{Kostov2016}	\\
Kepler-1661 b	&	Tr	&\citet{Socia2020AJ....159...94S}	\\
Kepler-34 b	&	Tr	&\citet{Welsh2012Natur.481..475W}	\\
Kepler-35 b	&	Tr	&\citet{Welsh2012Natur.481..475W}	\\
Kepler-38 b	&	Tr	&\citet{Orosz2012ApJ...758...87O}	\\
Kepler-413 b	&	Tr	&\citet{Kostov2014ApJ...784...14K}	\\
Kepler-453 b	&	Tr	&\citet{Welsh2015ApJ...809...26W}	\\
Kepler-47 b	&	Tr	&\citet{Orosz2012Sci...337.1511O}	\\
Kepler-47 c	&	Tr	&\citet{Orosz2012Sci...337.1511O}	\\
Kepler-47 d	&	Tr	&\citet{Orosz2019AJ....157..174O}	\\
PH1 b	&	Tr	&\citet{Schwamb2013ApJ...768..127S}	\\
TIC 172900988 b	&	Tr	&\citet{Kostov2021AJ....162..234K}	\\
TOI-1338 b	&	Tr	&\citet{Kostov2020AJ....159..253K}	\\
TOI-1338 c	&	RV	&\citet{Standing2023NatAs...7..702S}	\\
HD 202206 c	&	RV	&\citet{Correia2005AA...440..751C}	\\
BEBOP-3 b	&	RV	&\citet{Baycroft2025MNRAS.541.2801B}	\\
DE CVn b	&	ETV	&\citet{Han2018ApJ...868...53H}	\\
DP Leo b	&	ETV	&\citet{Qian2010ApJ...708L..66Q}	\\
HU Aqr b	&	ETV	&\citet{Qian2011MNRAS.414L..16Q}	\\
HU Aqr c	&	ETV	&\citet{Qian2011MNRAS.414L..16Q}	\\
Kepler-1660 b	&	ETV	&\citet{Goldberg2023MNRAS.525.4628G}	\\
Kepler-451 b	&	ETV	& \citet{Baran2015}	\\
Kepler-451 c	&	ETV	&\citet{Esmer2022MNRAS.511.5207E}	\\
Kepler-451 d	&	ETV	&\citet{Esmer2022MNRAS.511.5207E}	\\
MXB 1658-298 b	&	ETV	&\citet{Jain2017MNRAS.468L.118J}	\\
NN Ser c	&	ETV	&\citet{Beuermann2010AA...521L..60B}	\\
NN Ser d	&	ETV	&\citet{Beuermann2010AA...521L..60B}	\\
NSVS 14256825 b	&	ETV	&\citet{Zhu2019RAA....19..134Z}	\\
NY Vir b	&	ETV	&\citet{Qian2012ApJ...745L..23Q}	\\
NY Vir c	&	ETV	&\citet{Song2019AJ....157..184S}	\\
RR Cae b	&	ETV	&\citet{Qian2012MNRAS.422L..24Q}	\\
UZ For b	&	ETV	&\citet{Potter2011MNRAS.416.2202P}	\\
UZ For c	&	ETV	&\citet{Potter2011MNRAS.416.2202P}	\\
PSR B1620-26 b &	PTV	&\citet{Sigurdsson2003Sci...301..193S}	\\
OGLE-2007-BLG-349L c&ML& \citet{Bennett2016AJ....152..125B}\\
OGLE-2016-BLG-0613L b&ML& \citet{Han2017AJ....154..223H}\\
OGLE-2018-BLG-1700L b&ML&\citet{Han2020AJ....159...48H}\\
OGLE-2019-BLG-1470L c&ML&\citet{Kuang2022MNRAS.516.1704K}\\
OGLE-2023-BLG-0836L b&ML&\citet{Han2024AA...685A..16H}\\
2MASS J0103-5515 b&Im&\citet{Delorme2013AA...553L...5D}\\
2MASS J0249-0557 c&Im&\citet{Dupuy2018AJ....156...57D}\\
HD 284149 b&Im&\citet{Bonavita2017AA...608A.106B}\\
HIP 79098 b&Im&\citet{Janson2019AA...626A..99J}\\
ROXs 42 B b&Im&\citet{Currie2014ApJ...780L..30C}\\
Ross 458 c&Im&\citet{Burgasser2010ApJ...725.1405B}\\
SR 12 c&Im&\citet{Kuzuhara2011AJ....141..119K}\\
VHS J1256-1257 b&Im&\citet{Gauza2015ApJ...804...96G}\\
b Cen b&Im&\citet{Janson2021Natur.600..231J}\\
\enddata
\tablenotetext{a}{\scriptsize Method abbreviations: Tr = Transit; RV = Radial velocities; ETV = Eclipse timing variations; PTV = Pulsar timing variations; ML = Microlensing; Im = Imaging.}
\end{deluxetable}

\begin{table*}
\centering
\caption{Example results from Lomb–Scargle analyses of the eclipse timing variations (ETVs) of selected targets. Shown are binary and ETV periods, amplitudes, false alarm probabilities (FAPs), and derived companion masses. The complete dataset is provided in machine-readable format.}\label{tab:etv_results}
\begin{tabular}{lccccccccc}
\hline\hline
TIC & BJD$_0$ - 2457000 & $P_{\rm bin}$ [d] & $T_{\rm mag}$ & RUWE$^{a}$ & $P_{\rm ETV}$ [d] & Amp [min] & FAP & $m\sin i$ [$M_{\rm Jup}$] & $m$ [$M_{\rm Jup}$] \\
\hline
142979644 & 1687.545642 & 4.179031  & 11.62 & 1.101 & 534.96  & 0.33 & $2.5\times10^{-07}$ & 18.81  & 18.84 \\
166090445 & 1712.805666 & 5.324612  & 11.72 & 1.280 & 544.08  & 0.46 & $1.3\times10^{-03}$ & 25.88  & 25.90 \\
270648838 & 1521.746137 & 7.635327  & 10.64 & 1.134 & 158.30  & 0.14 & $4.5\times10^{-02}$ & 41.70  & 41.71 \\
66355834  & 1744.816358 & 10.369268 &  9.98 & 1.416 & 115.85  & 0.22 & $6.1\times10^{-01}$ & 90.22  & 90.26 \\
180412528 & 1326.289985 & 4.587005  & 10.76 & 1.312 & 78.05   & 0.30 & $2.0\times10^{-02}$ & 94.88  & 95.14 \\
350297040 & 1469.801653 & 10.873920 & 11.16 & 0.876 & 67.72   & 0.15 & $7.1\times10^{-02}$ & 96.26  & 96.33 \\
3816260   & 1390.527157 & 13.127871 & 10.09 & 1.046 & 67.15   & 0.15 & $1.0\times10^{-03}$ & 98.31  & 98.38 \\
140659980 & 1602.037316 & 9.440072  &  8.98 & 9.945 & 406.61  & 0.62 & $2.8\times10^{-02}$ & 120.58 & 120.85 \\
371706494 & 1827.605769 & 13.326036 & 11.63 & 1.004 & 60.34   & 0.23 & $2.7\times10^{-01}$ & 138.92 & 138.95 \\
148611095 & 1546.984106 & 4.837470  & 12.08 & 1.117 & 1774.51 & 4.05 & $3.7\times10^{-01}$ & 161.81 & 161.97 \\
\hline
\end{tabular}
\begin{flushleft}
$^{a}$ Renormalized Unit Weight Error \citep{RUWE2021A&A...649A...2L, Stassun2021ApJ...907L..33S}.
\end{flushleft}
\end{table*}

\begin{deluxetable}{rrr}
\tablecaption{Mid-eclipse times and errors from \texttt{allesfitter} models.\label{tab4}}
\tablehead{
\colhead{TIC} & \colhead{Mid$_{BJD}$} & \colhead{Cycle}
}
\startdata
 101395259&2458531.51984953$_{-0.00137407}^{+0.00135368}$ & 1 \\
 101395259&2458546.66920859$_{-0.00135829}^{+0.00148552}$ & 3 \\
 101395259&2458554.24388811$_{-0.00148822}^{+0.00140903}$ & 4 \\
 101395259&2458561.81856764$_{-0.00140932}^{+0.00142807}$ & 5 \\
 101395259&2458569.39324717$_{-0.00144174}^{+0.00139758}$ & 6 \\
\enddata
\tablecomments{Table \ref{tab4} is published in its entirety in the electronic 
edition of the {\it Astronomical Journal}.  A portion is shown here 
for guidance regarding its form and content.}
\end{deluxetable}

\begin{deluxetable}{rrrr}
\tablecaption{Mid-eclipse times and errors from the generalized Gaussian fits.\label{tab5}}
\tablehead{
\colhead{TIC} & \colhead{Type\tablenotemark{a}} & \colhead{Mid$_{BJD}$} & \colhead{Cycle}
}
\startdata
   3816260&sap& 2458387.76362195$\pm$0.00009764  & -0.5 \\
   3816260&sap& 2458390.52641895$\pm$0.00008267  &  0.0 \\
   3816260&sap& 2458400.89183172$\pm$0.00010196  &  0.5 \\
   3816260&sap& 2458403.65447792$\pm$0.00008064  &  1.0 \\
   3816260&sap& 2459451.14317993$\pm$0.00010055  & 80.5 \\
\enddata
\tablecomments{Table \ref{tab5} is published in its entirety in the electronic
edition of the {\it Astronomical Journal}.  A portion is shown here
for guidance regarding its form and content.}
\tablenotetext{a}{‘sap’ = Simple Aperture Photometry flux, ‘pdc’ = Presearch Data Conditioning flux, ‘ffi’ = flux extracted from Full Frame Images}
\end{deluxetable}

\onecolumngrid

\begin{figure*}
    \centering
    \includegraphics[width=1\columnwidth]{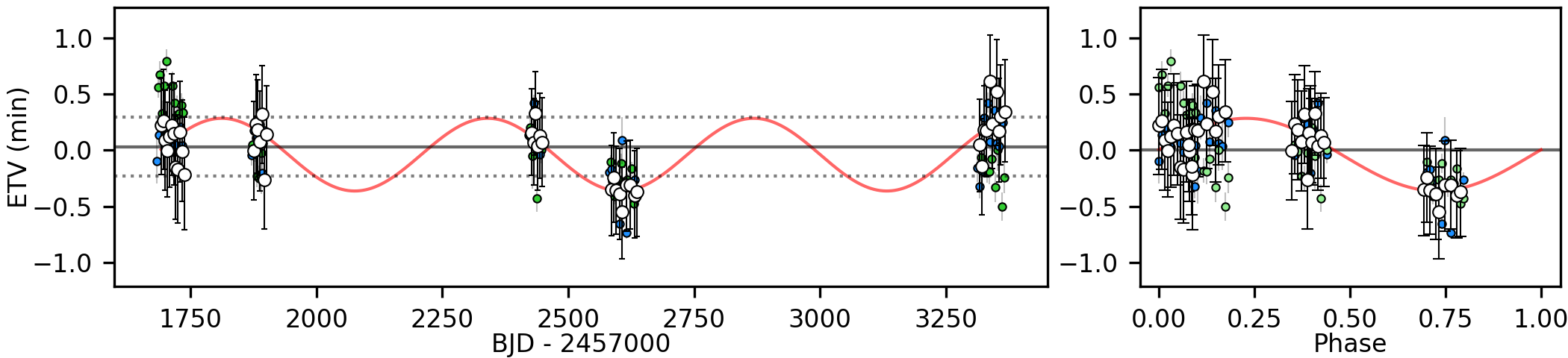}
    \caption{Eclipse Timing Variations (ETVs) of TIC 142979644 calculated with \texttt{allesfitter}. The left panel shows the timing data from \texttt{allesfitter} as white circles, with the red curve representing the model corresponding to the strongest peak in the Lomb–Scargle periodogram. The vertical gray dotted lines indicate the $\pm1\sigma$ intervals. For comparison, we overplotted the timing data from generalized Gaussian models, with primary eclipses shown as blue markers and secondary eclipses as green markers. The right panel presents the phase-folded ETVs with an arbitrary phase zero point for visualization.}
    \label{fig:etv_142979644}
\end{figure*}

\begin{figure*}
    \centering
    \includegraphics[width=0.8\columnwidth]{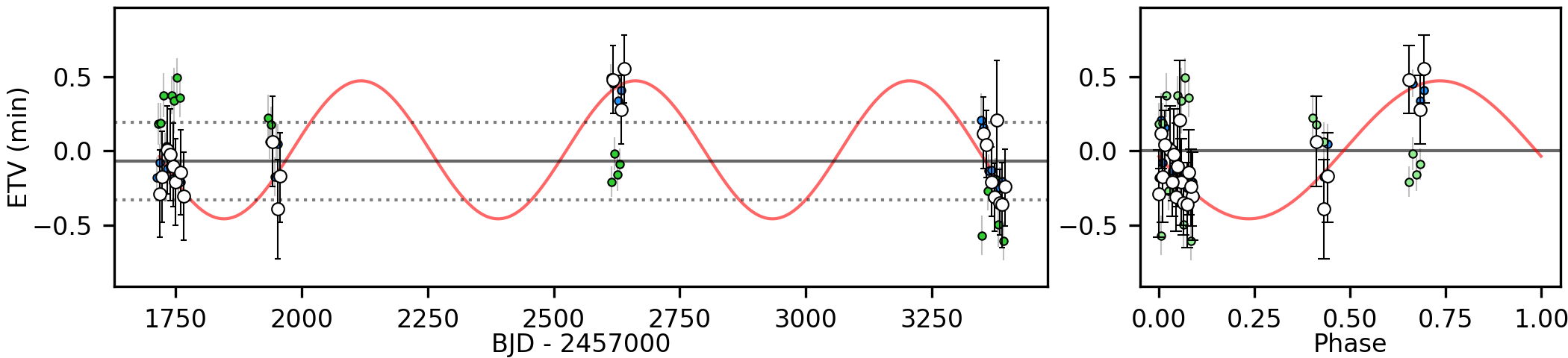}
    \caption{ETV of TIC 166090445. See caption of Figure \ref{fig:etv_142979644} for explanation of symbols and colors.}
    \label{fig:etv_166090445}
\end{figure*}

\begin{figure*}
    \centering
    \includegraphics[width=0.8\columnwidth]{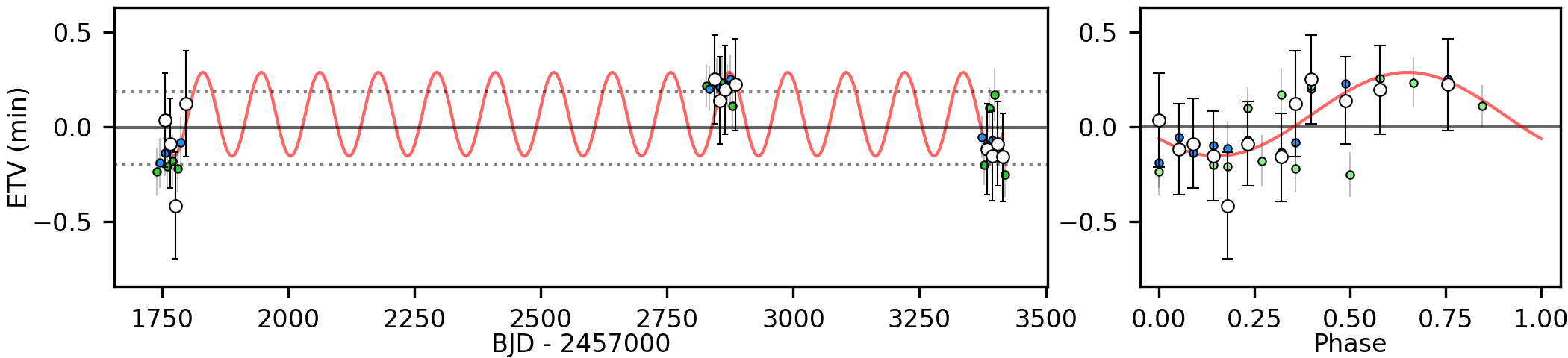}
    \caption{ETV of TIC 66355834. See caption of Figure \ref{fig:etv_142979644} for explanation of symbols and colors.}
    \label{fig:etv_66355834}
\end{figure*}

\begin{figure*}
    \centering
    \includegraphics[width=0.8\columnwidth]{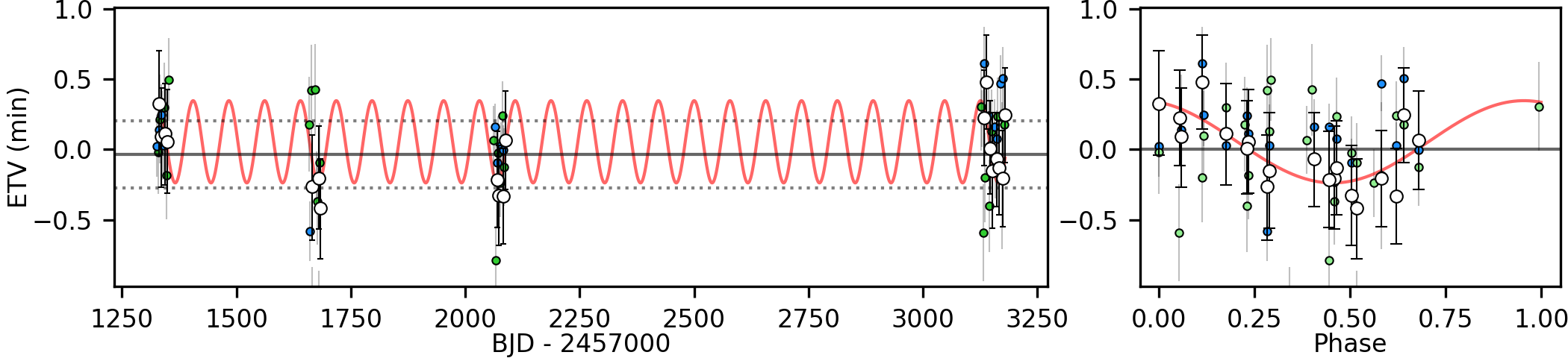}
    \caption{ETV of TIC 180412528. See caption of Figure \ref{fig:etv_142979644} for explanation of symbols and colors.}
    \label{fig:etv_180412528}
\end{figure*}

\begin{figure*}
    \centering
    \includegraphics[width=0.8\columnwidth]{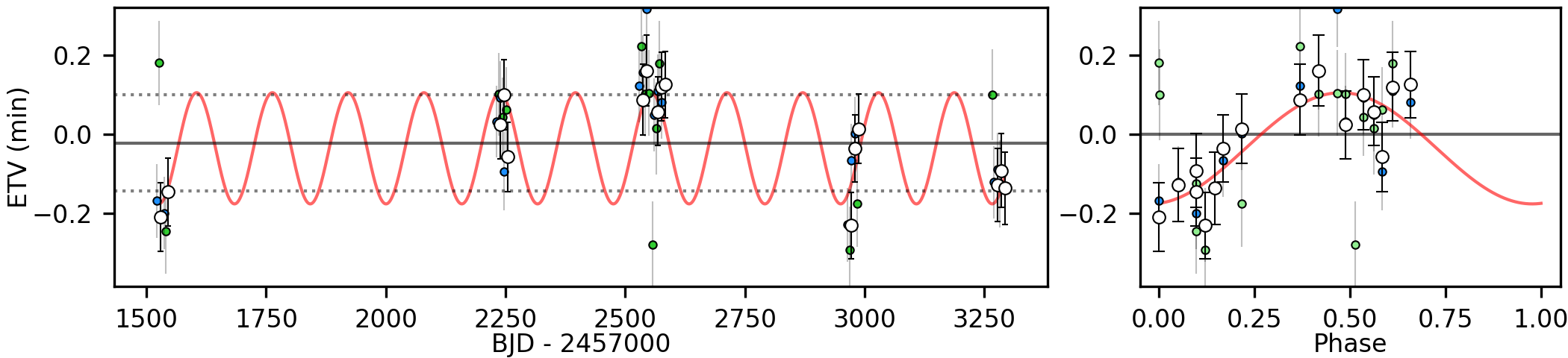}
    \caption{ETV of TIC 270648838. See caption of Figure \ref{fig:etv_142979644} for explanation of symbols and colors.}
    \label{fig:etv_270648838}
\end{figure*}

\begin{figure*}
    \centering
    \includegraphics[width=0.8\columnwidth]{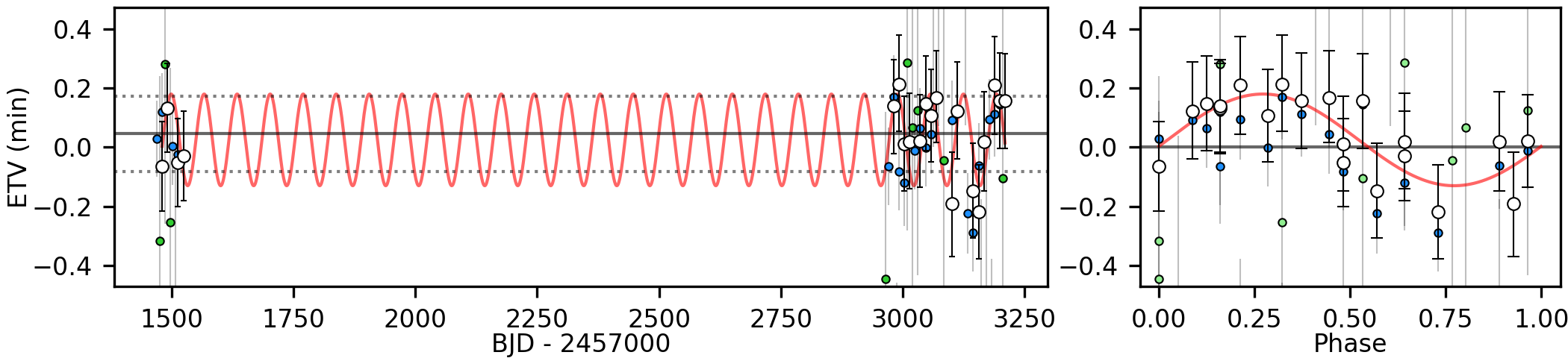}
    \caption{ETV of TIC 350297040. See caption of Figure \ref{fig:etv_142979644} for explanation of symbols and colors.}
    \label{fig:etv_350297040}
\end{figure*}

\bibliography{etv_tess}{}

\begin{thebibliography}{}
\expandafter\ifx\csname natexlab\endcsname\relax\def\natexlab#1{#1}\fi
\providecommand{\url}[1]{\href{#1}{#1}}
\providecommand{\dodoi}[1]{doi:~\href{http://doi.org/#1}{\nolinkurl{#1}}}
\providecommand{\doeprint}[1]{\href{http://ascl.net/#1}{\nolinkurl{http://ascl.net/#1}}}
\providecommand{\doarXiv}[1]{\href{https://arxiv.org/abs/#1}{\nolinkurl{https://arxiv.org/abs/#1}}}

\bibitem[{{Arancibia-Rojas} {et~al.}(2024){Arancibia-Rojas}, {Zorotovic}, {Vu{\v{c}}kovi{\'c}}, {Bobrick}, {Vos}, \& {Piraino-Cerda}}]{Arancibia-Rojas2024MNRAS.52711184A}
{Arancibia-Rojas}, E., {Zorotovic}, M., {Vu{\v{c}}kovi{\'c}}, M., {et~al.} 2024, \mnras, 527, 11184, \dodoi{10.1093/mnras/stad3891}

\bibitem[{{Armstrong} {et~al.}(2014){Armstrong}, {Osborn}, {Brown}, {Faedi}, {G{\'o}mez Maqueo Chew}, {Martin}, {Pollacco}, \& {Udry}}]{Armstrong2014MNRAS.444.1873A}
{Armstrong}, D.~J., {Osborn}, H.~P., {Brown}, D.~J.~A., {et~al.} 2014, \mnras, 444, 1873, \dodoi{10.1093/mnras/stu1570}

\bibitem[{{Astropy Collaboration} {et~al.}(2013){Astropy Collaboration}, {Robitaille}, {Tollerud}, {Greenfield}, {Droettboom}, {Bray}, {Aldcroft}, {Davis}, {Ginsburg}, {Price-Whelan}, {Kerzendorf}, {Conley}, {Crighton}, {Barbary}, {Muna}, {Ferguson}, {Grollier}, {Parikh}, {Nair}, {Unther}, {Deil}, {Woillez}, {Conseil}, {Kramer}, {Turner}, {Singer}, {Fox}, {Weaver}, {Zabalza}, {Edwards}, {Azalee Bostroem}, {Burke}, {Casey}, {Crawford}, {Dencheva}, {Ely}, {Jenness}, {Labrie}, {Lim}, {Pierfederici}, {Pontzen}, {Ptak}, {Refsdal}, {Servillat}, \& {Streicher}}]{astropy:2013}
{Astropy Collaboration}, {Robitaille}, T.~P., {Tollerud}, E.~J., {et~al.} 2013, \aap, 558, A33, \dodoi{10.1051/0004-6361/201322068}

\bibitem[{{Astropy Collaboration} {et~al.}(2018){Astropy Collaboration}, {Price-Whelan}, {Sip{\H{o}}cz}, {G{\"u}nther}, {Lim}, {Crawford}, {Conseil}, {Shupe}, {Craig}, {Dencheva}, {Ginsburg}, {Vand erPlas}, {Bradley}, {P{\'e}rez-Su{\'a}rez}, {de Val-Borro}, {Aldcroft}, {Cruz}, {Robitaille}, {Tollerud}, {Ardelean}, {Babej}, {Bach}, {Bachetti}, {Bakanov}, {Bamford}, {Barentsen}, {Barmby}, {Baumbach}, {Berry}, {Biscani}, {Boquien}, {Bostroem}, {Bouma}, {Brammer}, {Bray}, {Breytenbach}, {Buddelmeijer}, {Burke}, {Calderone}, {Cano Rodr{\'\i}guez}, {Cara}, {Cardoso}, {Cheedella}, {Copin}, {Corrales}, {Crichton}, {D'Avella}, {Deil}, {Depagne}, {Dietrich}, {Donath}, {Droettboom}, {Earl}, {Erben}, {Fabbro}, {Ferreira}, {Finethy}, {Fox}, {Garrison}, {Gibbons}, {Goldstein}, {Gommers}, {Greco}, {Greenfield}, {Groener}, {Grollier}, {Hagen}, {Hirst}, {Homeier}, {Horton}, {Hosseinzadeh}, {Hu}, {Hunkeler}, {Ivezi{\'c}}, {Jain}, {Jenness}, {Kanarek}, {Kendrew}, {Kern}, {Kerzendorf}, {Khvalko}, {King}, {Kirkby}, {Kulkarni},
  {Kumar}, {Lee}, {Lenz}, {Littlefair}, {Ma}, {Macleod}, {Mastropietro}, {McCully}, {Montagnac}, {Morris}, {Mueller}, {Mumford}, {Muna}, {Murphy}, {Nelson}, {Nguyen}, {Ninan}, {N{\"o}the}, {Ogaz}, {Oh}, {Parejko}, {Parley}, {Pascual}, {Patil}, {Patil}, {Plunkett}, {Prochaska}, {Rastogi}, {Reddy Janga}, {Sabater}, {Sakurikar}, {Seifert}, {Sherbert}, {Sherwood-Taylor}, {Shih}, {Sick}, {Silbiger}, {Singanamalla}, {Singer}, {Sladen}, {Sooley}, {Sornarajah}, {Streicher}, {Teuben}, {Thomas}, {Tremblay}, {Turner}, {Terr{\'o}n}, {van Kerkwijk}, {de la Vega}, {Watkins}, {Weaver}, {Whitmore}, {Woillez}, {Zabalza}, \& {Astropy Contributors}}]{astropy:2018}
{Astropy Collaboration}, {Price-Whelan}, A.~M., {Sip{\H{o}}cz}, B.~M., {et~al.} 2018, \aj, 156, 123, \dodoi{10.3847/1538-3881/aabc4f}

\bibitem[{{Astropy Collaboration} {et~al.}(2022){Astropy Collaboration}, {Price-Whelan}, {Lim}, {Earl}, {Starkman}, {Bradley}, {Shupe}, {Patil}, {Corrales}, {Brasseur}, {N{"o}the}, {Donath}, {Tollerud}, {Morris}, {Ginsburg}, {Vaher}, {Weaver}, {Tocknell}, {Jamieson}, {van Kerkwijk}, {Robitaille}, {Merry}, {Bachetti}, {G{"u}nther}, {Aldcroft}, {Alvarado-Montes}, {Archibald}, {B{'o}di}, {Bapat}, {Barentsen}, {Baz{'a}n}, {Biswas}, {Boquien}, {Burke}, {Cara}, {Cara}, {Conroy}, {Conseil}, {Craig}, {Cross}, {Cruz}, {D'Eugenio}, {Dencheva}, {Devillepoix}, {Dietrich}, {Eigenbrot}, {Erben}, {Ferreira}, {Foreman-Mackey}, {Fox}, {Freij}, {Garg}, {Geda}, {Glattly}, {Gondhalekar}, {Gordon}, {Grant}, {Greenfield}, {Groener}, {Guest}, {Gurovich}, {Handberg}, {Hart}, {Hatfield-Dodds}, {Homeier}, {Hosseinzadeh}, {Jenness}, {Jones}, {Joseph}, {Kalmbach}, {Karamehmetoglu}, {Ka{l}uszy{'n}ski}, {Kelley}, {Kern}, {Kerzendorf}, {Koch}, {Kulumani}, {Lee}, {Ly}, {Ma}, {MacBride}, {Maljaars}, {Muna}, {Murphy}, {Norman}, {O'Steen},
  {Oman}, {Pacifici}, {Pascual}, {Pascual-Granado}, {Patil}, {Perren}, {Pickering}, {Rastogi}, {Roulston}, {Ryan}, {Rykoff}, {Sabater}, {Sakurikar}, {Salgado}, {Sanghi}, {Saunders}, {Savchenko}, {Schwardt}, {Seifert-Eckert}, {Shih}, {Jain}, {Shukla}, {Sick}, {Simpson}, {Singanamalla}, {Singer}, {Singhal}, {Sinha}, {Sip{H{o}}cz}, {Spitler}, {Stansby}, {Streicher}, {{{S}}umak}, {Swinbank}, {Taranu}, {Tewary}, {Tremblay}, {Val-Borro}, {Van Kooten}, {Vasovi{'c}}, {Verma}, {de Miranda Cardoso}, {Williams}, {Wilson}, {Winkel}, {Wood-Vasey}, {Xue}, {Yoachim}, {Zhang}, {Zonca}, \& {Astropy Project Contributors}}]{astropy:2022}
{Astropy Collaboration}, {Price-Whelan}, A.~M., {Lim}, P.~L., {et~al.} 2022, \apj, 935, 167, \dodoi{10.3847/1538-4357/ac7c74}

\bibitem[{{Ba{\c{s}}t{\"u}rk} {et~al.}(2023){Ba{\c{s}}t{\"u}rk}, {Esmer}, {Demir}, \& {Selam}}]{Basturk2023BSRSL..9211197B}
{Ba{\c{s}}t{\"u}rk}, {\"O}., {Esmer}, E.~M., {Demir}, E., \& {Selam}, S.~O. 2023, Bulletin de la Societe Royale des Sciences de Liege, 92, 11197, \dodoi{10.25518/0037-9565.11197}

\bibitem[{{Baraffe} {et~al.}(2002){Baraffe}, {Chabrier}, {Allard}, \& {Hauschildt}}]{Baraffe2002A&A...382..563B}
{Baraffe}, I., {Chabrier}, G., {Allard}, F., \& {Hauschildt}, P.~H. 2002, \aap, 382, 563, \dodoi{10.1051/0004-6361:20011638}

\bibitem[{{Baran} {et~al.}(2015){Baran}, {Zola}, {Blokesz}, {{\O}stensen}, \& {Silvotti}}]{Baran2015}
{Baran}, A.~S., {Zola}, S., {Blokesz}, A., {{\O}stensen}, R.~H., \& {Silvotti}, R. 2015, \aap, 577, A146, \dodoi{10.1051/0004-6361/201425392}

\bibitem[{{Baycroft} {et~al.}(2025){Baycroft}, {Santerne}, {Triaud}, {Heidari}, {Sebastian}, {Davis}, {Correia}, {Sairam}, {Freckelton}, {Adamson}, {Boisse}, {Coleman}, {Dransfield}, {Faria}, {Grouffal}, {Hara}, {H{\'e}brard}, {Kunovac}, {Martin}, {Maxted}, {Nelson}, {Scott}, {Scutt}, \& {Standing}}]{Baycroft2025MNRAS.541.2801B}
{Baycroft}, T.~A., {Santerne}, A., {Triaud}, A. H.~M.~J., {et~al.} 2025, \mnras, 541, 2801, \dodoi{10.1093/mnras/staf1184}

\bibitem[{{Bear} \& {Soker}(2014)}]{Bear2014MNRAS.444.1698B}
{Bear}, E., \& {Soker}, N. 2014, \mnras, 444, 1698, \dodoi{10.1093/mnras/stu1529}

\bibitem[{{Benedict} \& {Harrison}(2017)}]{Benedict2017AJ....153..258B}
{Benedict}, G.~F., \& {Harrison}, T.~E. 2017, \aj, 153, 258, \dodoi{10.3847/1538-3881/aa6d59}

\bibitem[{{Bennett} {et~al.}(2016){Bennett}, {Rhie}, {Udalski}, {Gould}, {Tsapras}, {Kubas}, {Bond}, {Greenhill}, {Cassan}, {Rattenbury}, {Boyajian}, {Luhn}, {Penny}, {Anderson}, {Abe}, {Bhattacharya}, {Botzler}, {Donachie}, {Freeman}, {Fukui}, {Hirao}, {Itow}, {Koshimoto}, {Li}, {Ling}, {Masuda}, {Matsubara}, {Muraki}, {Nagakane}, {Ohnishi}, {Oyokawa}, {Perrott}, {Saito}, {Sharan}, {Sullivan}, {Sumi}, {Suzuki}, {Tristram}, {Yonehara}, {Yock}, {MOA Collaboration}, {Szyma{\'n}ski}, {Soszy{\'n}ski}, {Ulaczyk}, {Wyrzykowski}, {OGLE Collaboration}, {Allen}, {DePoy}, {Gal-Yam}, {Gaudi}, {Han}, {Monard}, {Ofek}, {Pogge}, {{\ensuremath{\mu}}FUN Collaboration}, {Street}, {Bramich}, {Dominik}, {Horne}, {Snodgrass}, {Steele}, {Robonet Collaboration}, {Albrow}, {Bachelet}, {Batista}, {Beaulieu}, {Brillant}, {Caldwell}, {Cole}, {Coutures}, {Dieters}, {Dominis Prester}, {Donatowicz}, {Fouqu{\'e}}, {Hundertmark}, {J{\o}rgensen}, {Kains}, {Kane}, {Marquette}, {Menzies}, {Pollard}, {Ranc}, {Sahu}, {Wambsganss}, {Williams},
  {Zub}, \& {PLANET Collaboration}}]{Bennett2016AJ....152..125B}
{Bennett}, D.~P., {Rhie}, S.~H., {Udalski}, A., {et~al.} 2016, \aj, 152, 125, \dodoi{10.3847/0004-6256/152/5/125}

\bibitem[{{Beuermann} {et~al.}(2010){Beuermann}, {Hessman}, {Dreizler}, {Marsh}, {Parsons}, {Winget}, {Miller}, {Schreiber}, {Kley}, {Dhillon}, {Littlefair}, {Copperwheat}, \& {Hermes}}]{Beuermann2010AA...521L..60B}
{Beuermann}, K., {Hessman}, F.~V., {Dreizler}, S., {et~al.} 2010, \aap, 521, L60, \dodoi{10.1051/0004-6361/201015472}

\bibitem[{{Bonavita} {et~al.}(2016){Bonavita}, {Desidera}, {Thalmann}, {Janson}, {Vigan}, {Chauvin}, \& {Lannier}}]{Bonavita2016A&A...593A..38B}
{Bonavita}, M., {Desidera}, S., {Thalmann}, C., {et~al.} 2016, \aap, 593, A38, \dodoi{10.1051/0004-6361/201628231}

\bibitem[{{Bonavita} {et~al.}(2017){Bonavita}, {D'Orazi}, {Mesa}, {Fontanive}, {Desidera}, {Messina}, {Daemgen}, {Gratton}, {Vigan}, {Bonnefoy}, {Zurlo}, {Antichi}, {Avenhaus}, {Baruffolo}, {Baudino}, {Beuzit}, {Boccaletti}, {Bruno}, {Buey}, {Carbillet}, {Cascone}, {Chauvin}, {Claudi}, {De Caprio}, {Fantinel}, {Farisato}, {Feldt}, {Galicher}, {Giro}, {Gry}, {Hagelberg}, {Incorvaia}, {Janson}, {Jaquet}, {Lagrange}, {Langlois}, {Lannier}, {Le Coroller}, {Lessio}, {Ligi}, {Maire}, {Meyer}, {Menard}, {Perrot}, {Peretti}, {Petit}, {Ramos}, {Roux}, {Salasnich}, {Salter}, {Samland}, {Scuderi}, {Schlieder}, {Surez}, {Turatto}, \& {Weber}}]{Bonavita2017AA...608A.106B}
{Bonavita}, M., {D'Orazi}, V., {Mesa}, D., {et~al.} 2017, \aap, 608, A106, \dodoi{10.1051/0004-6361/201731003}

\bibitem[{{Borkovits} {et~al.}(2016){Borkovits}, {Hajdu}, {Sztakovics}, {Rappaport}, {Levine}, {B{\'\i}r{\'o}}, \& {Klagyivik}}]{Borkovits2016MNRAS.455.4136B}
{Borkovits}, T., {Hajdu}, T., {Sztakovics}, J., {et~al.} 2016, \mnras, 455, 4136, \dodoi{10.1093/mnras/stv2530}

\bibitem[{{Borkovits} {et~al.}(2015){Borkovits}, {Rappaport}, {Hajdu}, \& {Sztakovics}}]{Borkovits2015MNRAS.448..946B}
{Borkovits}, T., {Rappaport}, S., {Hajdu}, T., \& {Sztakovics}, J. 2015, \mnras, 448, 946, \dodoi{10.1093/mnras/stv015}

\bibitem[{{Bours} {et~al.}(2016){Bours}, {Marsh}, {Parsons}, {Dhillon}, {Ashley}, {Bento}, {Breedt}, {Butterley}, {Caceres}, {Chote}, {Copperwheat}, {Hardy}, {Hermes}, {Irawati}, {Kerry}, {Kilkenny}, {Littlefair}, {McAllister}, {Rattanasoon}, {Sahman}, {Vu{\v{c}}kovi{\'c}}, \& {Wilson}}]{Bours2016MNRAS.460.3873B}
{Bours}, M.~C.~P., {Marsh}, T.~R., {Parsons}, S.~G., {et~al.} 2016, \mnras, 460, 3873, \dodoi{10.1093/mnras/stw1203}

\bibitem[{{Burgasser} {et~al.}(2010){Burgasser}, {Simcoe}, {Bochanski}, {Saumon}, {Mamajek}, {Cushing}, {Marley}, {McMurtry}, {Pipher}, \& {Forrest}}]{Burgasser2010ApJ...725.1405B}
{Burgasser}, A.~J., {Simcoe}, R.~A., {Bochanski}, J.~J., {et~al.} 2010, \apj, 725, 1405, \dodoi{10.1088/0004-637X/725/2/1405}

\bibitem[{{Burke} {et~al.}(2020){Burke}, {Levine}, {Fausnaugh}, {Vanderspek}, {Barclay}, {Libby-Roberts}, {Morris}, {Sipocz}, {Owens}, {Feinstein}, \& {Camacho}}]{TESSPointBurke2020ascl.soft03001B}
{Burke}, C.~J., {Levine}, A., {Fausnaugh}, M., {et~al.} 2020, {TESS-Point: High precision TESS pointing tool}, Astrophysics Source Code Library.
\newblock \doeprint{2003.001}

\bibitem[{{Ca{\~n}as} {et~al.}(2022){Ca{\~n}as}, {Mahadevan}, {Bender}, {Salazar Rivera}, {Monson}, {Beard}, {Lubin}, {Robertson}, {Gupta}, {Cochran}, {Fredrick}, {Hearty}, {Jones}, {Kanodia}, {Lin}, {Ninan}, {Ramsey}, {Schwab}, \& {Stef{\'a}nsson}}]{Canas2022}
{Ca{\~n}as}, C.~I., {Mahadevan}, S., {Bender}, C.~F., {et~al.} 2022, \aj, 163, 89, \dodoi{10.3847/1538-3881/ac415f}

\bibitem[{{Chabrier}(2003)}]{Chabrier2003}
{Chabrier}, G. 2003, \pasp, 115, 763, \dodoi{10.1086/376392}

\bibitem[{{Chavez} {et~al.}(2015){Chavez}, {Georgakarakos}, {Prodan}, {Reyes-Ruiz}, {Aceves}, {Betancourt}, \& {Perez-Tijerina}}]{Chavez2015MNRAS.446.1283C}
{Chavez}, C.~E., {Georgakarakos}, N., {Prodan}, S., {et~al.} 2015, \mnras, 446, 1283, \dodoi{10.1093/mnras/stu2142}

\bibitem[{{Chen} {et~al.}(2023){Chen}, {Lubow}, {Martin}, \& {Nixon}}]{Chen2023MNRAS.521.5033C}
{Chen}, C., {Lubow}, S.~H., {Martin}, R.~G., \& {Nixon}, C.~J. 2023, \mnras, 521, 5033, \dodoi{10.1093/mnras/stad739}

\bibitem[{{Chen} {et~al.}(2024){Chen}, {Martin}, {Lubow}, \& {Nixon}}]{Chen2024ApJ...961L...5C}
{Chen}, C., {Martin}, R.~G., {Lubow}, S.~H., \& {Nixon}, C.~J. 2024, \apjl, 961, L5, \dodoi{10.3847/2041-8213/ad17c5}

\bibitem[{{Coleman}(2024)}]{Coleman2024MNRAS.530..630C}
{Coleman}, G. A.~L. 2024, \mnras, 530, 630, \dodoi{10.1093/mnras/stae903}

\bibitem[{{Coleman} {et~al.}(2024){Coleman}, {Nelson}, {Triaud}, \& {Standing}}]{Coleman2024MNRAS.527..414C}
{Coleman}, G. A.~L., {Nelson}, R.~P., {Triaud}, A. H.~M.~J., \& {Standing}, M.~R. 2024, \mnras, 527, 414, \dodoi{10.1093/mnras/stad3216}

\bibitem[{{Correia} {et~al.}(2016){Correia}, {Bou{\'e}}, \& {Laskar}}]{Correia2016CeMDA.126..189C}
{Correia}, A. C.~M., {Bou{\'e}}, G., \& {Laskar}, J. 2016, Celestial Mechanics and Dynamical Astronomy, 126, 189, \dodoi{10.1007/s10569-016-9709-9}

\bibitem[{{Correia} {et~al.}(2005){Correia}, {Udry}, {Mayor}, {Laskar}, {Naef}, {Pepe}, {Queloz}, \& {Santos}}]{Correia2005AA...440..751C}
{Correia}, A.~C.~M., {Udry}, S., {Mayor}, M., {et~al.} 2005, \aap, 440, 751, \dodoi{10.1051/0004-6361:20042376}

\bibitem[{{Cunningham} {et~al.}(2024){Cunningham}, {Tremblay}, \& {W. O'Brien}}]{Cunningham2024MNRAS.527.3602C}
{Cunningham}, T., {Tremblay}, P.-E., \& {W. O'Brien}, M. 2024, \mnras, 527, 3602, \dodoi{10.1093/mnras/stad3275}

\bibitem[{{Currie} {et~al.}(2014){Currie}, {Daemgen}, {Debes}, {Lafreniere}, {Itoh}, {Jayawardhana}, {Ratzka}, \& {Correia}}]{Currie2014ApJ...780L..30C}
{Currie}, T., {Daemgen}, S., {Debes}, J., {et~al.} 2014, \apjl, 780, L30, \dodoi{10.1088/2041-8205/780/2/L30}

\bibitem[{{Czekala} {et~al.}(2019){Czekala}, {Chiang}, {Andrews}, {Jensen}, {Torres}, {Wilner}, {Stassun}, \& {Macintosh}}]{Czekala2019ApJ...883...22C}
{Czekala}, I., {Chiang}, E., {Andrews}, S.~M., {et~al.} 2019, \apj, 883, 22, \dodoi{10.3847/1538-4357/ab287b}

\bibitem[{{Deeg}(2020)}]{Deeg2020Galax...9....1D}
{Deeg}, H.~J. 2020, Galaxies, 9, 1, \dodoi{10.3390/galaxies9010001}

\bibitem[{{Deeg} \& {Tingley}(2017)}]{Deeg2017A&A...599A..93D}
{Deeg}, H.~J., \& {Tingley}, B. 2017, \aap, 599, A93, \dodoi{10.1051/0004-6361/201629350}

\bibitem[{{Delorme} {et~al.}(2013){Delorme}, {Gagn{\'e}}, {Girard}, {Lagrange}, {Chauvin}, {Naud}, {Lafreni{\`e}re}, {Doyon}, {Riedel}, {Bonnefoy}, \& {Malo}}]{Delorme2013AA...553L...5D}
{Delorme}, P., {Gagn{\'e}}, J., {Girard}, J.~H., {et~al.} 2013, \aap, 553, L5, \dodoi{10.1051/0004-6361/201321169}

\bibitem[{{Doyle} {et~al.}(2011){Doyle}, {Carter}, {Fabrycky}, {Slawson}, {Howell}, {Winn}, {Orosz}, {P{\v{r}}sa}, {Welsh}, {Quinn}, {Latham}, {Torres}, {Buchhave}, {Marcy}, {Fortney}, {Shporer}, {Ford}, {Lissauer}, {Ragozzine}, {Rucker}, {Batalha}, {Jenkins}, {Borucki}, {Koch}, {Middour}, {Hall}, {McCauliff}, {Fanelli}, {Quintana}, {Holman}, {Caldwell}, {Still}, {Stefanik}, {Brown}, {Esquerdo}, {Tang}, {Furesz}, {Geary}, {Berlind}, {Calkins}, {Short}, {Steffen}, {Sasselov}, {Dunham}, {Cochran}, {Boss}, {Haas}, {Buzasi}, \& {Fischer}}]{Doyle2011Sci...333.1602D}
{Doyle}, L.~R., {Carter}, J.~A., {Fabrycky}, D.~C., {et~al.} 2011, Science, 333, 1602, \dodoi{10.1126/science.1210923}

\bibitem[{{Dupuy} {et~al.}(2018){Dupuy}, {Liu}, {Allers}, {Biller}, {Kratter}, {Mann}, {Shkolnik}, {Kraus}, \& {Best}}]{Dupuy2018AJ....156...57D}
{Dupuy}, T.~J., {Liu}, M.~C., {Allers}, K.~N., {et~al.} 2018, \aj, 156, 57, \dodoi{10.3847/1538-3881/aacbc2}

\bibitem[{{Eker} {et~al.}(2018){Eker}, {Bak{\i}{\c{s}}}, {Bilir}, {Soydugan}, {Steer}, {Soydugan}, {Bak{\i}{\c{s}}}, {Ali{\c{c}}avu{\c{s}}}, {Aslan}, \& {Alpsoy}}]{Eker2018MNRAS.479.5491E}
{Eker}, Z., {Bak{\i}{\c{s}}}, V., {Bilir}, S., {et~al.} 2018, \mnras, 479, 5491, \dodoi{10.1093/mnras/sty1834}

\bibitem[{{Esmer} {et~al.}(2021){Esmer}, {Ba{\c{s}}t{\"u}rk}, {Hinse}, {Selam}, \& {Correia}}]{Esmer2021A&A...648A..85E}
{Esmer}, E.~M., {Ba{\c{s}}t{\"u}rk}, {\"O}., {Hinse}, T.~C., {Selam}, S.~O., \& {Correia}, A. C.~M. 2021, \aap, 648, A85, \dodoi{10.1051/0004-6361/202038640}

\bibitem[{{Esmer} {et~al.}(2023){Esmer}, {Ba{\c{s}}t{\"u}rk}, \& {Selam}}]{Esmer2023MNRAS.525.6050E}
{Esmer}, E.~M., {Ba{\c{s}}t{\"u}rk}, {\"O}., \& {Selam}, S.~O. 2023, \mnras, 525, 6050, \dodoi{10.1093/mnras/stad2648}

\bibitem[{{Esmer} {et~al.}(2024){Esmer}, {Ba{\c{s}}t{\"u}rk}, {Selam}, {Akar}, {Sertkan}, {G{\"u}ler}, \& {Yal{\c{c}}{\i}nkaya}}]{Esmer2024CoSka..54b.228E}
{Esmer}, E.~M., {Ba{\c{s}}t{\"u}rk}, {\"O}., {Selam}, S.~O., {et~al.} 2024, Contributions of the Astronomical Observatory Skalnate Pleso, 54, 228, \dodoi{10.31577/caosp.2024.54.2.228}

\bibitem[{{Esmer} {et~al.}(2022){Esmer}, {Ba{\c{s}}t{\"u}rk}, {Selam}, \& {Ali{\c{s}}}}]{Esmer2022MNRAS.511.5207E}
{Esmer}, E.~M., {Ba{\c{s}}t{\"u}rk}, {\"O}., {Selam}, S.~O., \& {Ali{\c{s}}}, S. 2022, \mnras, 511, 5207, \dodoi{10.1093/mnras/stac357}

\bibitem[{{Fleming} {et~al.}(2018){Fleming}, {Barnes}, {Graham}, {Luger}, \& {Quinn}}]{Fleming2018ApJ...858...86F}
{Fleming}, D.~P., {Barnes}, R., {Graham}, D.~E., {Luger}, R., \& {Quinn}, T.~R. 2018, \apj, 858, 86, \dodoi{10.3847/1538-4357/aabd38}

\bibitem[{Foreman-Mackey(2016)}]{Foreman-Mackey2016_Corner}
Foreman-Mackey, D. 2016, Journal of Open Source Software, 1, 24, \dodoi{10.21105/joss.00024}

\bibitem[{{Foucart} \& {Lai}(2013)}]{Foucart2013ApJ...764..106F}
{Foucart}, F., \& {Lai}, D. 2013, \apj, 764, 106, \dodoi{10.1088/0004-637X/764/1/106}

\bibitem[{{Gaia Collaboration} {et~al.}(2016){Gaia Collaboration}, {Prusti}, {de Bruijne}, {Brown}, {Vallenari}, {Babusiaux}, {Bailer-Jones}, {Bastian}, {Biermann}, {Evans}, {Eyer}, {Jansen}, {Jordi}, {Klioner}, {Lammers}, {Lindegren}, {Luri}, {Mignard}, {Milligan}, {Panem}, {Poinsignon}, {Pourbaix}, {Randich}, {Sarri}, {Sartoretti}, {Siddiqui}, {Soubiran}, {Valette}, {van Leeuwen}, {Walton}, {Aerts}, {Arenou}, {Cropper}, {Drimmel}, {H{\o}g}, {Katz}, {Lattanzi}, {O'Mullane}, {Grebel}, {Holland}, {Huc}, {Passot}, {Bramante}, {Cacciari}, {Casta{\~n}eda}, {Chaoul}, {Cheek}, {De Angeli}, {Fabricius}, {Guerra}, {Hern{\'a}ndez}, {Jean-Antoine-Piccolo}, {Masana}, {Messineo}, {Mowlavi}, {Nienartowicz}, {Ord{\'o}{\~n}ez-Blanco}, {Panuzzo}, {Portell}, {Richards}, {Riello}, {Seabroke}, {Tanga}, {Th{\'e}venin}, {Torra}, {Els}, {Gracia-Abril}, {Comoretto}, {Garcia-Reinaldos}, {Lock}, {Mercier}, {Altmann}, {Andrae}, {Astraatmadja}, {Bellas-Velidis}, {Benson}, {Berthier}, {Blomme}, {Busso}, {Carry}, {Cellino}, {Clementini},
  {Cowell}, {Creevey}, {Cuypers}, {Davidson}, {De Ridder}, {de Torres}, {Delchambre}, {Dell'Oro}, {Ducourant}, {Fr{\'e}mat}, {Garc{\'\i}a-Torres}, {Gosset}, {Halbwachs}, {Hambly}, {Harrison}, {Hauser}, {Hestroffer}, {Hodgkin}, {Huckle}, {Hutton}, {Jasniewicz}, {Jordan}, {Kontizas}, {Korn}, {Lanzafame}, {Manteiga}, {Moitinho}, {Muinonen}, {Osinde}, {Pancino}, {Pauwels}, {Petit}, {Recio-Blanco}, {Robin}, {Sarro}, {Siopis}, {Smith}, {Smith}, {Sozzetti}, {Thuillot}, {van Reeven}, {Viala}, {Abbas}, {Abreu Aramburu}, {Accart}, {Aguado}, {Allan}, {Allasia}, {Altavilla}, {{\'A}lvarez}, {Alves}, {Anderson}, {Andrei}, {Anglada Varela}, {Antiche}, {Antoja}, {Ant{\'o}n}, {Arcay}, {Atzei}, {Ayache}, {Bach}, {Baker}, {Balaguer-N{\'u}{\~n}ez}, {Barache}, {Barata}, {Barbier}, {Barblan}, {Baroni}, {Barrado y Navascu{\'e}s}, {Barros}, {Barstow}, {Becciani}, {Bellazzini}, {Bellei}, {Bello Garc{\'\i}a}, {Belokurov}, {Bendjoya}, {Berihuete}, {Bianchi}, {Bienaym{\'e}}, {Billebaud}, {Blagorodnova}, {Blanco-Cuaresma}, {Boch},
  {Bombrun}, {Borrachero}, {Bouquillon}, {Bourda}, {Bouy}, {Bragaglia}, {Breddels}, {Brouillet}, {Br{\"u}semeister}, {Bucciarelli}, {Budnik}, {Burgess}, {Burgon}, {Burlacu}, {Busonero}, {Buzzi}, {Caffau}, {Cambras}, {Campbell}, {Cancelliere}, {Cantat-Gaudin}, {Carlucci}, {Carrasco}, {Castellani}, {Charlot}, {Charnas}, {Charvet}, {Chassat}, {Chiavassa}, {Clotet}, {Cocozza}, {Collins}, {Collins}, {Costigan}, {Crifo}, {Cross}, {Crosta}, {Crowley}, {Dafonte}, {Damerdji}, {Dapergolas}, {David}, {David}, {De Cat}, {de Felice}, {de Laverny}, {De Luise}, {De March}, {de Martino}, {de Souza}, {Debosscher}, {del Pozo}, {Delbo}, {Delgado}, {Delgado}, {di Marco}, {Di Matteo}, {Diakite}, {Distefano}, {Dolding}, {Dos Anjos}, {Drazinos}, {Dur{\'a}n}, {Dzigan}, {Ecale}, {Edvardsson}, {Enke}, {Erdmann}, {Escolar}, {Espina}, {Evans}, {Eynard Bontemps}, {Fabre}, {Fabrizio}, {Faigler}, {Falc{\~a}o}, {Farr{\`a}s Casas}, {Faye}, {Federici}, {Fedorets}, {Fern{\'a}ndez-Hern{\'a}ndez}, {Fernique}, {Fienga}, {Figueras}, {Filippi},
  {Findeisen}, {Fonti}, {Fouesneau}, {Fraile}, {Fraser}, {Fuchs}, {Furnell}, {Gai}, {Galleti}, {Galluccio}, {Garabato}, {Garc{\'\i}a-Sedano}, {Gar{\'e}}, {Garofalo}, {Garralda}, {Gavras}, {Gerssen}, {Geyer}, {Gilmore}, {Girona}, {Giuffrida}, {Gomes}, {Gonz{\'a}lez-Marcos}, {Gonz{\'a}lez-N{\'u}{\~n}ez}, {Gonz{\'a}lez-Vidal}, {Granvik}, {Guerrier}, {Guillout}, {Guiraud}, {G{\'u}rpide}, {Guti{\'e}rrez-S{\'a}nchez}, {Guy}, {Haigron}, {Hatzidimitriou}, {Haywood}, {Heiter}, {Helmi}, {Hobbs}, {Hofmann}, {Holl}, {Holland}, {Hunt}, {Hypki}, {Icardi}, {Irwin}, {Jevardat de Fombelle}, {Jofr{\'e}}, {Jonker}, {Jorissen}, {Julbe}, {Karampelas}, {Kochoska}, {Kohley}, {Kolenberg}, {Kontizas}, {Koposov}, {Kordopatis}, {Koubsky}, {Kowalczyk}, {Krone-Martins}, {Kudryashova}, {Kull}, {Bachchan}, {Lacoste-Seris}, {Lanza}, {Lavigne}, {Le Poncin-Lafitte}, {Lebreton}, {Lebzelter}, {Leccia}, {Leclerc}, {Lecoeur-Taibi}, {Lemaitre}, {Lenhardt}, {Leroux}, {Liao}, {Licata}, {Lindstr{\o}m}, {Lister}, {Livanou}, {Lobel}, {L{\"o}ffler},
  {L{\'o}pez}, {Lopez-Lozano}, {Lorenz}, {Loureiro}, {MacDonald}, {Magalh{\~a}es Fernandes}, {Managau}, {Mann}, {Mantelet}, {Marchal}, {Marchant}, {Marconi}, {Marie}, {Marinoni}, {Marrese}, {Marschalk{\'o}}, {Marshall}, {Mart{\'\i}n-Fleitas}, {Martino}, {Mary}, {Matijevi{\v{c}}}, {Mazeh}, {McMillan}, {Messina}, {Mestre}, {Michalik}, {Millar}, {Miranda}, {Molina}, {Molinaro}, {Molinaro}, {Moln{\'a}r}, {Moniez}, {Montegriffo}, {Monteiro}, {Mor}, {Mora}, {Morbidelli}, {Morel}, {Morgenthaler}, {Morley}, {Morris}, {Mulone}, {Muraveva}, {Musella}, {Narbonne}, {Nelemans}, {Nicastro}, {Noval}, {Ord{\'e}novic}, {Ordieres-Mer{\'e}}, {Osborne}, {Pagani}, {Pagano}, {Pailler}, {Palacin}, {Palaversa}, {Parsons}, {Paulsen}, {Pecoraro}, {Pedrosa}, {Pentik{\"a}inen}, {Pereira}, {Pichon}, {Piersimoni}, {Pineau}, {Plachy}, {Plum}, {Poujoulet}, {Pr{\v{s}}a}, {Pulone}, {Ragaini}, {Rago}, {Rambaux}, {Ramos-Lerate}, {Ranalli}, {Rauw}, {Read}, {Regibo}, {Renk}, {Reyl{\'e}}, {Ribeiro}, {Rimoldini}, {Ripepi}, {Riva}, {Rixon},
  {Roelens}, {Romero-G{\'o}mez}, {Rowell}, {Royer}, {Rudolph}, {Ruiz-Dern}, {Sadowski}, {Sagrist{\`a} Sell{\'e}s}, {Sahlmann}, {Salgado}, {Salguero}, {Sarasso}, {Savietto}, {Schnorhk}, {Schultheis}, {Sciacca}, {Segol}, {Segovia}, {Segransan}, {Serpell}, {Shih}, {Smareglia}, {Smart}, {Smith}, {Solano}, {Solitro}, {Sordo}, {Soria Nieto}, {Souchay}, {Spagna}, {Spoto}, {Stampa}, {Steele}, {Steidelm{\"u}ller}, {Stephenson}, {Stoev}, {Suess}, {S{\"u}veges}, {Surdej}, {Szabados}, {Szegedi-Elek}, {Tapiador}, {Taris}, {Tauran}, {Taylor}, {Teixeira}, {Terrett}, {Tingley}, {Trager}, {Turon}, {Ulla}, {Utrilla}, {Valentini}, {van Elteren}, {Van Hemelryck}, {van Leeuwen}, {Varadi}, {Vecchiato}, {Veljanoski}, {Via}, {Vicente}, {Vogt}, {Voss}, {Votruba}, {Voutsinas}, {Walmsley}, {Weiler}, {Weingrill}, {Werner}, {Wevers}, {Whitehead}, {Wyrzykowski}, {Yoldas}, {{\v{Z}}erjal}, {Zucker}, {Zurbach}, {Zwitter}, {Alecu}, {Allen}, {Allende Prieto}, {Amorim}, {Anglada-Escud{\'e}}, {Arsenijevic}, {Azaz}, {Balm}, {Beck}, {Bernstein},
  {Bigot}, {Bijaoui}, {Blasco}, {Bonfigli}, {Bono}, {Boudreault}, {Bressan}, {Brown}, {Brunet}, {Bunclark}, {Buonanno}, {Butkevich}, {Carret}, {Carrion}, {Chemin}, {Ch{\'e}reau}, {Corcione}, {Darmigny}, {de Boer}, {de Teodoro}, {de Zeeuw}, {Delle Luche}, {Domingues}, {Dubath}, {Fodor}, {Fr{\'e}zouls}, {Fries}, {Fustes}, {Fyfe}, {Gallardo}, {Gallegos}, {Gardiol}, {Gebran}, {Gomboc}, {G{\'o}mez}, {Grux}, {Gueguen}, {Heyrovsky}, {Hoar}, {Iannicola}, {Isasi Parache}, {Janotto}, {Joliet}, {Jonckheere}, {Keil}, {Kim}, {Klagyivik}, {Klar}, {Knude}, {Kochukhov}, {Kolka}, {Kos}, {Kutka}, {Lainey}, {LeBouquin}, {Liu}, {Loreggia}, {Makarov}, {Marseille}, {Martayan}, {Martinez-Rubi}, {Massart}, {Meynadier}, {Mignot}, {Munari}, {Nguyen}, {Nordlander}, {Ocvirk}, {O'Flaherty}, {Olias Sanz}, {Ortiz}, {Osorio}, {Oszkiewicz}, {Ouzounis}, {Palmer}, {Park}, {Pasquato}, {Peltzer}, {Peralta}, {P{\'e}turaud}, {Pieniluoma}, {Pigozzi}, {Poels}, {Prat}, {Prod'homme}, {Raison}, {Rebordao}, {Risquez}, {Rocca-Volmerange}, {Rosen},
  {Ruiz-Fuertes}, {Russo}, {Sembay}, {Serraller Vizcaino}, {Short}, {Siebert}, {Silva}, {Sinachopoulos}, {Slezak}, {Soffel}, {Sosnowska}, {Strai{\v{z}}ys}, {ter Linden}, {Terrell}, {Theil}, {Tiede}, {Troisi}, {Tsalmantza}, {Tur}, {Vaccari}, {Vachier}, {Valles}, {Van Hamme}, {Veltz}, {Virtanen}, {Wallut}, {Wichmann}, {Wilkinson}, {Ziaeepour}, \& {Zschocke}}]{GaiaCollaboration2016A&A...595A...1G}
{Gaia Collaboration}, {Prusti}, T., {de Bruijne}, J.~H.~J., {et~al.} 2016, \aap, 595, A1, \dodoi{10.1051/0004-6361/201629272}

\bibitem[{{Gaia Collaboration} {et~al.}(2023){Gaia Collaboration}, {Vallenari}, {Brown}, {Prusti}, {de Bruijne}, {Arenou}, {Babusiaux}, {Biermann}, {Creevey}, {Ducourant}, {Evans}, {Eyer}, {Guerra}, {Hutton}, {Jordi}, {Klioner}, {Lammers}, {Lindegren}, {Luri}, {Mignard}, {Panem}, {Pourbaix}, {Randich}, {Sartoretti}, {Soubiran}, {Tanga}, {Walton}, {Bailer-Jones}, {Bastian}, {Drimmel}, {Jansen}, {Katz}, {Lattanzi}, {van Leeuwen}, {Bakker}, {Cacciari}, {Casta{\~n}eda}, {De Angeli}, {Fabricius}, {Fouesneau}, {Fr{\'e}mat}, {Galluccio}, {Guerrier}, {Heiter}, {Masana}, {Messineo}, {Mowlavi}, {Nicolas}, {Nienartowicz}, {Pailler}, {Panuzzo}, {Riclet}, {Roux}, {Seabroke}, {Sordo}, {Th{\'e}venin}, {Gracia-Abril}, {Portell}, {Teyssier}, {Altmann}, {Andrae}, {Audard}, {Bellas-Velidis}, {Benson}, {Berthier}, {Blomme}, {Burgess}, {Busonero}, {Busso}, {C{\'a}novas}, {Carry}, {Cellino}, {Cheek}, {Clementini}, {Damerdji}, {Davidson}, {de Teodoro}, {Nu{\~n}ez Campos}, {Delchambre}, {Dell'Oro}, {Esquej},
  {Fern{\'a}ndez-Hern{\'a}ndez}, {Fraile}, {Garabato}, {Garc{\'\i}a-Lario}, {Gosset}, {Haigron}, {Halbwachs}, {Hambly}, {Harrison}, {Hern{\'a}ndez}, {Hestroffer}, {Hodgkin}, {Holl}, {Jan{\ss}en}, {Jevardat de Fombelle}, {Jordan}, {Krone-Martins}, {Lanzafame}, {L{\"o}ffler}, {Marchal}, {Marrese}, {Moitinho}, {Muinonen}, {Osborne}, {Pancino}, {Pauwels}, {Recio-Blanco}, {Reyl{\'e}}, {Riello}, {Rimoldini}, {Roegiers}, {Rybizki}, {Sarro}, {Siopis}, {Smith}, {Sozzetti}, {Utrilla}, {van Leeuwen}, {Abbas}, {{\'A}brah{\'a}m}, {Abreu Aramburu}, {Aerts}, {Aguado}, {Ajaj}, {Aldea-Montero}, {Altavilla}, {{\'A}lvarez}, {Alves}, {Anders}, {Anderson}, {Anglada Varela}, {Antoja}, {Baines}, {Baker}, {Balaguer-N{\'u}{\~n}ez}, {Balbinot}, {Balog}, {Barache}, {Barbato}, {Barros}, {Barstow}, {Bartolom{\'e}}, {Bassilana}, {Bauchet}, {Becciani}, {Bellazzini}, {Berihuete}, {Bernet}, {Bertone}, {Bianchi}, {Binnenfeld}, {Blanco-Cuaresma}, {Blazere}, {Boch}, {Bombrun}, {Bossini}, {Bouquillon}, {Bragaglia}, {Bramante}, {Breedt},
  {Bressan}, {Brouillet}, {Brugaletta}, {Bucciarelli}, {Burlacu}, {Butkevich}, {Buzzi}, {Caffau}, {Cancelliere}, {Cantat-Gaudin}, {Carballo}, {Carlucci}, {Carnerero}, {Carrasco}, {Casamiquela}, {Castellani}, {Castro-Ginard}, {Chaoul}, {Charlot}, {Chemin}, {Chiaramida}, {Chiavassa}, {Chornay}, {Comoretto}, {Contursi}, {Cooper}, {Cornez}, {Cowell}, {Crifo}, {Cropper}, {Crosta}, {Crowley}, {Dafonte}, {Dapergolas}, {David}, {David}, {de Laverny}, {De Luise}, {De March}, {De Ridder}, {de Souza}, {de Torres}, {del Peloso}, {del Pozo}, {Delbo}, {Delgado}, {Delisle}, {Demouchy}, {Dharmawardena}, {Di Matteo}, {Diakite}, {Diener}, {Distefano}, {Dolding}, {Edvardsson}, {Enke}, {Fabre}, {Fabrizio}, {Faigler}, {Fedorets}, {Fernique}, {Fienga}, {Figueras}, {Fournier}, {Fouron}, {Fragkoudi}, {Gai}, {Garcia-Gutierrez}, {Garcia-Reinaldos}, {Garc{\'\i}a-Torres}, {Garofalo}, {Gavel}, {Gavras}, {Gerlach}, {Geyer}, {Giacobbe}, {Gilmore}, {Girona}, {Giuffrida}, {Gomel}, {Gomez}, {Gonz{\'a}lez-N{\'u}{\~n}ez},
  {Gonz{\'a}lez-Santamar{\'\i}a}, {Gonz{\'a}lez-Vidal}, {Granvik}, {Guillout}, {Guiraud}, {Guti{\'e}rrez-S{\'a}nchez}, {Guy}, {Hatzidimitriou}, {Hauser}, {Haywood}, {Helmer}, {Helmi}, {Sarmiento}, {Hidalgo}, {Hilger}, {H{\l}adczuk}, {Hobbs}, {Holland}, {Huckle}, {Jardine}, {Jasniewicz}, {Jean-Antoine Piccolo}, {Jim{\'e}nez-Arranz}, {Jorissen}, {Juaristi Campillo}, {Julbe}, {Karbevska}, {Kervella}, {Khanna}, {Kontizas}, {Kordopatis}, {Korn}, {K{\'o}sp{\'a}l}, {Kostrzewa-Rutkowska}, {Kruszy{\'n}ska}, {Kun}, {Laizeau}, {Lambert}, {Lanza}, {Lasne}, {Le Campion}, {Lebreton}, {Lebzelter}, {Leccia}, {Leclerc}, {Lecoeur-Taibi}, {Liao}, {Licata}, {Lindstr{\o}m}, {Lister}, {Livanou}, {Lobel}, {Lorca}, {Loup}, {Madrero Pardo}, {Magdaleno Romeo}, {Managau}, {Mann}, {Manteiga}, {Marchant}, {Marconi}, {Marcos}, {Marcos Santos}, {Mar{\'\i}n Pina}, {Marinoni}, {Marocco}, {Marshall}, {Martin Polo}, {Mart{\'\i}n-Fleitas}, {Marton}, {Mary}, {Masip}, {Massari}, {Mastrobuono-Battisti}, {Mazeh}, {McMillan}, {Messina}, {Michalik},
  {Millar}, {Mints}, {Molina}, {Molinaro}, {Moln{\'a}r}, {Monari}, {Mongui{\'o}}, {Montegriffo}, {Montero}, {Mor}, {Mora}, {Morbidelli}, {Morel}, {Morris}, {Muraveva}, {Murphy}, {Musella}, {Nagy}, {Noval}, {Oca{\~n}a}, {Ogden}, {Ordenovic}, {Osinde}, {Pagani}, {Pagano}, {Palaversa}, {Palicio}, {Pallas-Quintela}, {Panahi}, {Payne-Wardenaar}, {Pe{\~n}alosa Esteller}, {Penttil{\"a}}, {Pichon}, {Piersimoni}, {Pineau}, {Plachy}, {Plum}, {Poggio}, {Pr{\v{s}}a}, {Pulone}, {Racero}, {Ragaini}, {Rainer}, {Raiteri}, {Rambaux}, {Ramos}, {Ramos-Lerate}, {Re Fiorentin}, {Regibo}, {Richards}, {Rios Diaz}, {Ripepi}, {Riva}, {Rix}, {Rixon}, {Robichon}, {Robin}, {Robin}, {Roelens}, {Rogues}, {Rohrbasser}, {Romero-G{\'o}mez}, {Rowell}, {Royer}, {Ruz Mieres}, {Rybicki}, {Sadowski}, {S{\'a}ez N{\'u}{\~n}ez}, {Sagrist{\`a} Sell{\'e}s}, {Sahlmann}, {Salguero}, {Samaras}, {Sanchez Gimenez}, {Sanna}, {Santove{\~n}a}, {Sarasso}, {Schultheis}, {Sciacca}, {Segol}, {Segovia}, {S{\'e}gransan}, {Semeux}, {Shahaf}, {Siddiqui}, {Siebert},
  {Siltala}, {Silvelo}, {Slezak}, {Slezak}, {Smart}, {Snaith}, {Solano}, {Solitro}, {Souami}, {Souchay}, {Spagna}, {Spina}, {Spoto}, {Steele}, {Steidelm{\"u}ller}, {Stephenson}, {S{\"u}veges}, {Surdej}, {Szabados}, {Szegedi-Elek}, {Taris}, {Taylor}, {Teixeira}, {Tolomei}, {Tonello}, {Torra}, {Torra}, {Torralba Elipe}, {Trabucchi}, {Tsounis}, {Turon}, {Ulla}, {Unger}, {Vaillant}, {van Dillen}, {van Reeven}, {Vanel}, {Vecchiato}, {Viala}, {Vicente}, {Voutsinas}, {Weiler}, {Wevers}, {Wyrzykowski}, {Yoldas}, {Yvard}, {Zhao}, {Zorec}, {Zucker}, \& {Zwitter}}]{GaiaCollaboration2023A&A...674A...1G}
{Gaia Collaboration}, {Vallenari}, A., {Brown}, A.~G.~A., {et~al.} 2023, \aap, 674, A1, \dodoi{10.1051/0004-6361/202243940}

\bibitem[{{Gauza} {et~al.}(2015){Gauza}, {B{\'e}jar}, {P{\'e}rez-Garrido}, {Zapatero Osorio}, {Lodieu}, {Rebolo}, {Pall{\'e}}, \& {Nowak}}]{Gauza2015ApJ...804...96G}
{Gauza}, B., {B{\'e}jar}, V. J.~S., {P{\'e}rez-Garrido}, A., {et~al.} 2015, \apj, 804, 96, \dodoi{10.1088/0004-637X/804/2/96}

\bibitem[{{Getley} {et~al.}(2017){Getley}, {Carter}, {King}, \& {O'Toole}}]{Getley2017MNRAS.468.2932G}
{Getley}, A.~K., {Carter}, B., {King}, R., \& {O'Toole}, S. 2017, \mnras, 468, 2932, \dodoi{10.1093/mnras/stx604}

\bibitem[{{Getley} {et~al.}(2021){Getley}, {Carter}, {King}, \& {O'Toole}}]{Getley2021MNRAS.504.4291G}
---. 2021, \mnras, 504, 4291, \dodoi{10.1093/mnras/stab1207}

\bibitem[{{Goldberg} {et~al.}(2023){Goldberg}, {Fabrycky}, {Martin}, {Albrecht}, {Deeg}, \& {Nowak}}]{Goldberg2023MNRAS.525.4628G}
{Goldberg}, M., {Fabrycky}, D., {Martin}, D.~V., {et~al.} 2023, \mnras, 525, 4628, \dodoi{10.1093/mnras/stad2568}

\bibitem[{{Gorda} \& {Svechnikov}(1998)}]{Gorda1998}
{Gorda}, S.~Y., \& {Svechnikov}, M.~A. 1998, Astronomy Reports, 42, 793

\bibitem[{{Graham} {et~al.}(2021){Graham}, {Fleming}, \& {Barnes}}]{Graham2021A&A...650A.178G}
{Graham}, D.~E., {Fleming}, D.~P., \& {Barnes}, R. 2021, \aap, 650, A178, \dodoi{10.1051/0004-6361/202038940}

\bibitem[{{Guerrero} {et~al.}(2021){Guerrero}, {Seager}, {Huang}, {Vanderburg}, {Garcia Soto}, {Mireles}, {Hesse}, {Fong}, {Glidden}, {Shporer}, {Latham}, {Collins}, {Quinn}, {Burt}, {Dragomir}, {Crossfield}, {Vanderspek}, {Fausnaugh}, {Burke}, {Ricker}, {Daylan}, {Essack}, {G{\"u}nther}, {Osborn}, {Pepper}, {Rowden}, {Sha}, {Villanueva}, {Yahalomi}, {Yu}, {Ballard}, {Batalha}, {Berardo}, {Chontos}, {Dittmann}, {Esquerdo}, {Mikal-Evans}, {Jayaraman}, {Krishnamurthy}, {Louie}, {Mehrle}, {Niraula}, {Rackham}, {Rodriguez}, {Rowden}, {Sousa-Silva}, {Watanabe}, {Wong}, {Zhan}, {Zivanovic}, {Christiansen}, {Ciardi}, {Swain}, {Lund}, {Mullally}, {Fleming}, {Rodriguez}, {Boyd}, {Quintana}, {Barclay}, {Col{\'o}n}, {Rinehart}, {Schlieder}, {Clampin}, {Jenkins}, {Twicken}, {Caldwell}, {Coughlin}, {Henze}, {Lissauer}, {Morris}, {Rose}, {Smith}, {Tenenbaum}, {Ting}, {Wohler}, {Bakos}, {Bean}, {Berta-Thompson}, {Bieryla}, {Bouma}, {Buchhave}, {Butler}, {Charbonneau}, {Doty}, {Ge}, {Holman}, {Howard}, {Kaltenegger}, {Kane},
  {Kjeldsen}, {Kreidberg}, {Lin}, {Minsky}, {Narita}, {Paegert}, {P{\'a}l}, {Palle}, {Sasselov}, {Spencer}, {Sozzetti}, {Stassun}, {Torres}, {Udry}, \& {Winn}}]{Guerrero2021ApJS..254...39G}
{Guerrero}, N.~M., {Seager}, S., {Huang}, C.~X., {et~al.} 2021, \apjs, 254, 39, \dodoi{10.3847/1538-4365/abefe1}

\bibitem[{{G{\"u}nther} \& {Daylan}(2019)}]{allesfitter-code}
{G{\"u}nther}, M.~N., \& {Daylan}, T. 2019, {Allesfitter: Flexible Star and Exoplanet Inference From Photometry and Radial Velocity}, Astrophysics Source Code Library.
\newblock \doeprint{1903.003}

\bibitem[{{G{\"u}nther} \& {Daylan}(2021)}]{allesfitter-paper}
---. 2021, \apjs, 254, 13, \dodoi{10.3847/1538-4365/abe70e}

\bibitem[{{Haghighipour} \& {Kaltenegger}(2013)}]{Haghighipour2013ApJ...777..166H}
{Haghighipour}, N., \& {Kaltenegger}, L. 2013, \apj, 777, 166, \dodoi{10.1088/0004-637X/777/2/166}

\bibitem[{{Hamers} {et~al.}(2016){Hamers}, {Perets}, \& {Portegies Zwart}}]{Hamers2016MNRAS.455.3180H}
{Hamers}, A.~S., {Perets}, H.~B., \& {Portegies Zwart}, S.~F. 2016, \mnras, 455, 3180, \dodoi{10.1093/mnras/stv2447}

\bibitem[{{Han} {et~al.}(2017){Han}, {Udalski}, {Gould}, {Lee}, {Shvartzvald}, {Zang}, {Mao}, {Koz{\l}owski}, {Albrow}, {Chung}, {Hwang}, {Jung}, {Kim}, {Kim}, {Ryu}, {Shin}, {Yee}, {Zhu}, {Cha}, {Kim}, {Kim}, {Lee}, {Park}, {Kmtnet Collaboration}, {Skowron}, {Mr{\'o}z}, {Pietrukowicz}, {Poleski}, {Szyma{\'n}ski}, {Soszy{\'n}ski}, {Ulaczyk}, {Pawlak}, {Ogle Collaboration}, {Beichman}, {Bryden}, {Novati}, {Gaudi}, {Henderson}, {Howell}, {Jacklin}, {Ukirt Microlensing Team}, {Penny}, {Fouqu{\'e}}, {Wang}, \& {Cfht-K2C9 Microlensing Collaboration}}]{Han2017AJ....154..223H}
{Han}, C., {Udalski}, A., {Gould}, A., {et~al.} 2017, \aj, 154, 223, \dodoi{10.3847/1538-3881/aa9179}

\bibitem[{{Han} {et~al.}(2020){Han}, {Lee}, {Udalski}, {Gould}, {Bond}, {AUTHORS}, {Albrow}, {Chung}, {Hwang}, {Jung}, {Ryu}, {Shin}, {Shvartzvald}, {Yee}, {Zang}, {Cha}, {Kim}, {Kim}, {Kim}, {Lee}, {Lee}, {Park}, {Pogge}, {Jee}, {Kim}, {KMTNET COLLABORATION}, {Mr{\'o}z}, {Szyma{\'n}ski}, {Skowron}, {Poleski}, {Soszy{\'n}ski}, {Pietrukowicz}, {Koz{\l}owski}, {Ulaczyk}, {Rybicki}, {Iwanek}, {Wrona}, {OGLE COLLABORATION}, {Abe}, {Barry}, {Bennett}, {Bhattacharya}, {Donachie}, {Fujii}, {Fukui}, {Itow}, {Hirao}, {Kamei}, {Kondo}, {Koshimoto}, {Li}, {Matsubara}, {Muraki}, {Miyazaki}, {Nagakane}, {Ranc}, {Rattenbury}, {Suematsu}, {Sullivan}, {Sumi}, {Suzuki}, {Tristram}, {Yamakawa}, {Yonehara}, \& {THE MOA Collaboration}}]{Han2020AJ....159...48H}
{Han}, C., {Lee}, C.-U., {Udalski}, A., {et~al.} 2020, \aj, 159, 48, \dodoi{10.3847/1538-3881/ab5db9}

\bibitem[{{Han} {et~al.}(2024){Han}, {Udalski}, {Jung}, {Gould}, {Kim}, {Albrow}, {Chung}, {Hwang}, {Lee}, {Ryu}, {Shvartzvald}, {Shin}, {Yee}, {Yang}, {Zang}, {Cha}, {Kim}, {Kim}, {Lee}, {Lee}, {Park}, {Pogge}, {Mr{\'o}z}, {Mr{\'o}z}, {Szyma{\'n}ski}, {Skowron}, {Poleski}, {Soszy{\'n}ski}, {Pietrukowicz}, {Koz{\l}owski}, {Rybicki}, {Iwanek}, {Ulaczyk}, {Wrona}, \& {Gromadzki}}]{Han2024AA...685A..16H}
{Han}, C., {Udalski}, A., {Jung}, Y.~K., {et~al.} 2024, \aap, 685, A16, \dodoi{10.1051/0004-6361/202348791}

\bibitem[{{Han} {et~al.}(2018){Han}, {Qian}, {Zhu}, {Zhi}, {Dong}, {Soonthornthum}, {Poshyachinda}, {Sarotsakulchai}, {Fang}, {Wang}, \& {Voloshina}}]{Han2018ApJ...868...53H}
{Han}, Z.~T., {Qian}, S.~B., {Zhu}, L.~Y., {et~al.} 2018, \apj, 868, 53, \dodoi{10.3847/1538-4357/aae64d}

\bibitem[{Harris {et~al.}(2020)Harris, Millman, van~der Walt, Gommers, Virtanen, Cournapeau, Wieser, Taylor, Berg, Smith, Kern, Picus, Hoyer, van Kerkwijk, Brett, Haldane, del R{\'{i}}o, Wiebe, Peterson, G{\'{e}}rard-Marchant, Sheppard, Reddy, Weckesser, Abbasi, Gohlke, \& Oliphant}]{Numpy_harris2020array}
Harris, C.~R., Millman, K.~J., van~der Walt, S.~J., {et~al.} 2020, Nature, 585, 357, \dodoi{10.1038/s41586-020-2649-2}

\bibitem[{Hunter(2007)}]{Matplotlib_Hunter:2007}
Hunter, J.~D. 2007, Computing in Science \& Engineering, 9, 90, \dodoi{10.1109/MCSE.2007.55}

\bibitem[{{Jain} {et~al.}(2017){Jain}, {Paul}, {Sharma}, {Jaleel}, \& {Dutta}}]{Jain2017MNRAS.468L.118J}
{Jain}, C., {Paul}, B., {Sharma}, R., {Jaleel}, A., \& {Dutta}, A. 2017, \mnras, 468, L118, \dodoi{10.1093/mnrasl/slx039}

\bibitem[{{Janson} {et~al.}(2019){Janson}, {Asensio-Torres}, {Andr{\'e}}, {Bonnefoy}, {Delorme}, {Reffert}, {Desidera}, {Langlois}, {Chauvin}, {Gratton}, {Bohn}, {Eriksson}, {Marleau}, {Mamajek}, {Vigan}, \& {Carson}}]{Janson2019AA...626A..99J}
{Janson}, M., {Asensio-Torres}, R., {Andr{\'e}}, D., {et~al.} 2019, \aap, 626, A99, \dodoi{10.1051/0004-6361/201935687}

\bibitem[{{Janson} {et~al.}(2021){Janson}, {Gratton}, {Rodet}, {Vigan}, {Bonnefoy}, {Delorme}, {Mamajek}, {Reffert}, {Stock}, {Marleau}, {Langlois}, {Chauvin}, {Desidera}, {Ringqvist}, {Mayer}, {Viswanath}, {Squicciarini}, {Meyer}, {Samland}, {Petrus}, {Helled}, {Kenworthy}, {Quanz}, {Biller}, {Henning}, {Mesa}, {Engler}, \& {Carson}}]{Janson2021Natur.600..231J}
{Janson}, M., {Gratton}, R., {Rodet}, L., {et~al.} 2021, \nat, 600, 231, \dodoi{10.1038/s41586-021-04124-8}

\bibitem[{{Kane} \& {Hinkel}(2013)}]{Kane2013ApJ...762....7K}
{Kane}, S.~R., \& {Hinkel}, N.~R. 2013, \apj, 762, 7, \dodoi{10.1088/0004-637X/762/1/7}

\bibitem[{{Kostov} {et~al.}(2014){Kostov}, {McCullough}, {Carter}, {Deleuil}, {D{\'\i}az}, {Fabrycky}, {H{\'e}brard}, {Hinse}, {Mazeh}, {Orosz}, {Tsvetanov}, \& {Welsh}}]{Kostov2014ApJ...784...14K}
{Kostov}, V.~B., {McCullough}, P.~R., {Carter}, J.~A., {et~al.} 2014, \apj, 784, 14, \dodoi{10.1088/0004-637X/784/1/14}

\bibitem[{{Kostov} {et~al.}(2016){Kostov}, {Orosz}, {Welsh}, {Doyle}, {Fabrycky}, {Haghighipour}, {Quarles}, {Short}, {Cochran}, {Endl}, {Ford}, {Gregorio}, {Hinse}, {Isaacson}, {Jenkins}, {Jensen}, {Kane}, {Kull}, {Latham}, {Lissauer}, {Marcy}, {Mazeh}, {M{\"u}ller}, {Pepper}, {Quinn}, {Ragozzine}, {Shporer}, {Steffen}, {Torres}, {Windmiller}, \& {Borucki}}]{Kostov2016}
{Kostov}, V.~B., {Orosz}, J.~A., {Welsh}, W.~F., {et~al.} 2016, \apj, 827, 86, \dodoi{10.3847/0004-637X/827/1/86}

\bibitem[{{Kostov} {et~al.}(2020){Kostov}, {Orosz}, {Feinstein}, {Welsh}, {Cukier}, {Haghighipour}, {Quarles}, {Martin}, {Montet}, {Torres}, {Triaud}, {Barclay}, {Boyd}, {Briceno}, {Cameron}, {Correia}, {Gilbert}, {Gill}, {Gillon}, {Haqq-Misra}, {Hellier}, {Dressing}, {Fabrycky}, {Furesz}, {Jenkins}, {Kane}, {Kopparapu}, {Hod{\v{z}}i{\'c}}, {Latham}, {Law}, {Levine}, {Li}, {Lintott}, {Lissauer}, {Mann}, {Mazeh}, {Mardling}, {Maxted}, {Eisner}, {Pepe}, {Pepper}, {Pollacco}, {Quinn}, {Quintana}, {Rowe}, {Ricker}, {Rose}, {Seager}, {Santerne}, {S{\'e}gransan}, {Short}, {Smith}, {Standing}, {Tokovinin}, {Trifonov}, {Turner}, {Twicken}, {Udry}, {Vanderspek}, {Winn}, {Wolf}, {Ziegler}, {Ansorge}, {Barnet}, {Bergeron}, {Huten}, {Pappa}, \& {van der Straeten}}]{Kostov2020AJ....159..253K}
{Kostov}, V.~B., {Orosz}, J.~A., {Feinstein}, A.~D., {et~al.} 2020, \aj, 159, 253, \dodoi{10.3847/1538-3881/ab8a48}

\bibitem[{{Kostov} {et~al.}(2021){Kostov}, {Powell}, {Orosz}, {Welsh}, {Cochran}, {Collins}, {Endl}, {Hellier}, {Latham}, {MacQueen}, {Pepper}, {Quarles}, {Sairam}, {Torres}, {Wilson}, {Bergeron}, {Boyce}, {Bieryla}, {Buchheim}, {Ben Christiansen}, {Ciardi}, {Collins}, {Conti}, {Dixon}, {Guerra}, {Haghighipour}, {Herman}, {Hintz}, {Howard}, {Jensen}, {Kielkopf}, {Kruse}, {Law}, {Martin}, {Maxted}, {Montet}, {Murgas}, {Nelson}, {Olmschenk}, {Otero}, {Quimby}, {Richmond}, {Schwarz}, {Shporer}, {Stassun}, {Stephens}, {Triaud}, {Ulowetz}, {Walter}, {Wiley}, {Wood}, {Yenawine}, {Agol}, {Barclay}, {Beatty}, {Boisse}, {Caldwell}, {Christiansen}, {Col{\'o}n}, {Deleuil}, {Doyle}, {Fausnaugh}, {F{\H{u}}r{\'e}sz}, {Gilbert}, {H{\'e}brard}, {James}, {Jenkins}, {Kane}, {Kidwell}, {Kopparapu}, {Li}, {Lissauer}, {Lund}, {Majewski}, {Mazeh}, {Quinn}, {Quintana}, {Ricker}, {Rodriguez}, {Rowe}, {Santerne}, {Schlieder}, {Seager}, {Standing}, {Stevens}, {Ting}, {Vanderspek}, \& {Winn}}]{Kostov2021AJ....162..234K}
{Kostov}, V.~B., {Powell}, B.~P., {Orosz}, J.~A., {et~al.} 2021, \aj, 162, 234, \dodoi{10.3847/1538-3881/ac223a}

\bibitem[{{Kostov} {et~al.}(2022){Kostov}, {Powell}, {Rappaport}, {Borkovits}, {Gagliano}, {Jacobs}, {Kristiansen}, {LaCourse}, {Omohundro}, {Orosz}, {Schmitt}, {Schwengeler}, {Terentev}, {Torres}, {Barclay}, {Friedman}, {Kruse}, {Olmschenk}, {Vanderburg}, \& {Welsh}}]{Kostov2022ApJS..259...66K}
{Kostov}, V.~B., {Powell}, B.~P., {Rappaport}, S.~A., {et~al.} 2022, \apjs, 259, 66, \dodoi{10.3847/1538-4365/ac5458}

\bibitem[{{Kostov} {et~al.}(2025){Kostov}, {Powell}, {Fornear}, {Di Fraia}, {Gagliano}, {Jacobs}, {de Lambilly}, {Durantini Luca}, {Majewski}, {Omohundro}, {Orosz}, {Rappaport}, {Salik}, {Short}, {Welsh}, {Alexandrov}, {da Silva}, {Dunning}, {G{\"u}hne}, {Huten}, {Hyogo}, {Iannone}, {Lee}, {Magliano}, {Sharma}, {Tarr}, {Yablonsky}, {Acharya}, {Adams}, {Barclay}, {Montet}, {Mullally}, {Olmschenk}, {Pr{\v{s}}a}, {Quintana}, {Wilson}, {Balcioglu}, {Kruse}, \& {The Eclipsing Binary Patrol Collaboration}}]{Kostov2025ApJS..279...50K}
{Kostov}, V.~B., {Powell}, B.~P., {Fornear}, A.~U., {et~al.} 2025, \apjs, 279, 50, \dodoi{10.3847/1538-4365/ade2d8}

\bibitem[{{Kruse} {et~al.}(2021){Kruse}, {Powell}, {Kostov}, {Schnittman}, \& {Quintana}}]{Kruse2021tsc2.confE.163K}
{Kruse}, E., {Powell}, B., {Kostov}, V., {Schnittman}, J., \& {Quintana}, E. 2021, in Posters from the TESS Science Conference II (TSC2), 163, \dodoi{10.5281/zenodo.5131355}

\bibitem[{{Kuang} {et~al.}(2022){Kuang}, {Zang}, {Jung}, {Udalski}, {Yang}, {Mao}, {Albrow}, {Chung}, {Gould}, {Han}, {Hwang}, {Ryu}, {Shin}, {Shvartzvald}, {Yee}, {Cha}, {Kim}, {Kim}, {Kim}, {Lee}, {Lee}, {Lee}, {Park}, {Pogge}, {Mr{\'o}z}, {Skowron}, {Poleski}, {Szyma{\'n}ski}, {Soszy{\'n}ski}, {Pietrukowicz}, {Koz{\l}owski}, {Ulaczyk}, {Rybicki}, {Iwanek}, {Wrona}, {Gromadzki}, {Wang}, {Huang}, \& {Zhu}}]{Kuang2022MNRAS.516.1704K}
{Kuang}, R., {Zang}, W., {Jung}, Y.~K., {et~al.} 2022, \mnras, 516, 1704, \dodoi{10.1093/mnras/stac2315}

\bibitem[{{Kuzuhara} {et~al.}(2011){Kuzuhara}, {Tamura}, {Ishii}, {Kudo}, {Nishiyama}, \& {Kandori}}]{Kuzuhara2011AJ....141..119K}
{Kuzuhara}, M., {Tamura}, M., {Ishii}, M., {et~al.} 2011, \aj, 141, 119, \dodoi{10.1088/0004-6256/141/4/119}

\bibitem[{Levenberg(1944)}]{Levenberg1944}
Levenberg, K. 1944, Quarterly of Applied Mathematics, 2, 164, \dodoi{10.1090/qam/10666}

\bibitem[{{Lightkurve Collaboration} {et~al.}(2018){Lightkurve Collaboration}, {Cardoso}, {Hedges}, {Gully-Santiago}, {Saunders}, {Cody}, {Barclay}, {Hall}, {Sagear}, {Turtelboom}, {Zhang}, {Tzanidakis}, {Mighell}, {Coughlin}, {Bell}, {Berta-Thompson}, {Williams}, {Dotson}, \& {Barentsen}}]{Lightkurve2018ascl.soft12013L}
{Lightkurve Collaboration}, {Cardoso}, J.~V.~d.~M., {Hedges}, C., {et~al.} 2018, {Lightkurve: Kepler and TESS time series analysis in Python}, Astrophysics Source Code Library.
\newblock \doeprint{1812.013}

\bibitem[{{Lindegren} {et~al.}(2021){Lindegren}, {Klioner}, {Hern{\'a}ndez}, {Bombrun}, {Ramos-Lerate}, {Steidelm{\"u}ller}, {Bastian}, {Biermann}, {de Torres}, {Gerlach}, {Geyer}, {Hilger}, {Hobbs}, {Lammers}, {McMillan}, {Stephenson}, {Casta{\~n}eda}, {Davidson}, {Fabricius}, {Gracia-Abril}, {Portell}, {Rowell}, {Teyssier}, {Torra}, {Bartolom{\'e}}, {Clotet}, {Garralda}, {Gonz{\'a}lez-Vidal}, {Torra}, {Abbas}, {Altmann}, {Anglada Varela}, {Balaguer-N{\'u}{\~n}ez}, {Balog}, {Barache}, {Becciani}, {Bernet}, {Bertone}, {Bianchi}, {Bouquillon}, {Brown}, {Bucciarelli}, {Busonero}, {Butkevich}, {Buzzi}, {Cancelliere}, {Carlucci}, {Charlot}, {Cioni}, {Crosta}, {Crowley}, {del Peloso}, {del Pozo}, {Drimmel}, {Esquej}, {Fienga}, {Fraile}, {Gai}, {Garcia-Reinaldos}, {Guerra}, {Hambly}, {Hauser}, {Jan{\ss}en}, {Jordan}, {Kostrzewa-Rutkowska}, {Lattanzi}, {Liao}, {Licata}, {Lister}, {L{\"o}ffler}, {Marchant}, {Masip}, {Mignard}, {Mints}, {Molina}, {Mora}, {Morbidelli}, {Murphy}, {Pagani}, {Panuzzo}, {Pe{\~n}alosa
  Esteller}, {Poggio}, {Re Fiorentin}, {Riva}, {Sagrist{\`a} Sell{\'e}s}, {Sanchez Gimenez}, {Sarasso}, {Sciacca}, {Siddiqui}, {Smart}, {Souami}, {Spagna}, {Steele}, {Taris}, {Utrilla}, {van Reeven}, \& {Vecchiato}}]{RUWE2021A&A...649A...2L}
{Lindegren}, L., {Klioner}, S.~A., {Hern{\'a}ndez}, J., {et~al.} 2021, \aap, 649, A2, \dodoi{10.1051/0004-6361/202039709}

\bibitem[{{Lomb}(1976)}]{Lomb1976Ap&SS..39..447L}
{Lomb}, N.~R. 1976, \apss, 39, 447, \dodoi{10.1007/BF00648343}

\bibitem[{{Marcadon} \& {Pr{\v{s}}a}(2024)}]{Marcadon2024arXiv240307694M}
{Marcadon}, F., \& {Pr{\v{s}}a}, A. 2024, arXiv e-prints, arXiv:2403.07694, \dodoi{10.48550/arXiv.2403.07694}

\bibitem[{Marquardt(1963)}]{Marquardtdoi:10.1137/0111030}
Marquardt, D.~W. 1963, Journal of the Society for Industrial and Applied Mathematics, 11, 431, \dodoi{10.1137/0111030}

\bibitem[{{Martin} \& {Triaud}(2014)}]{Martin2014A&A...570A..91M}
{Martin}, D.~V., \& {Triaud}, A. H.~M.~J. 2014, \aap, 570, A91, \dodoi{10.1051/0004-6361/201323112}

\bibitem[{{Martin} {et~al.}(2019){Martin}, {Triaud}, {Udry}, {Marmier}, {Maxted}, {Collier Cameron}, {Hellier}, {Pepe}, {Pollacco}, {S{\'e}gransan}, \& {West}}]{Martin2019A&A...624A..68M}
{Martin}, D.~V., {Triaud}, A. H.~M.~J., {Udry}, S., {et~al.} 2019, \aap, 624, A68, \dodoi{10.1051/0004-6361/201833669}

\bibitem[{{Marzari} \& {Thebault}(2019)}]{Marzari2019Galax...7...84M}
{Marzari}, F., \& {Thebault}, P. 2019, Galaxies, 7, 84, \dodoi{10.3390/galaxies7040084}

\bibitem[{{Mikul{\'a}{\v{s}}ek} {et~al.}(2014){Mikul{\'a}{\v{s}}ek}, {Chrastina}, {Li{\v{s}}ka}, {Zejda}, {Jan{\'\i}k}, {Zhu}, \& {Qian}}]{Mikulasek2014CoSka..43..382M}
{Mikul{\'a}{\v{s}}ek}, Z., {Chrastina}, M., {Li{\v{s}}ka}, J., {et~al.} 2014, Contributions of the Astronomical Observatory Skalnate Pleso, 43, 382

\bibitem[{{Mitnyan} {et~al.}(2024){Mitnyan}, {Borkovits}, {Czavalinga}, {Rappaport}, {P{\'a}l}, {Powell}, \& {Hajdu}}]{Mitnyan2024A&A...685A..43M}
{Mitnyan}, T., {Borkovits}, T., {Czavalinga}, D.~R., {et~al.} 2024, \aap, 685, A43, \dodoi{10.1051/0004-6361/202348909}

\bibitem[{{Moharana} {et~al.}(2024){Moharana}, {He{\l}miniak}, {Marcadon}, {Pawar}, {Pawar}, {Garczy{\'n}ski}, {Per{\l}a}, {Koz{\l}owski}, {Sybilski}, {Ratajczak}, \& {Konacki}}]{Moharana2024MNRAS.527...53M}
{Moharana}, A., {He{\l}miniak}, K.~G., {Marcadon}, F., {et~al.} 2024, \mnras, 527, 53, \dodoi{10.1093/mnras/stad3117}

\bibitem[{Mordasini(2018)}]{Mordasini2018}
Mordasini, C. 2018, Planetary Population Synthesis (Cham: Springer International Publishing), 2425--2474, \dodoi{10.1007/978-3-319-55333-7_143}

\bibitem[{Newville {et~al.}(2025)Newville, Otten, Nelson, Stensitzki, Ingargiola, Allan, Fox, Carter, \& Rawlik}]{newville_2025_16175987}
Newville, M., Otten, R., Nelson, A., {et~al.} 2025, LMFIT: Non-Linear Least-Squares Minimization and Curve-Fitting for Python, 1.3.4,  Zenodo, \dodoi{10.5281/zenodo.16175987}

\bibitem[{{Orosz} {et~al.}(2012{\natexlab{a}}){Orosz}, {Welsh}, {Carter}, {Brugamyer}, {Buchhave}, {Cochran}, {Endl}, {Ford}, {MacQueen}, {Short}, {Torres}, {Windmiller}, {Agol}, {Barclay}, {Caldwell}, {Clarke}, {Doyle}, {Fabrycky}, {Geary}, {Haghighipour}, {Holman}, {Ibrahim}, {Jenkins}, {Kinemuchi}, {Li}, {Lissauer}, {Pr{\v{s}}a}, {Ragozzine}, {Shporer}, {Still}, \& {Wade}}]{Orosz2012ApJ...758...87O}
{Orosz}, J.~A., {Welsh}, W.~F., {Carter}, J.~A., {et~al.} 2012{\natexlab{a}}, \apj, 758, 87, \dodoi{10.1088/0004-637X/758/2/87}

\bibitem[{{Orosz} {et~al.}(2012{\natexlab{b}}){Orosz}, {Welsh}, {Carter}, {Fabrycky}, {Cochran}, {Endl}, {Ford}, {Haghighipour}, {MacQueen}, {Mazeh}, {Sanchis-Ojeda}, {Short}, {Torres}, {Agol}, {Buchhave}, {Doyle}, {Isaacson}, {Lissauer}, {Marcy}, {Shporer}, {Windmiller}, {Barclay}, {Boss}, {Clarke}, {Fortney}, {Geary}, {Holman}, {Huber}, {Jenkins}, {Kinemuchi}, {Kruse}, {Ragozzine}, {Sasselov}, {Still}, {Tenenbaum}, {Uddin}, {Winn}, {Koch}, \& {Borucki}}]{Orosz2012Sci...337.1511O}
---. 2012{\natexlab{b}}, Science, 337, 1511, \dodoi{10.1126/science.1228380}

\bibitem[{{Orosz} {et~al.}(2019){Orosz}, {Welsh}, {Haghighipour}, {Quarles}, {Short}, {Mills}, {Satyal}, {Torres}, {Agol}, {Fabrycky}, {Jontof-Hutter}, {Windmiller}, {M{\"u}ller}, {Hinse}, {Cochran}, {Endl}, {Ford}, {Mazeh}, \& {Lissauer}}]{Orosz2019AJ....157..174O}
{Orosz}, J.~A., {Welsh}, W.~F., {Haghighipour}, N., {et~al.} 2019, \aj, 157, 174, \dodoi{10.3847/1538-3881/ab0ca0}

\bibitem[{pandas~development team(2024)}]{the_pandas_development_team_2024_10697587}
pandas~development team, T. 2024, pandas-dev/pandas: Pandas, v2.2.1,  Zenodo, \dodoi{10.5281/zenodo.10697587}

\bibitem[{{Papageorgiou} {et~al.}(2021){Papageorgiou}, {Catelan}, {Christopoulou}, {Drake}, \& {Djorgovski}}]{Papageorgiou2021MNRAS.503.2979P}
{Papageorgiou}, A., {Catelan}, M., {Christopoulou}, P.-E., {Drake}, A.~J., \& {Djorgovski}, S.~G. 2021, \mnras, 503, 2979, \dodoi{10.1093/mnras/stab646}

\bibitem[{{Potter} {et~al.}(2011){Potter}, {Romero-Colmenero}, {Ramsay}, {Crawford}, {Gulbis}, {Barway}, {Zietsman}, {Kotze}, {Buckley}, {O'Donoghue}, {Siegmund}, {McPhate}, {Welsh}, \& {Vallerga}}]{Potter2011MNRAS.416.2202P}
{Potter}, S.~B., {Romero-Colmenero}, E., {Ramsay}, G., {et~al.} 2011, \mnras, 416, 2202, \dodoi{10.1111/j.1365-2966.2011.19198.x}

\bibitem[{{Pribulla } {et~al.}(2012){Pribulla }, {Va{\v{n}}ko}, {Ammler-von Eiff}, {Andreev}, {Aslant{\"u}rk}, {Awadalla}, {Balu{\v{d}}ansk{\'y}}, {Bonanno}, {Bo{\v{z}}i{\'c}}, {Catanzaro}, {{\c{C}}elik}, {Christopoulou}, {Covino}, {Cusano}, {Dimitrov}, {Dubovsk{\'y}}, {Eigmueller}, {Esmer}, {Frasca}, {Hamb{\'a}lek}, {Hanna}, {Hanslmeier}, {Kalomeni}, {Kjurkchieva}, {Krushevska}, {Kudzej}, {Kundra}, {Kuznyetsova}, {Lee}, {Leitzinger}, {Maciejewski}, {Moldovan}, {Morais}, {Mugrauer}, {Neuh{\"a}user}, {Niedzielski}, {Odert}, {Ohlert}, {{\"O}zavc{\i}}, {Papageorgiou}, {Parimucha}, {Poddan{\'y}}, {Pop}, {Raetz}, {Raetz}, {Romanyuk}, {Ru{\v{z}}djak}, {Schulz}, {{\c{S}}enavc{\i}}, {Srdoc}, {Szalai}, {Sz{\'e}kely}, {Sudar}, {Tezcan}, {T{\"o}r{\"u}n}, {Turcu}, {Vince}, \& {Zejda}}]{Pribulla2012AN....333..754P}
{Pribulla }, T., {Va{\v{n}}ko}, M., {Ammler-von Eiff}, M., {et~al.} 2012, Astronomische Nachrichten, 333, 754, \dodoi{10.1002/asna.201211722}

\bibitem[{{Pr{\v{s}}a} {et~al.}(2022){Pr{\v{s}}a}, {Kochoska}, {Conroy}, {Eisner}, {Hey}, {IJspeert}, {Kruse}, {Fleming}, {Johnston}, {Kristiansen}, {LaCourse}, {Mortensen}, {Pepper}, {Stassun}, {Torres}, {Abdul-Masih}, {Chakraborty}, {Gagliano}, {Guo}, {Hambleton}, {Hong}, {Jacobs}, {Jones}, {Kostov}, {Lee}, {Omohundro}, {Orosz}, {Page}, {Powell}, {Rappaport}, {Reed}, {Schnittman}, {Schwengeler}, {Shporer}, {Terentev}, {Vanderburg}, {Welsh}, {Caldwell}, {Doty}, {Jenkins}, {Latham}, {Ricker}, {Seager}, {Schlieder}, {Shiao}, {Vanderspek}, \& {Winn}}]{Prsa2022}
{Pr{\v{s}}a}, A., {Kochoska}, A., {Conroy}, K.~E., {et~al.} 2022, \apjs, 258, 16, \dodoi{10.3847/1538-4365/ac324a}

\bibitem[{{Qian} {et~al.}(2010){Qian}, {Liao}, {Zhu}, \& {Dai}}]{Qian2010ApJ...708L..66Q}
{Qian}, S.~B., {Liao}, W.~P., {Zhu}, L.~Y., \& {Dai}, Z.~B. 2010, \apjl, 708, L66, \dodoi{10.1088/2041-8205/708/1/L66}

\bibitem[{{Qian} {et~al.}(2012{\natexlab{a}}){Qian}, {Liu}, {Zhu}, {Dai}, {Fern{\'a}ndez Laj{\'u}s}, \& {Baume}}]{Qian2012MNRAS.422L..24Q}
{Qian}, S.~B., {Liu}, L., {Zhu}, L.~Y., {et~al.} 2012{\natexlab{a}}, \mnras, 422, L24, \dodoi{10.1111/j.1745-3933.2012.01228.x}

\bibitem[{{Qian} {et~al.}(2012{\natexlab{b}}){Qian}, {Zhu}, {Dai}, {Fern{\'a}ndez-Laj{\'u}s}, {Xiang}, \& {He}}]{Qian2012ApJ...745L..23Q}
{Qian}, S.~B., {Zhu}, L.~Y., {Dai}, Z.~B., {et~al.} 2012{\natexlab{b}}, \apjl, 745, L23, \dodoi{10.1088/2041-8205/745/2/L23}

\bibitem[{{Qian} {et~al.}(2011){Qian}, {Liu}, {Liao}, {Li}, {Zhu}, {Dai}, {He}, {Zhao}, {Zhang}, \& {Li}}]{Qian2011MNRAS.414L..16Q}
{Qian}, S.~B., {Liu}, L., {Liao}, W.~P., {et~al.} 2011, \mnras, 414, L16, \dodoi{10.1111/j.1745-3933.2011.01045.x}

\bibitem[{{Quarles} {et~al.}(2018){Quarles}, {Satyal}, {Kostov}, {Kaib}, \& {Haghighipour}}]{Quarles2018ApJ...856..150Q}
{Quarles}, B., {Satyal}, S., {Kostov}, V., {Kaib}, N., \& {Haghighipour}, N. 2018, \apj, 856, 150, \dodoi{10.3847/1538-4357/aab264}

\bibitem[{{Rein} \& {Liu}(2012)}]{Rein2012_REBOUND}
{Rein}, H., \& {Liu}, S.~F. 2012, \aap, 537, A128, \dodoi{10.1051/0004-6361/201118085}

\bibitem[{{Rein} \& {Spiegel}(2015)}]{Rein2015_IAS15}
{Rein}, H., \& {Spiegel}, D.~S. 2015, \mnras, 446, 1424, \dodoi{10.1093/mnras/stu2164}

\bibitem[{{Ricker} {et~al.}(2015){Ricker}, {Winn}, {Vanderspek}, {Latham}, {Bakos}, {Bean}, {Berta-Thompson}, {Brown}, {Buchhave}, {Butler}, {Butler}, {Chaplin}, {Charbonneau}, {Christensen-Dalsgaard}, {Clampin}, {Deming}, {Doty}, {De Lee}, {Dressing}, {Dunham}, {Endl}, {Fressin}, {Ge}, {Henning}, {Holman}, {Howard}, {Ida}, {Jenkins}, {Jernigan}, {Johnson}, {Kaltenegger}, {Kawai}, {Kjeldsen}, {Laughlin}, {Levine}, {Lin}, {Lissauer}, {MacQueen}, {Marcy}, {McCullough}, {Morton}, {Narita}, {Paegert}, {Palle}, {Pepe}, {Pepper}, {Quirrenbach}, {Rinehart}, {Sasselov}, {Sato}, {Seager}, {Sozzetti}, {Stassun}, {Sullivan}, {Szentgyorgyi}, {Torres}, {Udry}, \& {Villasenor}}]{Ricker2015JATIS...1a4003R}
{Ricker}, G.~R., {Winn}, J.~N., {Vanderspek}, R., {et~al.} 2015, Journal of Astronomical Telescopes, Instruments, and Systems, 1, 014003, \dodoi{10.1117/1.JATIS.1.1.014003}

\bibitem[{{Scargle}(1982)}]{Scargle1982ApJ...263..835S}
{Scargle}, J.~D. 1982, \apj, 263, 835, \dodoi{10.1086/160554}

\bibitem[{{Schanche} {et~al.}(2019){Schanche}, {Collier Cameron}, {Almenara}, {Alsubai}, {Anderson}, {Armstrong}, {Barkaoui}, {Barros}, {Bochi{\'n}ski}, {Bonomo}, {Bouchy}, {Brown}, {Burdanov}, {Busuttil}, {Deleuil}, {Delrez}, {Faedi}, {Gillon}, {Hay}, {Hebb}, {H{\'e}brard}, {Jehin}, {Kolb}, {Maxted}, {Miller}, {Nielsen}, {Pollacco}, {Pozuelos}, {Queloz}, {Relles}, {Smalley}, {Triaud}, {Udry}, {West}, \& {Wheatley}}]{Schanche2019}
{Schanche}, N., {Collier Cameron}, A., {Almenara}, J.~M., {et~al.} 2019, \mnras, 488, 4905, \dodoi{10.1093/mnras/stz2064}

\bibitem[{Scholz \& Stephens(1987)}]{Scholz1987}
Scholz, F.~W., \& Stephens, M.~A. 1987, Journal of the American Statistical Association, 82, 918, \dodoi{10.2307/2288805}

\bibitem[{{Schwamb} {et~al.}(2013){Schwamb}, {Orosz}, {Carter}, {Welsh}, {Fischer}, {Torres}, {Howard}, {Crepp}, {Keel}, {Lintott}, {Kaib}, {Terrell}, {Gagliano}, {Jek}, {Parrish}, {Smith}, {Lynn}, {Simpson}, {Giguere}, \& {Schawinski}}]{Schwamb2013ApJ...768..127S}
{Schwamb}, M.~E., {Orosz}, J.~A., {Carter}, J.~A., {et~al.} 2013, \apj, 768, 127, \dodoi{10.1088/0004-637X/768/2/127}

\bibitem[{{Sigurdsson} {et~al.}(2003){Sigurdsson}, {Richer}, {Hansen}, {Stairs}, \& {Thorsett}}]{Sigurdsson2003Sci...301..193S}
{Sigurdsson}, S., {Richer}, H.~B., {Hansen}, B.~M., {Stairs}, I.~H., \& {Thorsett}, S.~E. 2003, Science, 301, 193, \dodoi{10.1126/science.1086326}

\bibitem[{{Simonetti} {et~al.}(2020){Simonetti}, {Vladilo}, {Silva}, \& {Sozzetti}}]{Simonetti2020ApJ...903..141S}
{Simonetti}, P., {Vladilo}, G., {Silva}, L., \& {Sozzetti}, A. 2020, \apj, 903, 141, \dodoi{10.3847/1538-4357/abc074}

\bibitem[{{Socia} {et~al.}(2020){Socia}, {Welsh}, {Orosz}, {Cochran}, {Endl}, {Quarles}, {Short}, {Torres}, {Windmiller}, \& {Yenawine}}]{Socia2020AJ....159...94S}
{Socia}, Q.~J., {Welsh}, W.~F., {Orosz}, J.~A., {et~al.} 2020, \aj, 159, 94, \dodoi{10.3847/1538-3881/ab665b}

\bibitem[{{Song} {et~al.}(2019){Song}, {Mai}, {Mutel}, {Pulley}, {Faillace}, \& {Watkins}}]{Song2019AJ....157..184S}
{Song}, S., {Mai}, X., {Mutel}, R.~L., {et~al.} 2019, \aj, 157, 184, \dodoi{10.3847/1538-3881/ab1139}

\bibitem[{{Standing} {et~al.}(2023){Standing}, {Sairam}, {Martin}, {Triaud}, {Correia}, {Coleman}, {Baycroft}, {Kunovac}, {Boisse}, {Cameron}, {Dransfield}, {Faria}, {Gillon}, {Hara}, {Hellier}, {Howard}, {Lane}, {Mardling}, {Maxted}, {Miller}, {Nelson}, {Orosz}, {Pepe}, {Santerne}, {Sebastian}, {Udry}, \& {Welsh}}]{Standing2023NatAs...7..702S}
{Standing}, M.~R., {Sairam}, L., {Martin}, D.~V., {et~al.} 2023, Nature Astronomy, 7, 702, \dodoi{10.1038/s41550-023-01948-4}

\bibitem[{{Stassun} \& {Torres}(2021)}]{Stassun2021ApJ...907L..33S}
{Stassun}, K.~G., \& {Torres}, G. 2021, \apjl, 907, L33, \dodoi{10.3847/2041-8213/abdaad}

\bibitem[{{Sybilski} {et~al.}(2010){Sybilski}, {Konacki}, \& {Koz{\l}owski}}]{Sybilski2010MNRAS.405..657S}
{Sybilski}, P., {Konacki}, M., \& {Koz{\l}owski}, S.~K. 2010, \mnras, 405, 657, \dodoi{10.1111/j.1365-2966.2010.16490.x}

\bibitem[{{Tofflemire} {et~al.}(2023){Tofflemire}, {Kraus}, {Mann}, {Newton}, {Gully-Santiago}, {Vanderburg}, {Waalkes}, {Berta-Thompson}, {Collins}, {Collins}, {Nielsen}, {Bouchy}, {Ziegler}, {Brice{\~n}o}, \& {Law}}]{Tofflemire2023}
{Tofflemire}, B.~M., {Kraus}, A.~L., {Mann}, A.~W., {et~al.} 2023, \aj, 165, 46, \dodoi{10.3847/1538-3881/aca60f}

\bibitem[{{VanderPlas}(2018)}]{VanderPlas2018ApJS..236...16V}
{VanderPlas}, J.~T. 2018, \apjs, 236, 16, \dodoi{10.3847/1538-4365/aab766}

\bibitem[{Virtanen {et~al.}(2020)Virtanen, Gommers, Oliphant, Haberland, Reddy, Cournapeau, Burovski, Peterson, Weckesser, Bright, {van der Walt}, Brett, Wilson, Millman, Mayorov, Nelson, Jones, Kern, Larson, Carey, Polat, Feng, Moore, {VanderPlas}, Laxalde, Perktold, Cimrman, Henriksen, Quintero, Harris, Archibald, Ribeiro, Pedregosa, {van Mulbregt}, \& {SciPy 1.0 Contributors}}]{2020SciPy-NMeth}
Virtanen, P., Gommers, R., Oliphant, T.~E., {et~al.} 2020, Nature Methods, 17, 261, \dodoi{10.1038/s41592-019-0686-2}

\bibitem[{{von Boetticher} {et~al.}(2017){von Boetticher}, {Triaud}, {Queloz}, {Gill}, {Lendl}, {Delrez}, {Anderson}, {Collier Cameron}, {Faedi}, {Gillon}, {G{\'o}mez Maqueo Chew}, {Hebb}, {Hellier}, {Jehin}, {Maxted}, {Martin}, {Pepe}, {Pollacco}, {S{\'e}gransan}, {Smalley}, {Udry}, \& {West}}]{vonBoetticher2017}
{von Boetticher}, A., {Triaud}, A. H.~M.~J., {Queloz}, D., {et~al.} 2017, \aap, 604, L6, \dodoi{10.1051/0004-6361/201731107}

\bibitem[{{von Boetticher} {et~al.}(2019){von Boetticher}, {Triaud}, {Queloz}, {Gill}, {Maxted}, {Almleaky}, {Anderson}, {Bouchy}, {Burdanov}, {Collier Cameron}, {Delrez}, {Ducrot}, {Faedi}, {Gillon}, {G{\'o}mez Maqueo Chew}, {Hebb}, {Hellier}, {Jehin}, {Lendl}, {Marmier}, {Martin}, {McCormac}, {Pepe}, {Pollacco}, {S{\'e}gransan}, {Smalley}, {Thompson}, {Turner}, {Udry}, {Van Grootel}, \& {West}}]{vonBoetticher2019}
---. 2019, \aap, 625, A150, \dodoi{10.1051/0004-6361/201834539}

\bibitem[{{Wang} \& {Liu}(2024)}]{Wang2024AJ....168...31W}
{Wang}, M.-T., \& {Liu}, H.-G. 2024, \aj, 168, 31, \dodoi{10.3847/1538-3881/ad4a60}

\bibitem[{{Welsh} {et~al.}(2012){Welsh}, {Orosz}, {Carter}, {Fabrycky}, {Ford}, {Lissauer}, {Pr{\v{s}}a}, {Quinn}, {Ragozzine}, {Short}, {Torres}, {Winn}, {Doyle}, {Barclay}, {Batalha}, {Bloemen}, {Brugamyer}, {Buchhave}, {Caldwell}, {Caldwell}, {Christiansen}, {Ciardi}, {Cochran}, {Endl}, {Fortney}, {Gautier}, {Gilliland}, {Haas}, {Hall}, {Holman}, {Howard}, {Howell}, {Isaacson}, {Jenkins}, {Klaus}, {Latham}, {Li}, {Marcy}, {Mazeh}, {Quintana}, {Robertson}, {Shporer}, {Steffen}, {Windmiller}, {Koch}, \& {Borucki}}]{Welsh2012Natur.481..475W}
{Welsh}, W.~F., {Orosz}, J.~A., {Carter}, J.~A., {et~al.} 2012, \nat, 481, 475, \dodoi{10.1038/nature10768}

\bibitem[{{Welsh} {et~al.}(2015){Welsh}, {Orosz}, {Short}, {Cochran}, {Endl}, {Brugamyer}, {Haghighipour}, {Buchhave}, {Doyle}, {Fabrycky}, {Hinse}, {Kane}, {Kostov}, {Mazeh}, {Mills}, {M{\"u}ller}, {Quarles}, {Quinn}, {Ragozzine}, {Shporer}, {Steffen}, {Tal-Or}, {Torres}, {Windmiller}, \& {Borucki}}]{Welsh2015ApJ...809...26W}
{Welsh}, W.~F., {Orosz}, J.~A., {Short}, D.~R., {et~al.} 2015, \apj, 809, 26, \dodoi{10.1088/0004-637X/809/1/26}

\bibitem[{{W}es {M}c{K}inney(2010)}]{Pandas_mckinney-proc-scipy-2010}
{W}es {M}c{K}inney. 2010, in {P}roceedings of the 9th {P}ython in {S}cience {C}onference, ed. {S}t\'efan van~der {W}alt \& {J}arrod {M}illman, 56 -- 61, \dodoi{10.25080/Majora-92bf1922-00a}

\bibitem[{{Windemuth} {et~al.}(2019){Windemuth}, {Agol}, {Carter}, {Ford}, {Haghighipour}, {Orosz}, \& {Welsh}}]{Windemuth2019MNRAS.490.1313W}
{Windemuth}, D., {Agol}, E., {Carter}, J., {et~al.} 2019, \mnras, 490, 1313, \dodoi{10.1093/mnras/stz2637}

\bibitem[{{Zhu} {et~al.}(2019){Zhu}, {Qian}, {Fern{\'a}ndez Laj{\'u}s}, {Wang}, \& {Li}}]{Zhu2019RAA....19..134Z}
{Zhu}, L.-Y., {Qian}, S.-B., {Fern{\'a}ndez Laj{\'u}s}, E., {Wang}, Z.-H., \& {Li}, L.-J. 2019, Research in Astronomy and Astrophysics, 19, 134, \dodoi{10.1088/1674-4527/19/9/134}

\bibitem[{{Zorotovic} \& {Schreiber}(2013)}]{Zorotovic2013A&A...549A..95Z}
{Zorotovic}, M., \& {Schreiber}, M.~R. 2013, \aap, 549, A95, \dodoi{10.1051/0004-6361/201220321}

\end{thebibliography}
\bibliographystyle{aasjournal}

\end{document}